\documentclass[%
reprint,
amsmath,amssymb,
aps,superscriptaddress
]{revtex4-2}

\usepackage{graphicx}
\usepackage{dcolumn}
\usepackage{bm}
\usepackage[T1]{fontenc}
\usepackage{color,soul}
\usepackage{tikz}
\usepackage{hyperref}
\usetikzlibrary{shapes,arrows,chains,matrix}
\usepackage{tikz-cd} 
\newtheorem{theorem}{Theorem}
\newtheorem{corollary}{Corollary}[theorem]
\newtheorem{lemma}{Lemma}

\begin{document}
	
	\preprint{APS/123-QED}
	
	
	\title{General teleportation channel in Fermionic Quantum Theory}

	\author{Sanam Khan}\email{sanamkhan@igcar.gov.in}\affiliation{Homi Bhabha National Institute, Training School Complex, Anushaktinagar, Mumbai 400094, India}\affiliation{Computer Division, Indira Gandhi Centre for Atomic Research, Kalpakkam 603102,India}
    
	\author{R. Jehadeesan}\affiliation{Computer Division, Indira Gandhi Centre for Atomic Research, Kalpakkam 603102,India}
	\author{Sibasish Ghosh}\email{sibasish@imsc.res.in}\affiliation{Optics \& Quantum Information Group, The Institute of Mathematical Sciences, C. I. T. Campus, Taramani, Chennai 600113, India}\affiliation{Homi Bhabha National Institute, Training School Complex, Anushaktinagar, Mumbai 400094, India}


	

	
	\begin{abstract}
		Quantum Teleportation is a very useful scheme for transferring quantum information. Given that the quantum information is encoded in a state of a system of distinguishable particles, and given that the shared bi-partite entangled state is also that of a system of distinguishable particles,  the {\it optimal teleportation fidelity} of the shared state is known to be $(F_{max}d+1)/(d+1)$ with $F_{max}$ being the `maximal singlet fraction' of the shared state. However, Parity Superselection Rule (PSSR) in Fermionic Quantum Theory (FQT) puts constraint on the allowed set of physical states and operations, and thereby, leads to a different notion of Quantum entanglement - locally accessible and locally inaccessible (topological correlation). In the present work, we derive an expression for the {\it optimal teleportation fidelity} of locally accessible entanglement preservation, given that the quantum information to be teleported is encoded in fermionic modes of dimension \(2^N \times 2^N\) using \(2^N \times 2^N\)-dim shared fermionic resource between the sender and receiver. To get the optimal teleportation fidelity in FQT, we introduce PSSR restricted twirling operations and establish fermionic state-channel isomorphism. Remarkably, we notice that the structure of the canonical form of twirl invariant fermionic shared state differs from that of the {\it isotropic state} -- the corresponding canonical invariant form for teleportation in Standard Quantum Theory (SQT). In this context, we also introduce restricted Clifford twirling operation that constitute the Unitary 2-design in case of FQT for experimentally validating such optimal average fidelity. Finally, we discuss the preservation of locally inaccessible entanglement for a class of fermionic teleportation channel.

	\end{abstract}
	
	\maketitle
	
	
	\section{\label{sec:level1}Introduction}
	
	Quantum Teleportation (QT) is a quantum information transmission task, first introduced by Bennett \textit{et al.} \cite{bennett1993} where the sender and the receiver utilizes a pre-shared Einstein-Podolsky-Rosen (EPR)-correlated pairs \cite{epr} of particles. The first experimental demonstration of QT was due to Bouwmeester \textit{et al.} \cite{Boumeester1997} where the resource state used is entangled photon generated using parametric down conversion. Apart from sending unknown quantum state of a particle, QT also find its application in quantum computation. Gottesman and  Chuang, in their seminal paper \cite{Gottesman1999}, showed the essence of QT towards achieving fault-tolerant quantum computing. Also, the success probability of quantum gate operations in the linear optics quantum computation (LOQC) setup can be increased using non-deterministically prepared entangled states and QT \cite{knill2001}. Recent advancement in QT \cite{Langenfeld2021} also finds it's application towards quantum networks \cite{Avis2023}. 
	
	The question of finding optimal QT fidelity was first answered by  Horodecki \textit{et al.} \cite{horodecki99_2} where the authors have shown that the optimal teleportation fidelity is related to the maximal singlet fraction of the resource state (generally, noisy) shared between sender and receiver. The authors utilized local twirling operations to reduce any bi-partite state to the form of isotropic state which is also physically isomorphic to a depolarizing channel in SQT -- the corresponding reduced channel under the local twirling on  channel operation. Such optimal teleportation has recently been used for benchmarking the performance of noisy intermediate-scale quantum (NISQ) devices \cite{NISQ2022}.
	
	In case, information, encoded in a system of indistinguishable particles, needs to be teleported efficiently, one generally needs to consider an `entangled' state of a bi-partite system consisting of a system of indistinguishable particles (see, for example, ref. \cite{Lohmann}). Entanglement of indistinguishable particles has been at the centre of debate \cite{BENATTI20201} since last two decades. Based on the definition of subsytem accessible to local observers, two different perspectives have been put forward -- \textit{particle local} and \textit{mode local}. In \textit{particle local} picture, it was pointed out that due to the symmetrized and antisymmetrized structures of the respective Hilbert spaces, the von-neumann entropy of the reduced density matrix obtained through partial trace, behaves differently for Bosons and Fermions \cite{Ghirardi2004}. However, this problem can be overcome in the \(C^{\ast}\)-algebraic formalism by substituting the partial trace operation with the restrictions of state to the corresponding subalgebras, and  applying Gelfand–Naimark–Segal (GNS) construction \cite{Balachandran}. Analysis of mode entanglement utilizes the second quantization method, and hence, the total number of particles need not be fixed. In mode local picture, the entanglement is defined relative to a given partitioning of subsystems. This can be demonstrated using Bogoliubov transformation \cite{zanardi}. Such mode local picture has also found it's application to quantify the quantum mutual information between orbitals in quantum chemistry \cite{Ding2021}.
	
	Fermionic entanglement has been extensively studied both in particle picture \cite{Ghirardi2004,Schliemann,Li2001,Iemini2013} as well as in mode picture \cite{moriya2002,wolf1,friis2013,Benatti14}. Fermionic entanglement as a resource has been utilised in various quantum information processing tasks such as fermionic teleportation, superdense coding, thermodynamic processes and quantum metrology \cite{f_teleport,Gigena17,Esposito23,Benatti14}. Experimental demonstration of fermionic teleportation has become a reality using Solid-state system \cite{Steffen2013,Qiao2020}. In mode picture -- which, we are going to use in the present work -- the set of allowed fermionic states and operations must adhere to the parity superselection rule (PSSR) \cite{SSRfriis,PhysRevA.104.032411,Dariano2014}. Practical demonstration of such systems is possible with Majorana zero modes(MZMs), which is believed to be found in fractional quantum hall system \cite{moore1991,Nayak2005,Nayak2013} and semiconductor nanowire \cite{Mourik2012,sarma2010}. In this regard, A toy model (also known as Kitaev chain) has been proposed that exhibit localized zero-energy Majorana modes using 1-D p-wave superconductivity \cite{Kitaev_2001}. It has been demonstrated that, the ground state of the Kitaev chain has two-fold degeneracy with zero and one fermion (which is created by combining two MZM localized at the end of the chain). Such degenerate ground states can by used to implement topological qubit with braiding for quantum gate operation \cite{mzm2015}.  Even though the experimental demonstration of MZM still remains challanging \cite{tanaka2024}, MZM based teleportation has been simulated using superconducting qubits \cite{pan2021}.

	In this work, we attempt to answer the question of finding out the `optimal teleportation fidelity' in FQT using the help of mode local picture and wedge product formalism \cite{PhysRevA.104.032411}. We follow the similar approach as that of finding the optimal teleportation fidelity in SQT \cite{horodecki99_2}. 
	However, due to local PSSR constraints on the observables, entanglement of fermionic resource state  can be categorized as accessible and inaccessible entanglement. The latter is known as topological correlation \cite{zhou2023}. From now on we will call the locally accessible entanglement as the fermionic entanglement.
	
	Here, we adhere to the operational notion of teleportation to understand the effect of fermionic resource state with entanglement and topological correlation on the optimal teleportation fidelity. We consider subsystem teleportation, where one of the two $N$-modes subsystems of an \(N \times N\) -mode arbitrary bi-partite fermionic state is being teleported, and the fidelity of this teleportation is measured operationally. In this regard, we introduce the notion of `entanglement fidelity' (to be defined later) for fermionic system. The main results of this paper include the canonical structure of shared fermionic state under PSSR-respected local twirling operations and optimal fidelity of fermionic teleportation in the presence of topological correlation, but considering the entanglement of the shared fermionic state (\(N \times N\)-mode) as a resource.
	
	The rest of the paper is structured as follows. In Sec.\ref{sec:level2}, we introduce the mathematical structure of fermionic Hibert space and the associated Fock space. We discuss the importance of PSSR in FQT, and the restrictions on states and operations.We also discuss the notion of Entanglement and topological correlation in FQT with PSSR. In Sec.\ref{sec:level3}, we prove the exact teleportation scheme for an \(N \times N\) -mode resource state. In Sec.\ref{sec:level4}, we derive the structure of the `reduced' fermionic state under local twirling operation by introducing a Haar measure over a PSSR-respected Unitary group. In Sec.\ref{sec:level5}, we construct an Unitary-2 design for this restricted Haar measure using PSSR-respected Clifford group operations. We prove the fermionic State-Channel Isomorphism in Sec.\ref{sec:level6} and introduce the notion of fermionic channel twirling operation. In Sec.\ref{sec:level7}, we define the average entanglement fidelity for fermionic channel and derive the optimal fermionic teleportation fidelity. In Sec.\ref{sec:level9} we discuss teleportation of topological correlation for a class of fermionic channel. We provide concluding remarks in  Sec.\ref{sec:level10}.

	\section{\label{sec:level2}PRELIMINARIES}
	\subsection{Fermionic Hilbert Space}
	For an $N$-mode fermionic system, consider $f_j$ and $f_j^{\dagger}$ to be the annihilation and creation operators respectively, satisfying the relations
	\[\{f_i,f_j\}=0 \ \ \ \{f_i^{\dagger},f_j^{\dagger}\}=0 \ \ \ \{f_i,f_j^{\dagger}\}=\delta_{ij}\mathbb{I} \]
	with $i,j\in \{1, 2, \ldots, N\} \equiv [1,N]$.
	
	In this work, we will utilize wedge-product Hilbert space instead of tensor-product Hilbert space for a bi-partite system. Details about the relevant properties of wedge-product Hilbert spaces can be found in \cite{PhysRevA.104.032411}. For example, if $\mathcal{H_{PQ}}$ denotes the wedge-product Hilbert space of $\mathcal{H_P}$ and $\mathcal{H_Q}$ then we write $\mathcal{H_{PQ}}=\mathcal{H_P} \wedge \mathcal{H_Q}$. 
	
	Consider the Hilbert space $\mathcal{H_A}$ of an $N$-mode fermionic system A with each mode spanning a 2-dimensional subspace. Let us assume that the basis for each of these subspaces be given by $\{|\Omega\rangle, |j\rangle\}$ where \[|j\rangle=f_j^{\dagger}|\Omega\rangle\] and $|\Omega\rangle$ being vacuum state (for $j = 1, 2, \ldots, N$). Extending this basis over two modes using wedge product of Hilbert spaces, we get $\{|\Omega\rangle, \ |i\rangle,\ |j\rangle,\ |i\rangle \wedge |j\rangle\}$ as a basis for $\mathcal{H}_{ij}=\mathcal{H}_i \wedge \mathcal{H}_j$. Here, $\mathcal{H}_i$ and $\mathcal{H}_j$ correspond to the Hilbert spaces of $i$th and $j$th mode respectively along with the constraint $i < j$.
	
	Thus, the global Hilbert space of an $N$-mode fermionic system is given by \[ \mathcal{H_A}=\mathcal{H}_1\wedge\mathcal{H}_2\wedge \dots  \wedge\mathcal{H}_N
	\]i.e. the wedge-product of all single mode subspaces. One can define a basis for $\mathcal{H_A}$ \cite{PhysRevA.104.032411} as \begin{eqnarray}
		\mathcal{B_A}=\{|j_1\rangle\wedge|j_2\rangle&&\wedge\dots \wedge |j_k\rangle \nonumber\\
		&&\ | \ j_1 <j_2 <\dots< j_k , \ k\in [0,N]\}.  
		\label{eq:1}
	\end{eqnarray}
	Each $j_1,j_2,\dots,j_k$ in Eq.~(\ref{eq:1}) is associated with a particular mode. Notice, $k=0$ gives us the state $|\Omega{\rangle}$. In terms of fermionic creation operators, the aforesaid basis elements can also be written as \[
	|j_1\rangle\wedge|j_2\rangle\wedge\dots \wedge |j_k\rangle=f_{j_1}^{\dagger}f_{j_2}^{\dagger} \dots f_{j_k}^{\dagger}|\Omega\rangle.
	\] Dimension of this $N$-mode fermionic Hilbert space $\mathcal{H}$ is $2^N$.
	
	As per Pauli exclusion principle, each mode can be occupied by maximum one fermion. Basis elements generated by even number of fermions are denoted by $|E_i\rangle$ while elements generated by odd number of fermions are denoted by $|O_i\rangle$ with \(i\) ranging from \(1\) to \(2^{N-1}\). In this new notation, the basis $\mathcal{B_A}$ can be rewritten as 
	\begin{equation}
		\mathcal{B_A^P}=\{|E_1\rangle,|E_2\rangle,\dots,|E_{2^{N-1}}\rangle,|O_1\rangle,|O_2\rangle,\dots,|O_{2^{N-1}}\rangle\}
	\end{equation}
	Here, $|E_1\rangle \equiv |\Omega\rangle$, i.e. the ground state.
	
	\subsection{PSSR and physical states}
	PSSR imposes constraint on the form of a physical fermionic state. It states that, coherent superposition of fermionic state with even number of fermions and fermionic state with odd number of fermions is prohibited. For example, states $f_i^{\dagger}|\Omega\rangle  +|\Omega\rangle$  is not allowed due to having superposition of odd and even number of particles respectively. PSSR for spinor field was initially discussed in the context of special relativity \cite{PhysRev.88.101}. Recently, PSSR has been imposed on the states of fermionic systems to remove the problem of non-symmetric bipartitions in terms of eigenvalues \cite{SSRfriis} using the spin-statistics connection. To understand this, consider a fermionic state of the form $f_i^{\dagger}|\Omega\rangle+|\Omega\rangle$. Now, fermions being spin-\(\frac{1}{2}\) particles, such states upon the action of \(2\pi\) rotation goes to the state $-f_i^{\dagger}|\Omega\rangle + |\Omega\rangle$. This state is orthogonal to the state before the \(2\pi\) rotation. Therefore, if one demands the invariance of physics (in this case, via indistinguishability of states) upon \(2\pi\) rotation, one has to apply PSSR constraints \cite{pssr1968}.  
	
	Howerver it was argued that the necessity of imposing PSSR is profoundly related to only two aspects, invariance of laws of physics for different observers and the no-signalling principle \cite{SSR2}. The latter is related to the microcausality principle for separable Hilbert space, which states that the physical operation performed simultaneously at two differnt locations must commute with each other. The arguement for imposing PSSR constraint goes as follows. Consider a fermionic system with mode bi-partition \(\mathcal{M_A}\) and \(\mathcal{M_B}\). For any odd operators \(O_A\) and \(O_B\) acting on \(\mathcal{M_A}\) and \(\mathcal{M_B}\) is an odd monomial of fermionic creation and annihilation operators from the respective bi-partition. One then have the anti-commutation relation between \(O_A\) and \(O_B\). However, no-signalling principle demands the operators to commute, which is only possible if the operators of any one of the subsystems is restricted to be even monomial via PSSR constraint. Extending, this PSSR constraint to other subsystem using the invariance of laws of physics to all observers, we get PSSR constraint in both the subsystems.
	
	PSSR has also been {\it derived} in the framework of operational probabilistic theories for fermionic quantum systems following a set of assumptions \cite{Dariano2014}.
	
	Now consider a mixed state $\rho$ of an $N$-mode fermionic system. Due to PSSR, the most general form of $\rho$ in the basis  $\mathcal{B_A^P}$ is given by \begin{equation}
		\rho=\sum_{i=1,j=1}^{i=2^{N-1}, j=2^{N-1}}A_{ij}|E_i\rangle\langle E_j| + B_{ij}|O_i\rangle\langle O_j|
	\end{equation}
	with suitable  on A, B so that $\rho$ satisfies positivity and the trace condition ${\rm Tr} {\rho} = 1$. 
	
	To describe a physical state of a $N\times M$  -mode bi-partite fermionic system, where $N$ corresponds to fermionic modes of subsystem A and $M$ corresponds to fermionic modes of subsystem B, we extend the Basis  $\mathcal{B_A^P}$ over the Hilbert space $\mathcal{H_A} \wedge \mathcal{H_B}$. Accordingly, the extended basis $\mathcal{B_{AB}^P}$ will have the form 
	
	\begin{eqnarray}
		\mathcal{B_{AB}^P}=&&\bigg\{|E_i^A\rangle \wedge |E_j^B\rangle,|E_i^A\rangle \wedge |O_j^B\rangle , |O_i^A\rangle \wedge |E_j^B\rangle,\nonumber\\
		&&  |O_i^A\rangle \wedge |O_j^B\rangle \ \bigg| \ |E_{i(j)}^{A(B)}\rangle, |O_{i(j)}^{A(B)}\rangle \in \mathcal{B_{A(B)}^P}\bigg\}
	\end{eqnarray}.

	with $i$ ranging from $1$ to $2^{N-1}$ and $j$ ranging from $1$ to $2^{M-1}$. Using this extended basis, one can then express a generic PSSR-respected density matrix $W$ as
	
	\begin{widetext}
		\begin{eqnarray}
			W=&&\sum_{ijkl}a_{ijkl}|E_i\rangle\langle E_k|\wedge|E_j\rangle\langle E_l|+b_{ijkl}|E_i\rangle \langle O_k|\wedge|E_j\rangle\langle O_l|+c_{i j kl}|O_i\rangle \langle E_k|\wedge|O_j\rangle\langle E_l| +d_{ijkl} |O_i\rangle \langle O_k| \wedge| O_j\rangle\langle O_l| \nonumber\\
			&&+ A_{ijkl}|E_i\rangle \langle E_k|\wedge|O_j\rangle\langle O_l|+B_{i j k l}| E_i\rangle\langle O_k| \wedge| O_j\rangle\langle E_l| + C_{i j k l}|O_i\rangle \langle  E_k|\wedge|E_{j}\rangle \langle O_l|+D_{i j k l}| O_i\rangle \langle O_k| \wedge| E_j\rangle\langle E_l| 
			\label{eq:13}
		\end{eqnarray}
		with suitable conditions on the coefficients $a,b,c,d,A,B,C$ and $D$ such that $W>0$ and $tr(W)=1$.
	\end{widetext}
	\subsection{Operations on fermionic states}
	As described in \cite{PhysRevA.104.032411}, the most generic PSSR-respected unitary operation assumes block-diagonal form $U_{BD}$ in the basis $B'=\{B_e,B_o\}$ due to no-signalling principle where, $B_e$ ($B_o$) is the basis of even-parity (odd-parity) states of the fermionic system. However, one can always embed any anti-block diagonal unitary $U_{ABD}$ into the form of a block-diagonal unitary using an ancilla system \cite{f_teleport}. We explicitly prove this fact in the following lemma for dimension $2^N$.
	
	\begin{lemma}
		\label{lm1}
		Any anti-block diagonal unitary in the basis $\mathcal{B_A^P} \equiv \{B_e,B_o\}$ can always be embedded into the form of a block-diagonal unitary using a single ancilla system.
	\end{lemma}
	
	\textit{Proof.} Consider a single mode fermionic ancilla system $C$, with creation and annihilation operators described by $f_c^{\dagger}$ and $f_c$ respectively. Hence the basis can be written as $\{|\Omega_c\rangle, f_c^{\dagger}|\Omega_c\rangle\} \equiv \{|E_c\rangle, |O_c\rangle\}$. Let the structure of $U_{BD}$ and $U_{ABD}$ in the basis $\mathcal{B_A^P}$ be given by
	\begin{eqnarray}
		U_{BD}=&\sum_{ij}u_{ij}|E_i\rangle\langle E_j|+v_{ij}|O_i\rangle\langle O_j| \label{eq:3} \\ 
		U_{ABD}=&\sum_{ij}u'_{ij}|E_i\rangle\langle O_j|+v'_{ij}|O_i\rangle\langle E_j|   
		\label{eq:4}
	\end{eqnarray}
	Consider the following unitary $U_c=(f_c^{\dagger}+ f_c)$ on the ancillary mode. Then, any unitary of the form $U_C \wedge U_{ABD}$ will have the form of $U_{BD}$. $\blacksquare$
	
	\subsection{Fermionic Observables}
	In this work, we consider the physically allowed fermionic observable to be PSSR-respected. For a bi-partite system $\mathcal{AB}$, let $\hat{O}_P^A$ denote the set of all physical observables for system $A$ and $\hat{O}_P^B$ denote the set of all physical observables for system $B$. Then any product observables for system $\mathcal{AB}$ denoted by $\hat{O}_P^A \wedge \hat{O}_P^B$ respects local PSSR. Here,  local PSSR is defined as the PSSR on the subsystems of a given fermionic bi-partite state. 
	
	In general, any observable for system $A$ can be expressed in the basis $\mathcal{B_A^P}$ as
	\begin{equation}
		\hat{O}_P^A=\sum_{ij}\alpha_{ij}|E_i^A\rangle\langle E_j^A|+\beta_{ij}|O_i^A\rangle\langle O_j^A| 
	\end{equation}
	with suitable conditions on the coefficients $\alpha$ and $\beta$.
	
	\subsection{\label{sec:level8}Entanglement and topological correlation in fermionic system}
	Entanglement for the system of indistinguishable particles has been a topic of debate since the last decade \cite{BENATTI20201}. Moreover, for indistinguishable fermions there exist different notions of entanglement \cite{wolf1}. Following the discussions of \cite{PhysRevA.104.032411}, we define the operational fermionic product state as follows
	
	\textit{Definition 1.} A fermionic state $\rho^{AB}$ of a bi-partite system $\mathcal{AB}$ is said be a product state, if for any PSSR-respected observable $\hat{O}_P^A$ and $\hat{O}_P^B$ of system $A$ and system $B$ respectively, there exist $\rho^{A(B)}$ such that the following condition
	\begin{equation}
		Tr(\rho^{AB}(\hat{O}_P^A \wedge \hat{O}_P^B))=Tr(\rho^{A}\hat{O}_P^A ) \ Tr(\rho^{B}\hat{O}_P^B )
		\label{eq:2}
	\end{equation}
	satisfied, where $\rho^{A(B)}$ is a legitimate PSSR-respected density matrix for system $A(B)$. Notice that, in general, for any operational product state ${\rho}^{AB}$, we have $\rho^{AB} \neq \rho^{A} \wedge \rho^{B}$ due to the absence of local discriminability \cite{Dariano2014,D'Ariano_2014_2}. However, the following is always true for any local PSSR-respected product state.
	
	\begin{lemma}
		\label{lm2}
		For any local PSSR-respected fermionic state  $\rho^{AB}$ of a bi-partite system,
		\begin{eqnarray}
			Tr(\rho^{AB}(\hat{O}_P^A \wedge \hat{O}_P^B))=&&Tr(\rho^{A}\hat{O}_P^A ) \ Tr(\rho^{B}\hat{O}_P^B ) \nonumber\\  
			&&\Longleftrightarrow \nonumber \\
			\rho^{AB}=&&\rho^{A} \wedge \rho^{B} \ \ \ \ \forall \hat{O}_P^{A(B)}
		\end{eqnarray}    
		with $\rho^{A(B)}$ being legitimate PSSR-respected density matrix for system $A(B)$.
	\end{lemma} 
	
	The proof of the lemma \ref{lm2} is given in Appendix \ref{sec:a1}.
	
	A \textit{separable} state of fermionic system $\mathcal{AB}$ is the one which can be represented as a convex combination of product fermionic states $\{ \rho_i^{AB}\}$:
	\begin{equation}
		\rho_{sep}^{AB}=\sum_i p_i \rho_i^{AB}
		\label{eq:5}
	\end{equation}
	Any state which is not \textit{separable}, is said to be an \textit{entangled} state of the fermionic system $\mathcal{AB}$.
	Alternatively, one can also define the fermionic entangled state via local projection respecting parity\cite{wolf2007,Esposito23}. Notice, that the entanglement of the locally projected state is locally accessible.
	
	\textit{Example:-} Consider two different fermionic bi-partite states given by \[|\Psi_{AB}\rangle=\frac{1}{\sqrt{2}}(|E_1,E_1\rangle+|E_2,E_2\rangle)\]
	and
	\[|\Phi_{AB}\rangle=\frac{1}{\sqrt{2}}(|E_1,E_1\rangle+|O_1,O_1\rangle)\]
	Notice, that the locally projected (respecting local-PSSR) version of the state  \(|\Psi_{AB}\rangle\) cannot be written as convex combination of product states. Hence, it is entangled. On the other hand, the locally projected (respecting local-PSSR) version of the state  \(|\Phi_{AB}\rangle\) admits a convex combination of product states. Hence, it is not entangled. 
	
	However, the correlation of the state \(|\Phi_{AB}\rangle\) can be understood via the notion of topological correlation \cite{zhou2023}. For SQT, any unknown bi-partite state can be determined by Alice and Bob performing local quantum operations and classical communication (LOCC). On the other hand, FQT is constrained by PSSR, resulting in the non-local tomographic properties \cite{D'Ariano_2014_2}. Hence, there exist fermionic states which cannot be tomogrammed locally. Considering this, we define the following set \begin{eqnarray}
		\Theta(\rho_{AB}^f):=&&\{\tilde{\rho}_{AB}^f|\text{tr}(\rho_{AB}^f(\hat{O}_P^A \wedge \hat{O}_P^B))=\text{tr}(\tilde{\rho}_{AB}^f(\hat{O}_P^A \wedge \hat{O}_P^B))\nonumber\\
		&&\forall \hat{O}_P^{A(B)}  \}
	\end{eqnarray}
	The inferred fermionic state via local bi-partite quantum state tomography can then be defined using maximum entropy estimate \cite{jaynes1,jaynes2} as \[f(\rho_{AB}^f)=\arg\max_{\tilde{\rho}_{AB}^f \in \Theta}S(\tilde{\rho}_{AB}^f)\]  where $S()$ defines von Neumann entropy. Subsequently, the topological correlation is defined as \begin{equation}\delta S(\rho_{AB}):=S(f(\rho_{AB}))-S(\rho_{AB})  \label{eq:32} \end{equation} 
	Also, note that the topological correlation defined for anyonic system can be used for fermionic system by considering the fusion of two Ising anyons giving vacuum and fermion \cite{zhou2023,BONDERSON2017}.
	
		
		
		Fermionic entanglement can also be seen to be invariant under block-diagonal/anti-diagonal local unitary. In this regard we mention the following lemma. 
		
		\begin{lemma}
			\label{lm3}
			Transformation of a separable fermionic state into an entangled fermionic state is forbidden under the operation of local anti-block diagonal unitary. In other words,
			\[(U_{ABD} \wedge U'_{ABD(BD)}) \ \rho_{sep}^{AB} \ (U_{ABD}^{\dagger} \wedge {U'_{ABD(BD)}}^{\dagger}) \in SEP\]
			where $SEP$ denotes the set of all separable fermionic states. Notice that, the case of \((U_{BD} \wedge U_{BD})\) is trivial.
		\end{lemma} 
		
		Proof of lemma \ref{lm3} is given in Appendix \ref{sec:a1}. \\

       \section*{Summary of Main Results}
        \begin{itemize}
            \item Extension of single mode fermionic teleportation to \(N \times N\) mode fermionic teleportation.
            \item Introducing the notion of topological correlation as a resource for fermionic subsystem teleportation.
            \item Derivation of the canonical \(U \wedge U^{\ast}\) twirl invariant fermionic state and prove that PSSR restricted Clifford group is an exact Unitary 2-design.
            \item Establishment of the fermionic State-Channel isomorphism.
            \item Introduction of operational fidelity of fermionic channel and derivation of optimal teleportation fidelity in the presence of PSSR.
        \end{itemize}
		
		\section{\label{sec:level3}Fermionic teleportation for N\(\times\)N mode systems}
		Teleportation of complex amplitudes of even-parity and odd-parity vectors at the same time is prohibited. This makes a difference between the teleportation protocol in SQT and FQT. However, incoherent mixtures of even-parity vectors and odd parity vectors are allowed. Considering this fact, two different notions of fermionic teleportation have been discussed in \cite{f_teleport}:

		\begin{figure}[b]
			\includegraphics[width=0.9\linewidth]{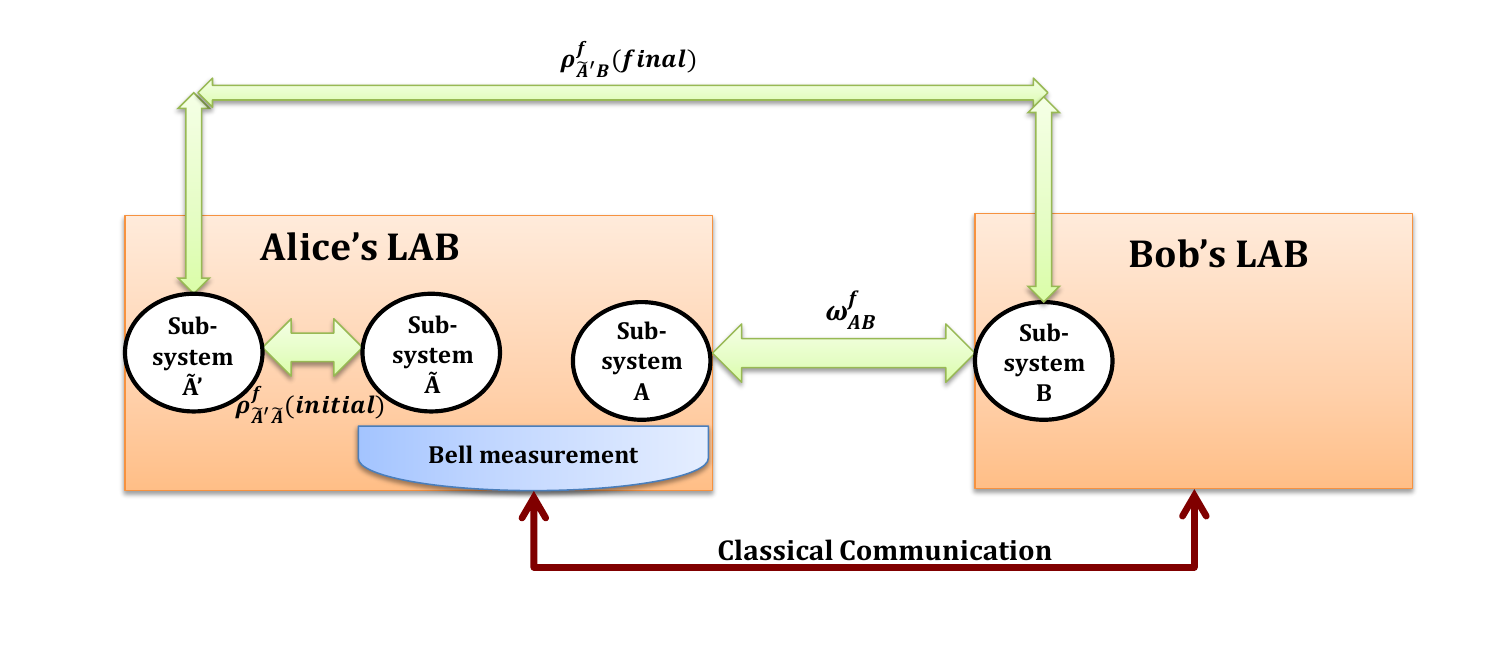}
			\caption{\label{fig:epsart} Teleportation in FQT. In this scheme \(\omega_{AB}^f\) is the resource state used to to teleport one half of the initial state \(\rho_{\tilde{A}\tilde{A}' }^f(initial)\) and produces the final state \(\rho_{B\tilde{A}'}^f(final)\).}
		\end{figure}
		
		\begin{enumerate}
			\item \textit{Fermionic single-mode exact teleportation} is associated with the procedure in which Alice starts with a bi-partite fermionic state \(|\psi_{\tilde{A}\tilde{A}'}\rangle\) (possibly entangled) and the information about the reduced state of Alice's side \(\rho_{\tilde{A}}\) is transferred to Bob's side such that the final state becomes \(|\psi_{B\tilde{A}'}\rangle\). This process utilizes a single copy of a two-mode fermionic state \(|\psi_{AB}\rangle\) given by \[|\psi_{AB}\rangle=\frac{1}{\sqrt{2}}(|E^A\rangle \wedge |E^B\rangle+|O^A\rangle \wedge |O^B\rangle)\]
			\item \textit{Fermionic two-mode teleportation} is associated with the procedure to transfer one qubit of information encoded in terms of complex amplitude \(\alpha\) and \(\beta\) (\(|\alpha|^2+|\beta|^2=1\)). However, to encode the information of \(\alpha\) and \(\beta\) using fermionic state, one need minimum two number of modes. Hence, one has to perform the \textit{Fermionic single-mode exact teleportation} twice to transmit the information encoded in terms of \(\alpha\) and \(\beta\). 
		\end{enumerate}
		Notice, that the entanglement of the shared resource for fermionic single-mode teleportation is inaccessible via PSSR respected LOCC. However, such states have non-zero topological correlation. Hence, single-mode teleportation demonstrates the resource value of topological correlation towards teleportation. On the other hand, in higher dimension, entanglement of shared resource state becomes accessible via PSSR respected LOCC.   
		
		In this paper, we generalize the concept of \textit{Fermionic single-mode teleportation} to N\(\times\)N mode system where we assume each fermionic system \(\tilde{A}', \tilde{A}, A\) and \(B\) consists of \(N\) fermionic modes and determine the optimal teleportation fidelity of entanglement preservation. The scheme is demonstrated in Fig.~\ref{fig:epsart} where information about the state of subsystem \(\tilde{A}\) is teleported to Bob using a pre-shared resource state \(\omega_{AB}^f\) between Alice and Bob. But before delve into the optimal fermionic teleportation, let us briefly discuss the teleportation protocol in d-dimensional Hilbert space as described in \cite{werner_teleport} in the context of teleporting subsystem.
		
		\subsection{Qudit Teleportation in SQT}
		In this QT protocol, four quantum systems are used. Consider the teleportation channel being applied to a part of the bi-partite state \(\rho_{ \tilde{A}' \tilde{A}}\) . Also, consider a pre-shared maximally entangled state \(|\omega \rangle \langle \omega |_{AB} \) between Alice and Bob. Hence the initial state can be written as 
		$\rho_{\tilde{A}' \tilde{A}} \otimes |\omega \rangle \langle \omega |_{AB}$. Then a successful teleportation is represented by the equation 
		\begin{eqnarray}
			&&\sum_{x\in X}\text{tr}((\rho_{\tilde{A}' \tilde{A}} \otimes |\omega \rangle \langle \omega |_{AB})(\hat{O}_{\tilde{A}'} \otimes M_x \otimes N_x (\hat{O}_B  ) )) \nonumber\\
			&&= \text{tr}(\rho_{ \tilde{A}' B} \hat{O}_{\tilde{A}'} \otimes \hat{O}_B )
			\label{eq:7}
		\end{eqnarray}
		with \(M_x= | \Psi_x\rangle\langle\Psi_x|_{\tilde{A} A}\) and \(\langle\Psi_x| \Psi_y\rangle = \delta_{xy}\). Here, \(\Psi_x\) constitutes the basis of maximally entangled states and \(X\) denotes a set of \(d^2\) elements.
		Also, \[N_x(\hat{O}_B )=U_x^{\dagger} (\hat{O}_B )U_x \] and 
		\[|\Psi_x\rangle\langle\Psi_x|=(U_x \otimes \mathbb{I})|\omega \rangle \langle \omega |(U_x^{\dagger} \otimes \mathbb{I})\] defines the \(U_x\) for each x. Hence, \(U_x\) also forms the bases of unitary operators. To prove Eq.~(\ref{eq:7}), let us proceed, as given in \cite{werner_teleport}, by choosing \(\rho_{\tilde{A}' \tilde{A}}=|\phi_2\rangle_{\tilde{A}'}\langle \phi_4|  \otimes |\phi_1\rangle_{\tilde{A}}\langle \phi_3|\), \( \hat{O}_B=|\psi_1\rangle_{B}\langle \psi_3|\) and \(\hat{O}_{\tilde{A}'}= |\psi_2\rangle_{\tilde{A}'}\langle \psi_4|\). Then, considering the \(x^{th} \) term in Eq.~(\ref{eq:7}) we get
		\begin{eqnarray}
			Term_x=&&\langle\phi_4 \otimes \phi_3 \otimes \omega | \psi_2 \otimes \Psi_x \otimes U_x^{\dagger} \psi_1\rangle \times \nonumber\\ && 
			\langle\psi_4 \otimes \Psi_x \otimes U_x^{\dagger} \psi_3 | \phi_2 \otimes \phi_1 \otimes \omega \rangle 
		\end{eqnarray}
		Now substituting \(|\Psi_x\rangle= (U_x \otimes \mathbb{I})|\omega \rangle\) and utilizing the inner product definition of tensor product Hilbert space we get 
		\[Term_x=d^{-2}\langle\phi_3|\psi_1\rangle\langle\psi_3|\phi_1\rangle\langle\phi_4|\psi_2\rangle\langle\psi_4|\phi_2\rangle\]
		Hence, summing over $x$ we get the desired result. 
		
		It is known that SQT satisfies local distinguishability of states \cite{dariano2010}. Hence, local operation and classical communications on the bi-partite system are sufficient to distinguish two different bi-partite states.
		
		\section{Optimal Teleportation in FQT} 
		While considering optimal teleportation in fermionic quantum theory, one has to consider the fact that the total correlation of the shared resource state has two part, entanglement and topological correlation as discussed in section ~\ref{sec:level8}. In this section, we consider the optimality of fermionic teleportation for entanglement preservation with entanglement as reource of the shared fermionic state. However, in the next section, we also discuss the effect of such optimal teleportation protocol on the input bi-partite state with topological correlation due to the presence of topological correlation in the shared resource state.   
		
		Following the approach by Horodecki et.al. \cite{horodecki99_2}, at first we determine the canonical form of fermionic bi-partite state invariant under local unitary twirling. We call them fermionic isotropic state. While deducing the family of twirl invariant state, we constructed the local-PSSR respected unitary 2-design by defining a restricted clifford group. Such unitary 2-designs are useful for experimentally validating the average fidelity of channel \cite{Dankert2009}. Establishing state-channel isomorphism for fermionic system, we show that any teleportation channel under twirling on channel operation reduces to a canonical form which is isomorphic to the fermionic isotropic state. Finally, the invariance of operational entanglement fidelity under channel twirling operation is used to derive the optimal teleportation fidelity.

		
		
		\subsection{\label{sec:level4}\(U \wedge U\) invariant fermionic state and reduction under unitary twirling}
		In this section we derive the family of fermionic states invariant under the $U \wedge U$ operations with $U$ satisfying Eq.~(\ref{eq:3}) and Eq.~(\ref{eq:4}). We follow the procedure mentioned in \cite{werner89,horodecki99}. Consider an operator $W$ on fermionic wedge product space of $N \times N $ mode system, given by, Eq.~(\ref{eq:13}). At first, let us impose the constraint 
		\begin{equation}
			(U_{BD} \wedge U_{BD}) \ W \ (U_{BD}^{\dagger} \wedge U_{BD}^{\dagger})=W 
			\label{eq:14}
		\end{equation}
		This gives us the first reduced form of $W$ with nonzero coefficients of type \(a_{iiii}\), \(a_{ijji}\), \(a_{ijij}\), \(d_{iiii}\), \(d_{ijji}\), \(d_{ijij}\), \(A_{iiii}\), \(A_{ijij}\), \(B_{iiii}\), \(B_{ijji}\), \(C_{iiii}\), \(C_{ijji}\), \(D_{iiii}\) and \(D_{ijij}\). For example, equating the coefficients of the basis \(|E_\alpha\rangle \wedge|E_\gamma\rangle\langle O_q|\wedge\langle O_s|\), we get the constraint equation \(b_{\alpha \gamma qs}=u_{\alpha \beta} u_{\gamma \delta} b_{\beta \delta pr} v_{qp}^{\ast}v_{sr}^{\ast}\). Now let us assume that, for a given unitary \(U_{BD}\), 
		\begin{eqnarray}
			u_{\alpha \beta}=0 \ \ \text{if} \ \ \alpha \neq \beta \ \ \text{and } \ \ u_{\alpha \beta}=1 \ \text{or} \ -1 \ \  \text{if} \ \ \alpha =\beta \nonumber\\  
			v_{\alpha \beta}=0 \ \ \text{if} \ \ \alpha \neq \beta \ \ \text{and } \ \ v_{\alpha \beta}=1 \ \text{or} \ -1 \ \  \text{if} \ \ \alpha =\beta   \nonumber
		\end{eqnarray}
		Hence, the only possible non-zero elements are \(b_{\alpha \alpha qq}\). But this can also be equated to zero by considering one of \(u_{\alpha \alpha}\) or \(v_{qq}\) equal to \(i\). Similarly, one can show that \(c_{\alpha \gamma qs}=0\) for all \(\alpha, \gamma, q\) and \(s\). 
		
		Now, equating the coefficients of the basis \(|E_\alpha\rangle \wedge|O_\gamma\rangle\langle E_q|\wedge\langle O_s|\), we get the constraint equation \(A_{\alpha \gamma qs}=u_{\alpha \beta} v_{\gamma \delta} A_{\beta \delta pr} u_{qp}^{\ast}v_{sr}^{\ast}\). With suitable choice of \(U_{BD}\), one can show that \(A_{\alpha q q \alpha}=0\) for all \(\alpha\) and \(q\). Proceeding similarly, one can show that, \(D_{\alpha q q \alpha}=0\), \(B_{\alpha q \alpha q}=0\) and \(C_{\alpha q \alpha q}=0\) for all \(\alpha, \gamma, q\) and \(s\). 
		
		Next, consider the permutation of basis in the even and odd parity sectors separately. This parity-respected permutation of basis can be obtained using block diagonal unitary operation \(U_{BD}\).  With this parity-respected permutation of basis in the even and odd sectors, along with the constraint equation (\ref{eq:14}) one can show that all the elements within the sets \(\{a_{ijij} | i \neq j\}\), \(\{a_{ijji} | i \neq j\}\), \(\{a_{iiii}\}\), \(\{d_{ijij} | i \neq j\}\), \(\{d_{ijji} | i \neq j\}\), \(\{d_{iiii}\}\), \(\{A_{ijij}\}\), \(\{B_{ijji}\}\), \(\{C_{ijji}\}\) and \(\{D_{ijij}\}\) are equal. Thus \(W\) depends on maximum 10 independent parameters. Now, consider the constraint equation
		\begin{equation}
			(U_{ABD} \wedge U_{ABD}) \ W \ (U_{ABD}^{\dagger} \wedge U_{ABD}^{\dagger})=W 
			\label{eq:15}
		\end{equation}
		This constraint equation (\ref{eq:15}) gives us further reduction of \(W\) in terms of number of independent parameters as follows: \(\{a_{ijij} | i \neq j\} = \{d_{ijij} | i \neq j\} = \{\tilde{a}^{E}\}\), \(\{a_{ijji} | i \neq j\} = \{d_{ijji} | i \neq j\} = \{\tilde{b}^{E}\}\), \(\{a_{iiii}\} = \{d_{iiii} \} = \{A^{E}\}\), \(\{A_{ijij}\} = \{D_{ijij} \} = \{a^{O}\}\) and \(\{B_{ijji}\} = \{C_{ijji} \} = \{b^{O}\}\) where, \(\{\tilde{a}^{E}\}\), \(\{\tilde{b}^{E}\}\), \(\{A^{E}\}\), \(\{a^{O}\}\) and \(\{b^{O}\}\) represents singleton set. Hence, one can write down \(W\), invariant under twirling operation as, 
		
		\begin{widetext}
			\begin{eqnarray}
				W=&&\tilde{a}^{E}\sum_{ij,i \neq j}|E_i\rangle \wedge|E_j\rangle\langle E_i|\wedge\langle E_j|+ |O_i\rangle \wedge | O_j\rangle\langle O_i| \wedge\langle O_j| + \tilde{b}^{E}\sum_{ij,i \neq j} |E_i\rangle \wedge|E_j\rangle\langle E_j|\wedge\langle E_i|+|O_i\rangle \wedge | O_j\rangle\langle O_j| \wedge\langle O_i| \nonumber\\
				&& + A^{E}\sum_{i}|E_i\rangle \wedge|E_i\rangle\langle E_i|\wedge\langle E_i| + |O_i\rangle \wedge | O_i\rangle\langle O_i| \wedge\langle O_i| 
				+ a^{O} \sum_{ij}|E_i\rangle \wedge\ |O_j\rangle\langle E_i|\wedge\langle O_j|+| O_i\rangle \wedge| E_j\rangle\langle O_i| \wedge\langle E_j| \nonumber\\
				&&+ b^{O} \sum_{ij} |E_i\rangle \wedge| O_j\rangle\langle O_j| \wedge\langle E_i| +|O_i\rangle \wedge|E_{j}\rangle \wedge \langle  E_j|\wedge\langle O_i|  
				\label{eq:16}
			\end{eqnarray}
			However, the sums with coefficients \(A^{E}, \tilde{a}^{E}\) and \(\tilde{b}^{E}\) in the R.H.S of Eq.~(\ref{eq:16}) are not invariant under \(U_{BD} \wedge U_{BD}\) twirling individually. Hence, \(A^{E}\) must depend on  \(\tilde{a}^{E}\) and \(\tilde{b}^{E}\). We define,
			\begin{eqnarray}
				\mathbb{I}^{|E\rangle \langle E|}=&&\sum_{ij}|E_{i}\rangle \wedge | E_{j}\rangle \langle E_{i}| \wedge |\langle E_{j} | + |O_{i}\rangle \wedge |O_{j}\rangle \langle O_{i}| \wedge |\langle O_{j} | \nonumber\\
				\mathbb{I}^{|O\rangle \langle O|}=&&\sum_{ij}|E_{i}\rangle \wedge| O_{j}\rangle \langle E_{i}| \wedge |\langle O_{j} | + |O_{i}\rangle \wedge |E_{j}\rangle \langle O_{i} |\wedge |\langle E_{j} | \nonumber\\
				H_{N}^{|E\rangle \langle E|}=&&\sum_{ij}|E_{i}\rangle \wedge | E_{j}\rangle \langle E_{j}| \wedge |\langle E_{i} | + |O_{i}\rangle \wedge |O_{j}\rangle \langle O_{j}| \wedge |\langle O_{i} | \nonumber\\
				H_{N}^{|O\rangle \langle O|}=&&\sum_{ij}|E_{i}\rangle \wedge| O_{j}\rangle \langle O_{j}| \wedge |\langle E_{i} | + |O_{i}\rangle \wedge |E_{j}\rangle \langle E_{j} |\wedge |\langle O_{i} | \nonumber
			\end{eqnarray}
			Thus, \(\mathbb{I}^{|E\rangle \langle E|}+ \mathbb{I}^{|O\rangle \langle O|} = \mathbb{I}\) and \( H_{N}^{|E\rangle \langle E|} +H_{N}^{|O\rangle \langle O|}= H_{N}\)
			
			\begin{lemma}
				\label{lm4}
				Any fermionic states invariant under the fermionic unitary operation \(U \wedge U\) is of the form
				\begin{equation}
					W_{werner}^{f}=a^{E} \mathbb{I}^{|E\rangle \langle E|} + a^{O} \mathbb{I}^{|O\rangle \langle O|} + b^{E} H_{N}^{|E\rangle \langle E|}+ b^{O} H_{N}^{|O\rangle \langle O|}
					\label{eq:17}
				\end{equation}
			\end{lemma} 
			
			\textit{Proof.} At first, we prove that \(W\) is invariant under the fermionic operation \(U \wedge U\). For this purpose, notice
			
			\begin{eqnarray}
				&&(U_{BD} \wedge U_{BD})(\mathbb{I}^{|E\rangle \langle E|})(U_{BD}^{\dagger} \wedge U_{BD}^{\dagger}) \nonumber\\
				&& = \Bigg(\sum_{efgh}p_{ef} p_{gh}|E_e^A\rangle \wedge|E_g^B\rangle\langle E_f^A|\wedge\langle E_h^B| + p_{ef} q_{gh}|E_e^A\rangle \wedge|O_g^B\rangle\langle E_f^A|\wedge\langle O_h^B| + q_{ef} p_{gh}|O_e^A\rangle \wedge|E_g^B\rangle\langle O_f^A|\wedge\langle E_h^B| \Bigg.\nonumber \\
				&&+ \Bigg. q_{ef} q_{gh}|O_e^A\rangle \wedge|O_g^B\rangle\langle O_f^A|\wedge\langle O_h^B| \Bigg) \Bigg(\mathbb{I}^{|E\rangle \langle E|}\Bigg)\Bigg(\sum_{ijkl}(p_{ij})^{\ast} (p_{kl})^{\ast}|E_j^A\rangle \wedge|E_l^B\rangle\langle E_i^A|\wedge\langle E_k^B| + (p_{ij})^{\ast} (q_{kl})^{\ast}|E_j^A\rangle \wedge|O_l^B\rangle\Bigg.\nonumber \\
				&&\langle E_i^A|\wedge\langle O_k^B|  + \Bigg.(q_{ij})^{\ast} (p_{kl})^{\ast}|O_j^A\rangle \wedge|E_l^B\rangle\langle O_i^A|\wedge\langle E_k^B| +  (q_{ij})^{\ast} (q_{kl})^{\ast}|O_j^A\rangle \wedge|O_l^B\rangle\langle O_i^A|\wedge\langle O_k^B| \Bigg)\nonumber \\
				&&= \mathbb{I}^{|E\rangle \langle E|} \nonumber
			\end{eqnarray}   
			where we have used the relation \(\sum_{f}p_{ef}(p_{if})^{\ast}=\delta_{ei}\) and \(\sum_{f}q_{ef}(q_{ei})^{\ast}=\delta_{ei}\). Similarly one can show that,\\
			\((U_{BD} \wedge U_{BD})(\mathbb{I}^{|O\rangle \langle O|})(U_{BD}^{\dagger} \wedge U_{BD}^{\dagger})= \mathbb{I}^{|O\rangle \langle O|} \). Same results can also be obtained for \(U_{ABD} \wedge U_{ABD}\) operation.
			
			Following the same steps one can show that, for \(U =\{U_{BD}, U_{ABD}\}\) 
			\begin{eqnarray}
				(U \wedge U)(H_{N}^{|E\rangle \langle E|})(U^{\dagger} \wedge U^{\dagger})= H_{N}^{|E\rangle \langle E|} \ \ \text{and} \ \ (U \wedge U)(H_{N}^{|O\rangle \langle O|})(U^{\dagger} \wedge U^{\dagger})= H_{N}^{|O\rangle \langle O|}
				\label{eq:29}
			\end{eqnarray}
			Hence,\(W_{werner}^{f}\) in Eq.~(\ref{eq:17}) is an invariant fermionic state under the fermionic operation \((U \wedge U)\). Notice, for \(A^{E}=\tilde{a}^{E} + \tilde{b}^{E}\), the \(W\) in  Eq.~(\ref{eq:16}) becomes equal to \(W_{werner}^{f}\). \(\blacksquare\)
		\end{widetext}
		
		We now show that, there exist a unique Haar measure on the restricted set \(U_{res}\equiv\{U_{BD},U_{ABD}\}\). In this regard, we provide the following lemma.
		
		\begin{lemma}
			\label{lm5}
			There exist a unique left invariat Haar measure on the restricted set \(U_{res}\equiv\{U_{BD},U_{ABD}\}\)
		\end{lemma} 
		
		Proof of the lemma is given in Appendix \ref{sec:a1}. Then one can easily show that the integral \[\int_{U_{res}} dU' (U' \wedge U') \rho (U'^{\dagger} \wedge U'^{\dagger})\] is invariant under the operation \(U \wedge U\). Hence the local twirling operation reduces any state to the invariant form given in lemma \ref{lm4}. 
		
		\begin{lemma}
			\label{lm6}
			The Overlap of an \(N \times N\)-mode fermionic state \(\rho^f\) with the operators \(\mathbb{I}^{|E\rangle \langle E|}+ \mathbb{I}^{|O\rangle \langle O|}\), \(\mathbb{I}^{|E\rangle \langle E|} - \mathbb{I}^{|O\rangle \langle O|} \), \(H^{|E\rangle \langle E|}_N\) and \(H^{|O\rangle \langle O|}_N\) remain invariant under the restricted Unitary group twirling operation on fermionic state. 
		\end{lemma}  
		
		Proof of lemma \ref{lm6} is given in Appendix \ref{sec:a3}. 
		
		\subsection{\label{sec:level5}Reduction of \(N \times N\) mode fermionic states under clifford twirling}
		Let us first introduce the fermionic Pauli operator acting on a single fermionic mode as follows 
		\begin{eqnarray}
			\sigma_z^{f}=&&|\Omega \rangle\langle\Omega|-f^{\dagger}|\Omega \rangle\langle\Omega|f \equiv |E\rangle\langle E|-|O\rangle\langle O| \nonumber \\
			\sigma_x^{f}=&&|\Omega \rangle\langle\Omega|f+f^{\dagger}|\Omega \rangle\langle\Omega| \equiv |E\rangle\langle O|+|O\rangle\langle E| \nonumber \\
			\sigma_y^{f}=&& -i|\Omega \rangle\langle\Omega|f+i f^{\dagger}|\Omega \rangle\langle\Omega| \equiv -i|E\rangle\langle O|+i|O\rangle\langle E| \nonumber
		\end{eqnarray}
		We define the fermionic Pauli group \cite{nielsen_chuang_2010} on 1-mode system  as 
		\[P_1 \equiv \{\pm\mathbb{I},\pm\sigma_x^{f},\pm\sigma_y^{f}, \pm \sigma_z^{f},\pm i\mathbb{I},\pm i\sigma_x^{f},\pm\ i\sigma_y^{f}, \pm i\sigma_z^{f}\}\] For \(N\)-mode system, the fermionic Pauli group \(P_{N}\) can be written by taking the \(N\)-fold wedge product of the fermionic Pauli operators along with identity element and multiplicative factor \(\{\pm1, \pm i\}\). We now define the Clifford group \(C_N\) as the normalizer of the fermionic Pauli group as mentioned in \cite{DiVincenzo_2002}. However, all the Clifford group elements are not legitimate fermionic operator due to it's non-block/non-antiblock diagonal structure in the basis \(\mathcal{B_A^{P}}\). 
		
		Consider, \(C^{res}_N\) as the restricted set with elements from Clifford group such that 
		\begin{eqnarray}
			C^{res}_{N}&&=\{C \in C_N| \Gamma^{BD} \in \Xi^{BD}  \implies C\Gamma^{BD}C^{\dagger} \in \Xi^{BD} \ \nonumber\\ 
			&&\text{and} \ \Gamma^{ABD} \in \Xi^{ABD}  \implies \ C\Gamma^{ABD}C^{\dagger} \in \Xi^{ABD}\} 
			\label{eq:18}
		\end{eqnarray}
		where the set \(\Xi^{BD}\) consists of block-diagonal elements of the fermionic Pauli group \(P_{N}\) and the set \(\Xi^{ABD}\) consists of antiblock-diagonal elements of the fermionic Pauli group \(P_{N}\) in the basis \(\mathcal{B_A^{P}}\). Using this restricted set \( C^{res}_{N}\) we define a restricted Clifford twirl operation as follows
		\begin{equation}
			\mathcal{T}_{C^{res}_{N}}[\rho^f]=\frac{1}{|C^{res}_{N}|}\sum_{c \in C^{res}_{N}} (c \wedge c)\rho^f( c^{\dagger} \wedge c^{\dagger})
			\label{eq:20}
		\end{equation}
		where \(\rho^f\) is any fermionic state of a \(N \times N\)-mode system.
		
		\begin{lemma}
			\label{lm7}
			All the elements of \(C^{res}_{N}\) are legitimate unitary operator in FQT. 
		\end{lemma} 
		
		The proof of lemma \ref{lm7} is given in Appendix \ref{sec:a3}.
		
		As in \cite{DiVincenzo_2002}, we will denote \(P_N^H\) as the hermitian subset of \(P_N\). We will now prove that restricted Clifford twirling operation reduces any \(N \times N\)-mode fermionic state into the form of Eq.~(\ref{eq:17}).
		
		\begin{theorem}
			\label{th1}
			The following statements are true:
			\begin{enumerate}
				\item \(\mathcal{T}_{C^{res}_{N}}\)=\(\mathcal{T}_{C^{res}_{N}}^{\dagger}\)
				\item \(\mathcal{T}_{C^{res}_{N}}[\mathbb{I}_N]=\mathbb{I}_N\) and \(\mathcal{T}_{C^{res}_{N}}[(\sigma_z^{f} \wedge \dots \wedge \sigma_z^{f})^{A} \wedge (\sigma_z^{f} \wedge \dots \wedge \sigma_z^{f})^{B}]=(\sigma_z^{f} \wedge \dots \wedge \sigma_z^{f})^{A} \wedge (\sigma_z^{f} \wedge \dots \wedge \sigma_z^{f})^{B}\)
				\item \(\mathcal{T}_{C^{res}_{N}}[H^{|E\rangle \langle E|}_N]=H^{|E\rangle \langle E|}_N\) and \(\mathcal{T}_{C^{res}_{N}}[H^{|O\rangle \langle O|}_N]=H^{|O\rangle \langle O|}_N\)
				\item For any \(N \times N\)-mode fermionic state \(\rho^f\), \(\mathcal{T}_{C^{res}_{N}}[\rho^f]\) becomes the linear combination of \(\mathbb{I}^{|E\rangle \langle E|}\), \(\mathbb{I}^{|O\rangle \langle O|}\), \(H^{|E\rangle \langle E|}_N\) and \(H^{|O\rangle \langle O|}_N\)
			\end{enumerate}
			where, \(\mathcal{T}_{C^{res}_{N}}^{\dagger}=\frac{1}{|C^{res}_{N}|}\sum_{c \in C^{res}_{N}} ( c^{\dagger} \wedge c^{\dagger})\rho^f(c \wedge c)\).
		\end{theorem} 
		
		Proof of Theorem \ref{th1} is given in Appendix \ref{sec:a3}.
		
		\begin{corollary}
			\label{col1}
			The Overlap of an \(N \times N\)-mode fermionic state \(\rho^f\) with the operators \(\mathbb{I}_N \wedge \mathbb{I}_N, \ (\sigma_z^{f} \wedge \dots \wedge \sigma_z^{f} )\wedge (\sigma_z^{f} \wedge \dots \wedge \sigma_z^{f}), \ H^{|E\rangle \langle E|}_N\) and \(H^{|O\rangle \langle O|}_N\) remains invariant under the restricted Clifford twirling operation on fermionic state.
		\end{corollary} 
		
		Hence, comparing Eq.~(\ref{eq:17}) and Theorem \ref{th1} we get that the restricted Clifford twirling operation reduces any fermionic state to the invariant form under \(U \wedge U\) operation as given in lemma \ref{lm4}.
		
		For any \(N \times N\)-mode fermionic state \(\rho^f\) which is invariant under \(U \wedge U\), \((\mathbb{I}_N \wedge \mathbb{T}_f)(\rho^f)\) is invariant under the operation \(U \wedge U^{\ast}\) with \(U^{\ast}\) denotes the complex conjugation of \(U\). Here, \(\mathbb{T}_f\) denotes the partial transpose operator (see Appendix \ref{sec:a3}).
		
		\begin{corollary}
			\label{col2}
			The following statements are true:
			\begin{enumerate}
				\item \(\mathcal{T}^{\ast}_{C^{res}_{N}}\)=\((\mathcal{T}^{\ast}_{C^{res}_{N}})^{\dagger}\)
				\item \(\mathcal{T}^{\ast}_{C^{res}_{N}}[\mathbb{I}_N]=\mathbb{I}_N\) and \(\mathcal{T}^{\ast}_{C^{res}_{N}}[(\sigma_z^{f} \wedge \dots \wedge \sigma_z^{f})^{A} \wedge (\sigma_z^{f} \wedge \dots \wedge \sigma_z^{f})^{B}]=(\sigma_z^{f} \wedge \dots \wedge \sigma_z^{f})^{A} \wedge (\sigma_z^{f} \wedge \dots \wedge \sigma_z^{f})^{B}\)
				\item \(\mathcal{T}^{\ast}_{C^{res}_{N}}[\Psi^{|E\rangle \langle E|}_N]=\Psi^{|E\rangle \langle E|}_N\) and \(\mathcal{T}^{\ast}_{C^{res}_{N}}[\Psi^{|E\rangle \langle O|}_N]=\Psi^{|E\rangle \langle O|}_N\)
				\item For any \(N \times N\)-mode fermionic state \(\rho^f\), \(\mathcal{T}^{\ast}_{C^{res}_{N}}[\rho^f]\) becomes a linear combination of \(\mathbb{I}^{|E\rangle \langle E|}\), \(\mathbb{I}^{|O\rangle \langle O|}\), \(\Psi^{|E\rangle \langle E|}_N\) and \(\Psi^{|E\rangle \langle O|}_N\)
			\end{enumerate}
			where we define\begin{eqnarray}
				\mathcal{T}^{\ast}_{C^{res}_{N}}[\rho^f]&&=\frac{1}{|C^{res}_{N}|}\sum_{c \in C^{res}_{N}} (c \wedge c^{\ast})\rho^f( c^{\dagger} \wedge (c^{\ast})^{\dagger}) \nonumber \\
				\Psi^{|E\rangle \langle E|}_N&&=(\mathbb{I}_N \wedge \mathbb{T}_f)H^{|E\rangle \langle E|}_N   \nonumber \\
				\Psi^{|E\rangle \langle O|}_N&&=(\mathbb{I}_N \wedge \mathbb{T}_f)H^{|O\rangle \langle O|}_N   \nonumber
			\end{eqnarray}
		\end{corollary} 
		
		\textit{Proof.} For a generic \(N \times N\)-mode fermionic state \(W\)given by Eq.~(\ref{eq:13}), it can be easily proved that \begin{eqnarray}
			&& (U \wedge U^{\ast})\{(\mathbb{I}_N \wedge \mathbb{T}_f)(W)\}(U^{\dagger} \wedge (U^{\ast})^{\dagger}) \nonumber \\
			&&= (\mathbb{I}_N \wedge \mathbb{T}_f)\{(U \wedge U)(W) (U^{\dagger} \wedge U^{\dagger})\} \nonumber
		\end{eqnarray}
		where, \(U \in \{U_{BD}, U_{ABD}\}\). Therefore, for any \(\rho^f\) \begin{eqnarray} 
			\mathcal{T}^{\ast}_{C^{res}_{N}}[\rho^f]= (\mathbb{I}_N \wedge \mathbb{T}_f) \mathcal{T}_{C^{res}_{N}}[(\mathbb{I}_N \wedge \mathbb{T}_f)\rho^f]
		\end{eqnarray}
		Hence statement (1)-(4) can be proved using the results of theorem \ref{th1}. \(\blacksquare\)
		
		We therefore, express the canonical twirl invariant fermionic state as 
		\begin{eqnarray}
			\omega_{noise}^f=&&\alpha(\mathbb{I}_N \wedge\mathbb{I}_N) + \alpha'(\mathbb{S}_z \wedge \mathbb{S}_z )\nonumber\\
			&&+ \beta (\Psi^{|E\rangle \langle E|}_N+\Upsilon \ \Psi^{|E\rangle \langle O|}_N)  \nonumber\\
			\label{eq:51}
		\end{eqnarray}
		where, \(\mathbb{S}_z:=(\sigma_z^{f} \wedge \dots \wedge \sigma_z^{f} )\) and \( p_1:=\frac{1}{2}(\alpha + \alpha')d^2, \ p_2:=\frac{1}{2}(\alpha - \alpha')d^2,\ F_f:=\beta d\) are probabilities. 
		Using Corollary \ref{col1} and Corollary \ref{col2} we define the following twirl invariant quantities\begin{eqnarray}
			I_1:=&&\text{tr}[\mathbb{I}^{|E\rangle \langle E|} \rho^f] ,\ 
			I_2:=\text{tr}[\mathbb{I}^{|O\rangle \langle O|} \rho^f] \nonumber\\
			I_3:=&&\text{tr}[\Psi^{|E\rangle \langle E|}_N \rho^f],\
			I_4:=\text{tr}[\Psi^{|E\rangle \langle O|}_N \rho^f]
			\nonumber\\
		\end{eqnarray}
		One can then easily rewrite \( p_1, \ p_2\) and \(F_f\) in terms of the twirl invariant quantities as \begin{eqnarray}
			F_f=&&\frac{I_3-2I_1/d}{d/2-2/d}, \ p_2=I_2 \ \ \text{and} \nonumber\\
			p_1=&&1-F_f-p_2=1-\frac{I_3-2I_1/d}{d/2-2/d}-I_2
			\label{eq:45}
		\end{eqnarray} 
		
		Remarkably, the structure of fermionic isotropic state \(\omega_{noise}^f\)  differs from the structure of isotropic state in SQT. One can argue that the difference in the exact part of the shared resource is tied to the existence of topological correlation. Indeed, for local-parity preserving resource states \((\Upsilon=0)\), \(f(\rho_{AB})\) would be equal to \(\rho_{AB}\) in Eq.(~\ref{eq:32}), thus giving us \(\delta S(\rho_{AB})=0\). Notice, any local-parity violation leads to inaccessible entanglement.  
		
		\subsection{\label{sec:level6}State-Channel isomorphism in FQT}
		Quantum operations for FQT in it's most general form or Fermionic Channel has already been introduced in \cite{PhysRevA.104.032411} where the equivalence of operator-sum representation and completely Positive map has been proved. In this section, following the procedure mentioned in \cite{horodecki99_2}, we will establish an isomorphism given by
		\begin{equation}
			\rho^f_{\mathcal{E}}=(\mathbb{I} \wedge \mathcal{E} )( |\psi^{+}\rangle \langle \psi^{+}|_{f})
		\end{equation}
		Here, \(|\psi^{+}\rangle \langle \psi^{+}|_{f}=1/d(\Psi^{|E\rangle \langle E|}_N + \Psi^{|E\rangle \langle O|}_N )\), represented in basis \(\mathcal{B_{AB}^P}\). Also, the first reduction of \(\rho^f_{\mathcal{E}}\) following fermionic partial tracing procedure gives \(\mathbb{I}/d\) with \(\mathcal{E}\) representing a fermionic channel. To prove the surjective property of this isomorphism, let us assume that the spectral decomposition of \(\rho^f_{\mathcal{E}}\) is given by 
		\[\rho^f_{\mathcal{E}}=\sum_m p_m^E |\psi^E_m\rangle \langle \psi^E_m|+ \sum_m p_m^O |\psi^O_m\rangle \langle \psi^O_m|\]
		where \(|\psi^E_m\rangle\) and \(|\psi^O_m\rangle\) represents eigenvectors of \(\rho^f_{\mathcal{E}}\) for all \(m\). Let, 
		\begin{eqnarray}
			|\psi^E_1\rangle=&&\sum_{ij}A_{ij}^{EE}|E_i\rangle \wedge |E_j\rangle + A_{ij}^{OO}|O_i\rangle \wedge |O_j\rangle \nonumber \\
			|\psi^O_1\rangle=&&\sum_{ij}A_{ij}^{EO}|E_i\rangle \wedge |O_j\rangle + A_{ij}^{OE}|O_i\rangle \wedge |E_j\rangle \nonumber
		\end{eqnarray}
		One can then represent \(|\psi^E_1\rangle\) in terms of a fermionic operator \(K_1^{BD}\)as
		\[|\psi^E_1\rangle=(\mathbb{I}\wedge K_1^{BD}) |\psi^{+}\rangle _{f}\]
		with, \(K_1^{BD}\) being block diagonal in the basis \(\mathcal{B^P_A}\). To see this, let us assume 
		\[K_1^{BD}=\sum_{ij}S_{ij}|E_i\rangle\langle E_j|+ T_{ij}|O_i\rangle\langle O_j|\]
		Then, \((\mathbb{I}\wedge K_1^{BD}) |\psi^{+}\rangle _{f}\) is given by \[= \frac{1}{\sqrt{d}}\sum_{ij}S_{ij}|E_j\rangle \wedge |E_i\rangle + T_{ij}|O_j\rangle \wedge |O_i\rangle\]
		Fixing \(\langle E_i|K_1^{BD}|E_j\rangle = \sqrt{d}A_{ji}^{EE}\) and \(\langle O_i|K_1^{BD}|O_j\rangle = \sqrt{d}A_{ji}^{OO}\) we get \(|\psi^E_1\rangle=(\mathbb{I}\wedge K_1^{BD}) |\psi^{+}\rangle _{f}\). Similarly one can find anti-block diagonal  \(K_1^{ABD}\) and show that \(|\psi^O_1\rangle=(\mathbb{I}\wedge K_1^{ABD}) |\psi^{+}\rangle _f \). Hence, \(\rho^f_{\mathcal{E}}\) can be written as
		\begin{eqnarray}
			\rho^f_{\mathcal{E}}=&&\sum_m p_m^E (\mathbb{I}\wedge K_m^{BD})( |\psi^{+}\rangle \langle \psi^{+}|_{f}) (\mathbb{I}\wedge {K_m^{BD}}^{\dagger})+  \nonumber \\
			&&p_m^O (\mathbb{I}\wedge K_m^{ABD})( |\psi^{+}\rangle \langle \psi^{+}|_{f}) (\mathbb{I}\wedge {K_m^{ABD}}^{\dagger}) \nonumber \\
			&&= (\mathbb{I} \wedge \mathcal{E} )( |\psi^{+}\rangle \langle \psi^{+}|_{f}) 
			\label{eq:26}
		\end{eqnarray}
		Then \((\mathbb{I} \wedge \mathcal{E} )\) is a positive map with \(\{K_m^{BD},K_m^{ABD}| \ \forall m\}\) representing the Kraus operator for the channel \(\mathcal{E}\). 
		
		To see the trace-preservion condition of the channel \(\mathcal{E}\), we need to show that the operator \[Z^f=\sum_{m}p_m^E {K_m^{BD}}^{\dagger}K_m^{BD}+ p_m^O {K_m^{ABD}}^{\dagger}K_m^{ABD}=\mathbb{I}\]. For this, recall that the consistency condition for partial trace in FQT \cite{friis2013} is given by 
		\[\text{tr}(\mathcal{O}_A \rho^f_A)=\text{tr}((\mathcal{O}_{A}\wedge \mathbb{I}_B) \rho^f_{AB})\]
		for any bi-partite fermionic states \( \rho^f_{AB}\) with reduced state \(\rho^f_A\). For \(\rho^f_A=\mathbb{I}/d\) and letting \(\rho^f_{AB}=\rho^f_{\mathcal{E}}\), we get
		\begin{eqnarray}
			&&\text{tr}(\mathcal{O}_A)=d\text{tr}((\mathcal{O}_{A}\wedge \mathbb{I}_B) \rho^f_{AB})=\nonumber \\
			&&d\sum_m p_m^E \text{tr} ((\mathcal{O}_{A}\wedge \mathbb{I}_B)(\mathbb{I}\wedge K_m^{BD})( |\psi^{+}\rangle \langle \psi^{+}|_{f}) (\mathbb{I}\wedge {K_m^{BD}}^{\dagger}))  \nonumber \\
			&&+p_m^O \text{tr} ((\mathcal{O}_{A}\wedge \mathbb{I}_B) (\mathbb{I}\wedge K_m^{ABD})( |\psi^{+}\rangle \langle \psi^{+}|_{f}) (\mathbb{I}\wedge {K_m^{ABD}}^{\dagger})) \nonumber \\
			&&=d\text{tr}(( |\psi^{+}\rangle \langle \psi^{+}|_{f})(\mathcal{O}_{A}\wedge Z^f)) \nonumber \\
			&&=d\text{tr}(( |\psi^{+}\rangle \langle \psi^{+}|_{f})(\mathcal{O}_{A}(Z^f)^{T}\wedge \mathbb{I}_B)) \nonumber \\
			&&=\text{tr}(\mathcal{O}_{A}(Z^f)^{T}) \ \ \ \ \ \forall \mathcal{O}_{A} \nonumber
		\end{eqnarray}
		where we have used the properties \((\mathbb{I} \wedge Z^f)|\psi^{+}\rangle_f=((Z^f)^{T}\wedge \mathbb{I})|\psi^{+}\rangle_f\) and partial trace of \(|\psi^{+}\rangle \langle \psi^{+}|_{f}\) is given by \(\mathbb{I}/d\). Hence, \((Z^f)^{T}=Z^f=\mathbb{I}\). This completes the proof of surjectivity. 
		
		To show the injectivity property of the isomorphism, let there exists two channels \(\mathcal{E}_1\) and \(\mathcal{E}_2\) such that
		\begin{eqnarray}
			&&(\mathbb{I} \wedge \mathcal{E}_1 )( |\psi^{+}\rangle \langle \psi^{+}|_{f})=(\mathbb{I} \wedge \mathcal{E}_2 )( |\psi^{+}\rangle \langle \psi^{+}|_{f})  \nonumber \\
			&&\implies (\mathbb{I} \wedge (\mathcal{E}_1- \mathcal{E}_2))( |\psi^{+}\rangle \langle \psi^{+}|_{f})=0
			\label{eq:27}
		\end{eqnarray}
		Let \((\mathcal{E}_1- \mathcal{E}_2)=\mathcal{E}'\). Notice, that \begin{eqnarray}
			&&(\mathbb{I} \wedge \mathcal{E}')( |\psi^{+}\rangle \langle \psi^{+}|_{f})=\frac{1}{d}\sum_{ij}|E_i\rangle \langle E_j| \wedge \mathcal{E}'(| E_i\rangle\langle E_j |)  \nonumber \\
			&&+ |O_i\rangle\langle O_j| \wedge  \mathcal{E}'(|O_i\rangle\langle O_j | )+ |E_i\rangle \langle O_j|\wedge  \mathcal{E}'(| E_i\rangle \langle O_j |)   + \nonumber \\
			&&|O_i\rangle  \langle E_j| \wedge  \mathcal{E}'(|O_i\rangle\langle E_j|)
			\label{eq:28}
		\end{eqnarray}
		Notice, under the action of the trace annhilating map \(\mathcal{E}'\) on the basis \(| E_i\rangle\langle E_j |\) is given by \(\gamma_{ijkl}| E_k\rangle\langle E_l |\) or \(\gamma_{ijkl}| O_k\rangle\langle O_l |\) due to PSSR restriction on most general operations. Similarly, one can find the action of \(\mathcal{E}'\) on other basis. Comparing, Eq.~(\ref{eq:27}) and Eq.~(\ref{eq:28}) one gets \(\mathcal{E}'=0\). This proves the injectivity.
		
		As in the case of SQT, in FQT also we define the restricted twirling operation on fermionic channel as follows: Alice applies unitary which is an element of restricted clifford set on one half of the fermionic bi-partite state and send it through the channel \(\mathcal{E}\) to Bob. Bob upon receiving the state, applies the inverse Unitary.  
		
		Now we prove the following lemma to be used while deriving optimal teleportation fidelity.	
		
		\begin{lemma}
			\label{lm8}
			For any fermionic channel \(\mathcal{E}\) we have the relation
			\[\rho_{\mathcal{T}^{\ast}_{C^{res}_{N}}[\mathcal{E}]}^f=\mathcal{T}^{\ast}_{C^{res}_{N}}[\rho_{\mathcal{E}}^f]\]
			where, \(\rho_{\mathcal{T}^{\ast}_{C^{res}_{N}}[\mathcal{E}]}^f=(\mathbb{I} \wedge \mathcal{T}^{\ast}_{C^{res}_{N}}[\mathcal{E}] )( |\psi^{+}\rangle \langle \psi^{+}|_{f})\) with \(\mathcal{T}^{\ast}_{C^{res}_{N}}[\mathcal{E}] \) representing restricted twirling operation on fermionic channel.
		\end{lemma} 
		
		\textit{Proof.} Given for FQT, \(\rho_{\mathcal{T}^{\ast}_{C^{res}_{N}}[\mathcal{E}]}^f\)\begin{eqnarray}
			&&=(\mathbb{I} \wedge \mathcal{T}^{\ast}_{C^{res}_{N}}[\mathcal{E}] )( |\psi^{+}\rangle \langle \psi^{+}|_{f}) \nonumber \\
			&&=\frac{1}{|C^{res}_{N}|}\sum_{c \in C^{res}_{N}} (\mathbb{I} \wedge c^{\dagger})[(\mathbb{I} \wedge \mathcal{E} )((\mathbb{I} \wedge c)|\psi^{+}\rangle \langle \psi^{+}|_{f}\nonumber\\
			&&(\mathbb{I} \wedge c^{\dagger}))](\mathbb{I} \wedge c) \nonumber \\
			&&=\frac{1}{|C^{res}_{N}|}\sum_{c \in C^{res}_{N}} (\mathbb{I} \wedge c^{\dagger})[(\mathbb{I} \wedge \mathcal{E} )((c^T \wedge \mathbb{I})|\psi^{+}\rangle \langle \psi^{+}|_{f}\nonumber\\
			&&(c^{\ast} \wedge \mathbb{I}))](\mathbb{I} \wedge c) \nonumber \\
			&&= \mathcal{T}^{\ast}_{C^{res}_{N}}[\rho_{\mathcal{E}}^f]
		\end{eqnarray}
		where we used the fact that, for all \(c \in C^{res}_{N}\) we also have, \(c^{\dagger} \in C^{res}_{N}\), thus the sum remain invariant under hermitian conjugate. \(\blacksquare\)
		\subsection{\label{sec:level7}Optimal teleportation fidelity}
		In the previous section we found that under restricted clifford twirling operation, any \(N \times N\)-mode fermionic state can be reduced to the canonical form of \begin{eqnarray}
			\omega_{noise}^f=(1-F_f)\Bigg(\frac{\mathbb{I}_N \wedge\mathbb{I}_N}{d^2} + \Upsilon_1\frac{\mathbb{S}_z \wedge \mathbb{S}_z}{d^2} \Bigg)+ F_f (\omega_{AB}^f)  \nonumber\\
			\label{eq:31}
		\end{eqnarray}
		where, \(\mathbb{S}_z=(\sigma_z^{f} \wedge \dots \wedge \sigma_z^{f} )\), \(1-F_f=\alpha d^2,\ F_f=\beta d\) and \(\Upsilon_1=\frac{p_1-p_2}{p_1+p_2}\) with \(\ -1\leq\Upsilon_1\leq 1\) (\(\Upsilon_1=0 \ \text{for} \ \ p_1=p_2=0\) ). Note that, \(p_1,\ p_2 \ \text{and} \ F_f\) can be expressed in terms of the twirl invariant quantities as specified in Eq.(~\ref{eq:45}).  The state \(\omega_{AB}^f\) is given by  \begin{equation}
			\omega_{AB}^f=\text{loc-p}(|\psi^{+}\rangle \langle \psi^{+}|_{f})+\Upsilon \overline{\text{loc-p}}(|\psi^{+}\rangle \langle \psi^{+}|_{f})
			\label{eq:30}
		\end{equation}
		where, \(\Upsilon\) is real and loc-p() indicates the local-parity preserving part, while \(\overline{\text{loc-p()}}\) represents local-parity violating part. Also, \(\alpha,\ \beta, \ \Upsilon_1\) can be expressed in terms of the twirl invariant overlap (see \textit{Corollary 1}). 
		
		In case of any fermionic channel \(\mathcal{E}\) in FQT we define the average entanglement fidelity of the channel as
		\begin{eqnarray}
			F(\mathbb{I} \wedge \mathcal{E}) =&&\int d\psi  \Bigg( \min_{\{E_m^{AB}\}} \sum_m \sqrt{\text{tr}(|\psi\rangle \langle \psi |E_m^{AB}) }\nonumber \\
			&&\sqrt{\text{tr}((\mathbb{I} \wedge \mathcal{E}(|\psi\rangle \langle \psi |))E_m^{AB})} \Bigg)^{2}
		\end{eqnarray} 
		where \(|\psi\rangle =\{|\psi_{O}\rangle ,|\psi_{E}\rangle \}\) represents fermionic bi-partite state with Parity of \(|\psi_{O}\rangle \) being odd and \(|\psi_{E}\rangle \langle\psi_{E} |\) represents fermionic bi-partite state with Parity of \(|\psi_{E}\rangle \) being even. Also, the integrals are calculated considering uniform measure over all even and odd states respectively. As the measurement is restricted by PSSR constrained LOCC, we consider here the local-parity preserving POVM elements \(\{E_m^{AB}\}\). 
		
		The entanglement fidelity defined here is a measure of the preservation of fermionic entanglement only. Hence, the optimality of the teleportation scheme considered in this paper is related to the preservation of entanglement in contrast to the topological correlation.   
		
		Notice, this average entanglement fidelity of the fermionic channel is invariant under the restricted twirling operation on fermionic channel as introduced in section ~\ref{sec:level5} (see Appendix \ref{sec:a4}). 
		
		Now we consider the fermionic subsystem teleportation scheme explained in section ~\ref{sec:level3}, with the shared resource \( \omega_{noise}^f\) (given in Eq.~(\ref{eq:31})).  
		
		\begin{lemma}
			\label{lm9}
			For any pure fermionic state \(|\psi\rangle \langle \psi |_{\tilde{A}\tilde{A}'}\), the output state of the fermionic teleportation channel with the teleportation scheme involving Bell state measurement and classical communication \cite{bennett1993} using shared resource \( \omega_{noise}^f\) is given by 
			\begin{eqnarray}
				\rho^f_{B\tilde{A}'}=&&\alpha d^2\Bigg(\frac{\mathbb{I}_N \wedge\rho^f_{\tilde{A}'} }{d} \pm \Upsilon_1\frac{(\mathbb{I}_N \wedge\rho^f_{\tilde{A}'} )(\mathbb{S}_z \wedge \mathbb{S}_z)}{d} \Bigg) \nonumber \\
				&&+\beta d \ \text{loc-p}(|\psi\rangle \langle \psi |_{B\tilde{A}'}) + \Upsilon\beta d \ \overline{\text{loc-p}}(|\psi\rangle \langle \psi |_{B\tilde{A}'}) \nonumber \\
				&& \equiv (1-F_f)(\rho^f_{B\tilde{A}'})_{noisy}+ F_f(\rho^f_{B\tilde{A}'})_{exact} \nonumber \\
				\label{eq:25}
			\end{eqnarray}  
			with \(\pm\) depending upon the parity of the input state \(|\psi\rangle \langle \psi |_{\tilde{A}\tilde{A}'}\). Here, \(\text{loc-p}(|\psi\rangle \langle \psi |)\) denotes the local-parity respecting part of \(|\psi\rangle \langle \psi |\) and \(\overline{\text{loc-p}}(|\psi\rangle \langle \psi |)\) denotes the local-parity violating part of \(|\psi\rangle \langle \psi |\). Here, we define \begin{eqnarray}
				(\rho^f_{B\tilde{A}'})_{noisy}=&&\Bigg(\frac{\mathbb{I}_N \wedge\rho^f_{\tilde{A}'}}{d} \pm \Upsilon_1\frac{(\mathbb{I}_N \wedge\rho^f_{\tilde{A}'})(\mathbb{S}_z \wedge \mathbb{S}_z)}{d} \Bigg) \nonumber \\
				(\rho^f_{B\tilde{A}'})_{exact}=&&\text{loc-p}(|\psi\rangle \langle \psi |_{B\tilde{A}'}) + \Upsilon \ \overline{\text{loc-p}}(|\psi\rangle \langle \psi |_{B\tilde{A}'}) \nonumber \\
			\end{eqnarray}
		\end{lemma} 
		
		The proof of lemma \ref{lm9} is given in Appendix \ref{sec:a5}.
		
		Notice, in FQT also, the mathematical state-channel isomorphism between the noisy fermionic resource state and the corresponding teleportation channel is physical. Indeed, applying the fermionic teleportation channel (whose action is given by lemma \ref{lm9} on one half of the state \(|\psi^{+}\rangle \langle \psi^{+}|_{f}\) outputs noisy fermionic resource state given by Eq.~(\ref{eq:31}). Hence using corollary \ref{col2} and lemma \ref{lm8} one can show that any Fermionic channel can be reduced to the form mentioned in lemma \ref{lm9} without affecting the entanglement fidelity. In other words, we have the following commutative diagram:
		\begin{center}
			\begin{tikzcd}[ampersand replacement = \&]
				\rho^{f}_{\mathcal{E}} \arrow[leftrightarrow, "iso"]{r} \arrow[d,"\mathcal{T}^{\ast}_{C^{res}_N}"] \& \mathcal{E} \arrow[d,"\mathcal{T}^{\ast}_{C^{res}_N}"] \\
				\omega_{noise}^f \arrow[leftrightarrow,"iso"]{r} \& \mathcal{E}_{\omega_{noise}^f}
			\end{tikzcd} 
		\end{center} 
		
		To calculate the teleportation fidelity of this teleportation channel, one can easily show that for the exact case, \(F(\mathbb{I} \wedge \mathcal{E}_{exact})=1\) (see Appendix \ref{sec:a5}).
		
		For the noisy part, we get, (see Appendix \ref{sec:a5}),\begin{eqnarray}
			F(\mathbb{I} \wedge \mathcal{E})_{noisy}=\frac{4(1+\Upsilon_1)}{d^2+4} \nonumber
		\end{eqnarray}
		
		Hence, for any fermionic channel \(\mathcal{E}\), the overall average entanglement fidelity of fermionic teleportation is given by \begin{eqnarray}
			F(\mathbb{I} \wedge \mathcal{E}) = \frac{4(1+\Upsilon_1)(1-F_f)}{d^2+4}+ F_f
			\label{eq:43}
		\end{eqnarray}
		where, \(F_f\) and \(\Upsilon_1\) are twirl invariant parameter. The invariance of \(F_f\) and \(\Upsilon_1\) under twirling on channel operation can be shown using lemma \ref{lm8} and lemma \ref{lm6}. Also note that, \(\Upsilon_1\) is related to the weight of identity operators in even and odd parity subspace. For \(\Upsilon_1 < 0\), Alice and Bob can apply local unitaries of the form \(U_{BD} \wedge U_{ABD}\) to change the sign of \(\Upsilon_1\), thus incresing the fidelity. Finally we mention the main result of this paper.
		\begin{theorem}
			\label{th2}
			The maximum subsystem teleportation fidelity of \(N \times N\)-mode fermionic system using the set of \(N \times N\)-mode shared resource state \(\rho^f\), characterized by the twirl invariant parameters \(F_f\) and \(\Upsilon_1\) whose optimum values \(|\Upsilon_1|_{opt}\) and \((F_f)_{max}\) can be achieved via LOCC, is given by \begin{eqnarray}
				F(\mathbb{I} \wedge \mathcal{E})_{max} =&& \frac{4(1+|\Upsilon_1|_{opt}))}{d^2+4} \\
				&&+(F_f)_{max} \left(1- \frac{4(1+|\Upsilon_1|_{opt})}{d^2+4}\right)\nonumber
			\end{eqnarray}

			The optimum values of the twirl invariant parameters can be obtained via fermionic local filtering operations as in the case of SQT \cite{horodecki2000,Verstraete2001}.
		\end{theorem}
		\textit{Proof.} The proof follows in the similar direction of \cite{horodecki99_2}. Consider an optimal fermionic teleportation channel via the given state \(\rho^f\). Such channel can always be twirled into a channel whose action is specified by Eq.(~\ref{eq:25}). We also know that such twirling on channel operation do not decrease the fidelity. Now apply this fermionic channel on the fermionic state \(|\psi^{+}\rangle \langle \psi^{+}|_{f}\) to get the set of states \(\omega_{noise}^f\) with fixed parameter \(F_f\) and \(\Upsilon_1\) and satisfying the equation \[F(\mathbb{I} \wedge \mathcal{E})_{max}=\frac{4(1+\Upsilon_1)(1-F_f)}{d^2+4}+ F_f\] One can always apply some LOCC to optimize the twirl invariant parameters, thus giving us \[F(\mathbb{I} \wedge \mathcal{E})_{max} \leq \frac{4(1+|\Upsilon_1|_{opt})(1-(F_f)_{max})}{d^2+4}+ (F_f)_{max}\] 
		To see the reverse inequality, consider the given state \(\rho\) whose twirl invariant parameters can be optimized via LOCC. Applying the twirling on state operation and performing standard fermionic teleportation, one gets the reverse inequality. $\blacksquare $
		
		We note that the teleportation fidelity dervied in this paper has been optimized over all such fermionic channels that preserves topological correlation. For example, the subsystem teleportation of states \(\psi_{even}\) and \(\psi_{odd}\) can only be possible if the fermionic channel preserves topological correlation. Indeed, one can further optimize the teleportation fidelity using local-parity projection of the resource state provided the state space is restricted to \(\{\psi_{EE},\psi_{EO},\psi_{OE},\psi_{OO}\}\) . However, such optimization would discard all fermionic channels preserving topological correlation.
		
		\section{\label{sec:level9}Teleportation of Topological correlation}
		In previous section we discussed the optimal subsystem teleportation that preserves the entanglement. Here, we attempt to answer the question of preserving topological correlation using topological correlation of the shared state as a resource.
		
		Here, we consider a class of shared resource state \[\omega_{AB}^f=\text{loc-p}(|\psi^{+}\rangle \langle \psi^{+}|_{f})+\Upsilon \overline{\text{loc-p}}(|\psi^{+}\rangle \langle \psi^{+}|_{f})\] and standard teleportation protocol, which is optimal for preserving entanglement. Then, using lemma \ref{lm9} the output state of the teleportation channel is given by \[\rho^f_{B\tilde{A}'}=\text{loc-p}(|\psi\rangle \langle \psi |_{B\tilde{A}'}) + \Upsilon \ \overline{\text{loc-p}}(|\psi\rangle \langle \psi |_{\tilde{A}'B}) \]
		Notice that, due to PSSR constraint on local observables, it is not possible for Alice and Bob to distinguish states with and without topological correlation via local bi-partite quantum state tomography. Hence, the average fidelity following \cite{nielsen_chuang_2010} in this case should be defined as \begin{equation}
			F_{topo}(\mathbb{I} \wedge \mathcal{E})=\int d\psi\left\langle \psi\left|\mathbb{I} \wedge \mathcal{E}(|\psi\rangle\langle\psi|)\right|\psi\right\rangle
			\label{eq:44}
		\end{equation} 
		Now, notice that the kraus operators of the teleportation channel are given by 
		\begin{eqnarray}
			K_1^{BD}=&&\sum_m|E_m\rangle\langle E_m|+ \Upsilon \ \sum_m |O_m\rangle\langle O_m| \nonumber\\ 
			K_2^{BD}=&&\sqrt{1-{\Upsilon}^2} \ \sum_m |O_m\rangle\langle O_m| \nonumber
		\end{eqnarray}
		One can immediately verify that, \(\sum_i K_i^{BD\dagger}K_i^{BD}=\mathbb{I}\) and \[\sum_i(\mathbb{I} \wedge K_i^{BD})|\psi\rangle\langle\psi|_{\tilde{A}'\tilde{A}}(\mathbb{I} \wedge K_i^{BD\dagger})=\rho^f_{\tilde{A}'B}\]
		Hence, the average fidelity \(F_{topo}(\mathbb{I} \wedge \mathcal{E})\) is given by \begin{equation}
			F_{topo}(\mathbb{I} \wedge \mathcal{E})=\sum_i\int d\psi |\langle\psi|(\mathbb{I} \wedge K_i^{BD})|\psi\rangle|^2
		\end{equation}
		where the averaging is done over random ensemble \(\{\psi_{even}, \psi_{odd}\}\).
		
		Notice that, the channel action can be compared to a phase damping channel \cite{nielsen_chuang_2010} which introduces decoherence between even (locally) and odd (locally) parity sector. 
		
		Also, for \(\Upsilon=1\), we have \(F_{topo}(\mathbb{I} \wedge \mathcal{E})=1\), thus reproducing the exact result of \cite{f_teleport}.

		\section{\label{sec:level10}Conclusion}
		In this work, we discussed subsystem teleportation channel in FQT that preserves two different types of correlation - topological correlation and fermionic entanglement. Towards this, we utilized a generic shared resource state having both topological correlation and fermionic entanglement and assessed the fidelity of the channel by entanglement fidelity measure. Due to local-PSSR constraint, we introduced operationally motivated entanglement fidelity measure with local parity-preserving POVMs. Although, such measures are suitable for the assessment of entanglement preservation, states that differ due to topological correlation cannot be distinguished with this measure.

		
		To derive an expression of optimal teleportation fidelity for entanglement preservation, at first we derived the fermionic invariant state  under local twirling operation and found that the structure of this invariant state differ from that of Werner state - the corresponding invariant state in SQT. Such differences in the resource state was attributed to the existence of non-trivial topological correlation. We introduced a Haar measure on the restricted Unitary group \(U_{res}\) and showed that any state can be reduced to the invariant form using restricted Unitary group twirling. We constitute the Unitary-2 design of the restricted Unitary group twirling by introducing a restricted Clifford set. Such unitary 2-designs are crucial for experimentally validating the average fidelity of channel. Introducing the partial transpose operation on fermionic modes, we derived the fermionic isotropic state - a linear combination of noise in even subspace, noise in odd subspace and the fermionic resource state with non-trivial topological correlation. However, such fermionic resource state  with non-trivial topological correlation was found to be equivalent to the exact teleportation channel for entanglement preservation irrespective of the topological correlation. We introduced the state-channel isomorphism and channel twirling operation in FQT, and derived the optimal teleportation fidelity. The fidelity expression depends only on the parameters invariant under the local-twirling operations. We note that, one can always increase the optimal teleportation fidelity via local filtering operation as in the case of SQT.
		
		Even though we have provided the optimal teleportation fidelity expression for entanglement preservations, such expression for preserving topological correlation still remains open. It would also be interesting to see if there exist a fermionic channel that would preserve both fermionic entanglement and topological correlation with optimal teleportation fidelity. Towards this, one may consider the optimal teleportation method where an optimal strategy is implemented in the receiver end \cite{HORODECKI199621,gong2024}.  In this work, however, we considered a class of resource states with non-trivial topological correlation and discussed the effect of teleportation channel. Even though such channels preserve entanglement exactly, they act like a phase damping channel which introduces decoherence between even parity (locally) and odd parity (locally) sector.  
		
		As mentioned earlier, a demonstration of our optimal teleportation protocol is possible with MZM \cite{inprep}. However, such realization would require implementation of the restricted clifford twirling group element using braiding of MZM followed by the implementation of standard teleportation protocol \cite{zhou2022}.In recent years, fermionic neutral atom based platform has gained considerable attention for simulating fermionic systems \cite{pnas2304294120}. In this context, error correction has also been proposed using Majorana code along with methods to implement braiding gate \cite{zkpl-hh28,schuckert2025}. Such platforms are suitable for the implementation of our optimal teleportation protocol \cite{inprep}.    
		
		In this paper, optimality of teleportation fidelity is restricted to the use of maximally entangled basis based measurement. However, as shown in case of SQT \cite{ghosal2025}, one can also extend the optimality of fermionic subsystem teleportation including non-maximal entangled measurement. Apart from FQT, one may also try to find the optimal teleportation fidelity for quantum information encoded in Bosonic modes. Also, general QT channel may find its application in Fermionic Quantum Computation in the NISQ-like settings. 
		
		Even though PSSR constraint is considered to be fundamental, a more easily realizable fermionic state in experiments comes with number superselection rule (NSSR). In this context, fermionic teleportation has been demonstrated using quantum dot spin which possess very high decoherence time \cite{galler2021,small1999}. The teleportation protocol utilizes a conditional tunneling process to achieve unit fidelity. However, extending such protocol with subsystem dimension \(^nC_{n/2}\) (where \(n/2\) denotes the number of quantum dots per subsystem) will require generalization of the conditional tunneling process. Such generalization would help one to specify the optimal fermionic teleportation in the context of quantum dot \cite{inprep}.

		\begin{acknowledgments}
			We wish to thank Aravinda S.  for fruitful discussions on Quantum Information of system of indistinguishable particles. We would also like to thank Sourin Das, Utkarsh Mishra and K. Shirish Rao for fruitful discussion on Majorana Zero Mode.
		\end{acknowledgments}
		\begin{widetext}
			\appendix
			\section{\label{sec:a1}Fermionic Hilbert space}
			
			\textit{Proof of lemma \ref{lm2}.} Let us assume, 
			\[\hat{O}_P^A=\sum_{ij}p_{ij}^A|E_i^A\rangle\langle E_j^A|+q_{ij}^A|O_i^A\rangle\langle O_j^A| \]
			denotes an observable of system $A$ and 
			\[\hat{O}_P^B=\sum_{ij}p_{ij}^B|E_i^B\rangle\langle E_j^B|+q_{ij}^B|O_i^B\rangle\langle O_j^B| \]
			denotes an observable of system $B$. We denote a local PSSR-respected fermionic state $ \rho_{local}^{AB}$ as
			
			\begin{eqnarray}
				\rho_{local}^{AB}=&&\sum_{abcd}A'_{abcd}|E_a^A\rangle \wedge|E_b^B\rangle\langle E_c^A|\wedge\langle E_d^B| + B'_{abcd}|E_a^A\rangle \wedge|O_b^B\rangle\langle E_c^A|\wedge\langle O_d^B|+ \nonumber\\&&C'_{abcd}|O_a^A\rangle \wedge|E_b^B\rangle\langle O_c^A|\wedge\langle E_d^B| + D'_{abcd}|O_a^A\rangle \wedge|O_b^B\rangle\langle O_c^A|\wedge\langle O_d^B| \nonumber
			\end{eqnarray}

			with suitable conditions on the coefficients $A',B',C'$ and $D'$ such that $\rho_{local}^{AB}>0$ and $tr(\rho_{local}^{AB})=1$. 
			
			Using the partial tracing procedure outlined in \cite{PhysRevA.104.032411} we obtain the reduced density matrix for system $A$ as
			\begin{eqnarray}
				\rho_{local}^{A}=&&\sum_{abc}A'_{abcb}|E_a^A\rangle \langle E_c^A|+ B'_{abcb}|E_a^A\rangle \langle E_c^A|+ C'_{abcb}|O_a^A\rangle \langle O_c^A|+ D'_{abcb}|O_a^A\rangle \langle O_c^A| \nonumber
			\end{eqnarray}
			and reduced density matrix for system $B$ as
			\begin{eqnarray}
				\rho_{local}^{B}=&&\sum_{abd}A'_{abad}|E_b^B\rangle \langle E_d^B|+ C'_{abad}|E_b^B\rangle \langle E_d^B|+ B'_{abad}|O_b^B\rangle \langle O_d^B|+ D'_{abad}|O_b^B\rangle \langle O_d^B| \nonumber
			\end{eqnarray}
			Then imposing the trace condition of product state (see Eq.~(\ref{eq:2})) and comparing coefficients of $p_{ca}^A \ p_{db}^B, \ p_{ca}^A \ q_{db}^B, \ q_{ca}^A \ p_{db}^B$ and $q_{ca}^A \ q_{db}^B$ we get
			\[ \rho_{local}^{AB}=\rho_{local}^{A} \wedge \rho_{local}^{B} \]
			
			Reverse implication is trivially satisfied. $\blacksquare$
			
			\textit{Proof of lemma \ref{lm3}.} Let us assume that the anti block-diagonal and block-diagonal unitary matrices acting on system $A$ and $B$ respectively is given by
			\[ U_{ABD}^A=\sum_{ij}p_{ij}|E_i^A\rangle\langle O_j^A|+q_{ij}|O_i^A\rangle\langle E_j^A| 
			\ \ \ \text{and} \ \ \
			{U'_{BD}}^B=\sum_{kl}r_{kl}|E_k^B\rangle\langle E_l^B|+s_{kl}|O_k^A\rangle\langle O_l^B| \] 
			Now consider a separable fermionic state $\rho_{sep}^{AB}$ given by Eq.~(\ref{eq:5}) with \[ \rho_i^{A}= \sum_s |\psi_s^i\rangle\langle \psi_s^i | \ \ \ \text{and} \ \ \ \rho_i^{B}= \sum_t |\phi_t^i\rangle\langle \phi_t^i | \] 
			with $|\psi_s^i\rangle, |\phi_s^i\rangle$ unnormalized. Without loss of generality, consider $|\psi_s^i\rangle, |\phi_s^i\rangle$ to be even and odd fermionic state respectively. Hence, expanding in the basis $\mathcal{B_{A(B)}^P}$ we get
			\begin{eqnarray}
				|\psi_s^i\rangle\langle \psi_s^i |=\sum_{y_sz_s}{\alpha_{y_s}^i}{(\alpha_{z_s}^i)}^{\ast}|E_{y_s}^A\rangle\langle E_{z_s}^A| \ \ \ \text{and} \ \ \
				|\phi_t^i\rangle\langle \phi_t^i |=\sum_{m_tn_t}{\beta_{m_t}^i}{(\beta_{n_t}^i)}^{\ast}|O_{m_t}^B\rangle\langle O_{n_t}^B| 
				\label{eq:6}
			\end{eqnarray}
			Then 
			
			\begin{eqnarray}
				&&(U^A_{ABD} \wedge {U'}^B_{BD}) \ \rho_{sep}^{AB} \ ({U^A_{ABD}}^{\dagger} \wedge {{U'}^B_{BD}}^{\dagger}) \nonumber \\
				&& = \Bigg(\sum_{efgh}p_{ef} r_{gh}|E_e^A\rangle \wedge|E_g^B\rangle\langle O_f^A|\wedge\langle E_h^B| + p_{ef} s_{gh}|E_e^A\rangle \wedge|O_g^B\rangle\langle O_f^A|\wedge\langle O_h^B| + q_{ef} r_{gh}|O_e^A\rangle \wedge|E_g^B\rangle\langle E_f^A|\wedge\langle E_h^B| \Bigg.\nonumber \\
				&&+ \Bigg. q_{ef} s_{gh}|O_e^A\rangle \wedge|O_g^B\rangle\langle E_f^A|\wedge\langle O_h^B| \Bigg) \rho_{sep}^{AB} \Bigg(\sum_{ijkl}(p_{ij})^{\ast} (r_{kl})^{\ast}|O_j^A\rangle \wedge|E_l^B\rangle\langle E_i^A|\wedge\langle E_k^B| + (p_{ij})^{\ast} (s_{kl})^{\ast}|O_j^A\rangle \wedge|O_l^B\rangle\langle E_i^A|\wedge\langle O_k^B| \Bigg.\nonumber \\
				&& + \Bigg.(q_{ij})^{\ast} (r_{kl})^{\ast}|E_j^A\rangle \wedge|E_l^B\rangle\langle O_i^A|\wedge\langle E_k^B| +  (q_{ij})^{\ast} (s_{kl})^{\ast}|E_j^A\rangle \wedge|O_l^B\rangle\langle O_i^A|\wedge\langle O_k^B| \Bigg)\nonumber \\
				&& \text{Considering the \(i^{th}\) term of \(\rho_{sep}^{AB}\) and dropping the index \(i\) in Eq.~(\ref{eq:6}) we get} \nonumber \\
				&&= \sum_{st}\Bigg( \sum_{ei}\sum_{y_sz_s}(q_{ey_s}\alpha_{y_s})(q_{iz_s}\alpha_{z_s})^{\ast}|O_e^A\rangle\langle O_i^A|\Bigg) \wedge  \Bigg(\sum_{gk}\sum_{m_tn_t}(s_{gm_t}\beta_{m_t})(s_{kn_t}\beta_{n_t})^{\ast}|O_g^B\rangle\langle O_k^B|\Bigg) \nonumber \\
				&& + \ \ \text{terms with non-local parity}  \nonumber \\
				&&= \Sigma^A \wedge \Sigma^B + \ \ \text{terms with local-PSSR violation} \ \ \ \in SEP
			\end{eqnarray}
			
			One can then obtain the desired result by taking convex combination over index \(i\). Similar result can be obtained by replacing block-diagonal unitary with anti block-diagonal for system \(B\). $\blacksquare$
			
			\textit{Proof of lemma \ref{lm5}.} It is well known that, given G be a \textit{locally compact group} then, there exists a
			\textit{left Haar measure} on G \cite{Alfsen1963, Gleason}. Also, for G be a \textit{locally compact group}, and given \(\mu\), \(\mu '\) be
			two \textit{left Haar measures} on G. Then, \(\mu=\alpha \mu'\) for some \(\alpha \in \mathbb{R}^{+}\). We also know that every closed subgroup of a locally compact group is a locally compact group \cite{Halmos1976}.
			We also know that the Unitary group is closed and bounded. Hence it is compact due to Heine-borel theorem. Hence, we only need to show that \(U_{res}\) is a closed subgroup of the Unitary group. The closure property of \(U_{res}\) can be proved using the property that\[U_{BD(ABD)}^n=\{U \in \mathbb{M}_{BD(ABD)}(n,\mathbb{C})|U^{\dagger}U=UU^{\dagger}=\mathbb{I}\}\]
			where \(\mathbb{M}_{BD(ABD)}(n,\mathbb{C})\) represents the block(anti-block)-diagonal subspace of \(\mathbb{M}(n,\mathbb{C})\). To see that \(U_{res}\) constitutes a subgroup of the Unitary gorup, notice that the identity element \(\mathbb{I}_N\) is also an element of \(U_{res}\). Also, \(U_{BD(ABD)}^{-1}=U_{BD(ABD)}^{\dagger}\), hence an element in \(U_{res}\). Also, note that matrix multiplication of \(U_{BD(ABD)}\) with \(U_{BD(ABD)}\) is also an element of \(U_{res}\). Hence, \(U_{res}\) is a closed subgroup of the Unitary group. Hence, there exist a unique left Haar measure on the set \(U_{res}\). \(\blacksquare\)

			\section{\label{sec:a3}Reduction under Restricted Unitary group and Restricted Clifford Twirling}
			
			\textit{Proof of lemma \ref{lm6}.} To check the ovarlap of a given state \(\rho^f\) under restricted Unitary group twirling, notice that \begin{eqnarray}
				\text{tr}\Bigg[H_{N}^{|E\rangle \langle E|} \int_{U_{res}}dU (U \wedge U) \rho (U^{\dagger} \wedge U^{\dagger})\Bigg]=\int_{U_{res}}dU\text{tr}[H_{N}^{|E\rangle \langle E|}  \rho ]=\text{tr}[H_{N}^{|E\rangle \langle E|}  \rho ] \nonumber
			\end{eqnarray}
			where we have used Eq.~(\ref{eq:29}). Similarly, one can show that overlap of \(\rho^f\) with the operators \(\mathbb{I}^{|E\rangle \langle E|}+ \mathbb{I}^{|O\rangle \langle O|}\), \(\mathbb{I}^{|E\rangle \langle E|} - \mathbb{I}^{|O\rangle \langle O|} \) and \(H^{|O\rangle \langle O|}_N\) remains invariant under the restricted Unitary group twirling operation. \(\blacksquare\)
			
			\textit{Proof of lemma \ref{lm7}.} We prove it by contradiction. Let \(U \in C^{res}_{N}\) and \(U\) is given by
			\[\sum_{ij}u_{ij}|E_i\rangle\langle E_j|+ v_{ij}|O_i\rangle\langle O_j| +u'_{ij}|E_i\rangle\langle O_j|+v'_{ij}|O_i\rangle\langle E_j|\]
			Also, from Eq.~(\ref{eq:18}) we get, for 
			\[\Gamma^{ABD} \in \Xi^{ABD}  \implies U\Gamma^{ABD}U^{\dagger} \in \Xi^{ABD} \] with
			
			\[\Gamma^{ABD}=\sum_{mn}A_{mn}|E_m\rangle\langle O_n|+B_{mn}|O_m\rangle\langle E_n| \] Hence substituting the \(U\) and \(\Gamma^{ABD}\) we get the constraint for all \(i,p\)
			\begin{eqnarray}
				\sum_{mq}u'_{im}B_{mq} (u_{pq})^{\ast} =0 \ \ \ \ &&\sum_{mq}v'_{im}A_{mq} (v_{pq})^{\ast} =0  \nonumber \\
				\sum_{mq}u_{im}A_{mq} (u'_{pq})^{\ast} =0 \ \ \ \ &&\sum_{mq}v_{im}B_{mq} (v'_{pq})^{\ast} =0 \nonumber 
			\end{eqnarray}
			Let us assume that for a given \(a,b,c\) and \(d\)
			\begin{equation}
				u'_{ab} \neq 0 \ \text{and} \ u_{cd} \neq 0
				\label{eq:19}
			\end{equation}
			
			Now consider a matrix \(M_f=|E_d\rangle\langle O_b|+|O_b\rangle\langle E_d|\). Notice that the set \( \Xi^{ABD}\) constitute the basis set (apart from the multiplicative phase factor) for all complex anti- block diagonal matrices of dimension \(2^n \times 2^n\). Hence,  \(M_f\) can be expanded as
			\[M_f=\sum_i \Gamma^{ABD}_i \ \ \ \text{with} \ \ \ \Gamma^{ABD} \in \Xi^{ABD}\]
			Hence, 
			\[U M_f U^{\dagger}=\sum_i U\Gamma^{ABD}_i U^{\dagger} \ \in \ \Xi^{ABD}\]
			But notice, \(U M_f U^{\dagger}\) has a nonzero element in the basis \(|E_a\rangle\langle E_c|\) due to the condition in Eq.~(\ref{eq:19}) - a contradiction. \(\blacksquare\)
			
			\textit{Proof of Theorem \ref{th1}.} Statement (1) can be proven by considering the adjoint of Eq.~(\ref{eq:20}). To prove Statement (2), notice that \(\mathcal{T}_{C^{res}_{N}}[\mathbb{I}_N]=\mathbb{I}_N\) is obvious from Eq.~(\ref{eq:20}). Also, notice that,
			\begin{eqnarray}
				U_{BD}\Bigg(\sigma_z^{f} \wedge \dots \wedge \sigma_z^{f}\Bigg)U_{BD}^{\dagger}=&&\Bigg(\sum_{ij}u_{ij}|E_i\rangle\langle E_j|+ v_{ij}|O_i\rangle\langle O_j|\Bigg)\Bigg(\sum_k |E_k\rangle\langle E_k |-|O_k\rangle\langle O_k | \Bigg) \nonumber \\
				&&\Bigg(\sum_{pq}(u_{pq})^{\ast}|E_q\rangle\langle E_p|+ (v_{pq})^{\ast}|O_q\rangle\langle O_p|\Bigg) \nonumber \\
				=&&\Bigg(\sum_{ik}u_{ik}|E_i\rangle\langle E_k|- v_{ik}|O_i\rangle\langle O_k|\Bigg)\Bigg(\sum_{pq}(u_{pq})^{\ast}|E_q\rangle\langle E_p|+ (v_{pq})^{\ast}|O_q\rangle\langle O_p|\Bigg) \nonumber \\
				=&&\sum_{ipk}u_{ik}(u_{pk})^{\ast}|E_i\rangle\langle E_p|-v_{ik}(v_{pk})^{\ast}|O_i\rangle\langle O_p|=\sigma_z^{f} \wedge \dots \wedge \sigma_z^{f} \nonumber
			\end{eqnarray}
			where we have utilised the fact that
			\begin{equation}
				\sigma_z^{f} \wedge \dots \wedge \sigma_z^{f}=\sum_k |E_k\rangle\langle E_k |-|O_k\rangle\langle O_k 
				\label{eq:23}
			\end{equation}
			by rewriting the wedge product of fermionic single mode basis in terms of the elements of \(\mathcal{B_A^P}\). Similarly, one can show that \(U_{ABD}\Bigg(\sigma_z^{f} \wedge \dots \wedge \sigma_z^{f}\Bigg)U_{ABD}^{\dagger}=-\Bigg(\sigma_z^{f} \wedge \dots \wedge \sigma_z^{f}\Bigg)\). Hence for all \(c \in C^{res}_{N}\)
			\[c(\sigma_z^{f} \wedge \dots \wedge \sigma_z^{f})^{A} c^{\dagger} \wedge c(\sigma_z^{f} \wedge \dots \wedge \sigma_z^{f})^{B} c^{\dagger}=(\sigma_z^{f} \wedge \dots \wedge \sigma_z^{f})^{A} \wedge (\sigma_z^{f} \wedge \dots \wedge \sigma_z^{f})^{B})\]
			
			One can then prove the statement (2) using Eq.~(\ref{eq:20}). Before proving statement (3), notice that \(H_N\) can be written as 
			\(H_N=K_N\sum_{P \in P_N^H}P \wedge P\) with \(K_N\) being a constant, value of which depends on the number of modes \(N\) of the fermionic system. To see this, consider a \(1 \times 1\)-mode fermionic state given by
			\begin{eqnarray}
				|\psi^{+}\rangle \langle \psi^{+}|_{f}&&=\frac{1}{2}[|E\rangle \wedge | E\rangle \langle E| \wedge \langle E | + |O\rangle \wedge |O\rangle \langle O| \wedge \langle O| + |E\rangle \wedge | E\rangle \langle O| \wedge \langle O | + |O\rangle \wedge |O\rangle \langle E| \wedge \langle E| ]\nonumber
			\end{eqnarray}
			Define, partial transpose operator \((\mathbb{I} \wedge \mathbb{T}_f)\) on the Fermionic Hilbert Space. Hence, \begin{eqnarray}
				&&(\mathbb{I} \wedge \mathbb{T}_f)( |\psi^{+}\rangle \langle \psi^{+}|_{f}) = \frac{1}{2}\Bigg[|E\rangle \wedge | E\rangle \langle E| \wedge \langle E | + |O\rangle \wedge |O\rangle \langle O| \wedge \langle O| + |E\rangle \wedge | O\rangle \langle O| \wedge \langle E | + |O\rangle \wedge |E\rangle \langle E| \wedge \langle O| \Bigg]
				\label{eq:21}
			\end{eqnarray}
			Taking, N-fold wedge product of Eq.~(\ref{eq:21}) and rewriting in terms of the elements of the basis \(\mathcal{B_A^P}\),we get 
			\begin{eqnarray}
				[(\mathbb{I} \wedge \mathbb{T}_f)( |\psi^{+}\rangle \langle \psi^{+}|_{f})]^{\wedge N} =&&\frac{1}{2^N}\Bigg[\sum_{ij}|E_{i}\rangle \wedge | E_{j}\rangle \langle E_{j}| \wedge \langle E_{i} | +|O_{i}\rangle \wedge |O_{j}\rangle \langle O_{j}| \wedge \langle O_{i} |\nonumber\\
				&&+|E_{i}\rangle \wedge| O_{j}\rangle \langle O_{j}| \wedge \langle E_{i} | +|O_{i}\rangle \wedge |E_{j}\rangle \langle E_{j} |\wedge \langle O_{i} | \Bigg] =\frac{1}{2^N}H_N
			\end{eqnarray}
			Also, notice that, 
			\begin{eqnarray}
				I \wedge I+ \sigma_x^f \wedge \sigma_x^f +\sigma_z^f \wedge \sigma_z^f +\sigma_y^f \wedge \sigma_y^f =&&2 \Bigg[|E\rangle \wedge | E\rangle \langle E| \wedge \langle E | + \nonumber \\
				&&|O\rangle \wedge |O\rangle \langle O| \wedge \langle O| + |E\rangle \wedge | O\rangle \langle O| \wedge \langle E | + |O\rangle \wedge |E\rangle \langle E| \wedge \langle O| \Bigg] \nonumber \\
				&&=4[(\mathbb{I} \wedge \mathbb{T}_f)( |\psi^{+}\rangle \langle \psi^{+}|_{f})] 
			\end{eqnarray}
			Therefore, one can express \(H_N\) as 
			\begin{eqnarray}
				H_N=&& 2^N [(\mathbb{I} \wedge \mathbb{T}_f)( |\psi^{+}\rangle \langle \psi^{+}|_{f})]^{\wedge N} = \frac{2^N}{4^N}[I \wedge I+ \sigma_x^f \wedge \sigma_x^f +\sigma_z^f \wedge \sigma_z^f +\sigma_y^f \wedge \sigma_y^f]^{\wedge N} \nonumber \\
				=&& \frac{1}{2}\frac{1}{2^N} \sum_{P \in P_N^H}P \wedge P
			\end{eqnarray}
			Now, \(H^{|E\rangle \langle E|}_N\) is defined as the even subspace projection of \(H_N\) while \(H^{|O\rangle \langle O|}_N\) is defined as the odd subspace projection of \(H_N\). Thus, \(H^{|E\rangle \langle E|}_N\) and \(H^{|O\rangle \langle O|}_N\) can be expressed as
			\begin{eqnarray}
				H^{|E\rangle \langle E|}_N =\frac{1}{2}\frac{1}{2^N}\sum_{\Gamma \in \Xi_{BD}^H} \Gamma \wedge \Gamma  \ \ \ \text{and} \ \ \
				H^{|O\rangle \langle O|}_N =\frac{1}{2}\frac{1}{2^N}\sum_{\Gamma \in \Xi_{ABD}^H} \Gamma \wedge \Gamma  
				\label{eq:22}
			\end{eqnarray}
			where, \(\Xi_{BD}^H\) and \(\Xi_{ABD}^H\) consist of block-diagonal and anti-block diagonal elements of \(P_N^H\) respectively. Then,
			\begin{eqnarray}
				\mathcal{T}_{C^{res}_{N}}[H^{|E\rangle \langle E|}_N]=&&\frac{1}{|C^{res}_{N}|}\sum_{c \in C^{res}_{N}} (c \wedge c)H^{|E\rangle \langle E|}_N( c^{\dagger} \wedge c^{\dagger}) =\frac{1}{2^{N+1}}\frac{1}{|C^{res}_{N}|}\sum_{c \in C^{res}_{N}} \sum_{\Gamma \in \Xi_{BD}^H} (c\Gamma c^{\dagger})\wedge (c\Gamma c^{\dagger}) \nonumber \\
				&&=H^{|E\rangle \langle E|}_N
			\end{eqnarray}
			as \(c \in C^{res}_{N}\) permutes the elements of \(\Xi_{BD}^H\). Similarly, for \(H^{|O\rangle \langle O|}_N\), one can prove that \(\mathcal{T}_{C^{res}_{N}}[H^{|O\rangle \langle O|}_N]=H^{|O\rangle \langle O|}_N\) using the fact that \(c \in C^{res}_{N}\) permutes the elements of \(\Xi_{ABD}^H\). This proves the statement (3).
			
			To prove statement (4), notice that, any generic fermionic density matrix (PSSR-respected) of \(N \times N\)-mode bi-partite fermionic system can be expressed as
			\begin{equation}
				\rho^f=\sum_{\Gamma_i, \Gamma_j \in \Xi_{BD}^H} A_{ij}\Gamma_i \wedge \Gamma_j + \sum_{\Gamma_i, \Gamma_j \in \Xi_{ABD}^H} B_{ij} \Gamma_i \wedge \Gamma_j
			\end{equation}
			Now, following the procedure mentioned in \cite{DiVincenzo_2002}, choose a basis \( \Gamma_1 \wedge \Gamma_2 \in \Xi_{BD}^H \wedge \Xi_{BD}^H\) such that \(\Gamma_1 \neq \Gamma_2\). Also, let us assume that \(\Gamma_1 \neq \mathbb{I}_N\) and \(\Gamma_1 \neq \sigma_z^{f} \wedge \dots \wedge \sigma_z^{f}\). Consider an element \(c' \in  C^{res}_{N}\) such that \(c'\Gamma_1 c'^{\dagger}=\sigma_z^{f} \wedge \mathbb{I} \wedge \dots \wedge \mathbb{I}\) and \(c'\Gamma_2 c'^{\dagger}=\gamma=\mathbb{I} \wedge \sigma_z^{f} \wedge \dots \wedge \mathbb{I}\) or \(\sigma_x^{f} \wedge \sigma_x^{f} \wedge \mathbb{I} \wedge \dots \wedge \mathbb{I}\) or \(\mathbb{I}_N\) or \(\pm (\sigma_z^{f} \wedge \dots \wedge \sigma_z^{f})\) depending upon the condition \([\Gamma_1, \Gamma_2]=0\) or \(\{\Gamma_1, \Gamma_2\}=0\) or \(\Gamma_2=\mathbb{I}_N\) or \(\Gamma_2=\sigma_z^{f} \wedge \dots \wedge \sigma_z^{f}\). Hence,
			\begin{eqnarray}
				\mathcal{T}_{C^{res}_{N}}[\Gamma_1 \wedge \Gamma_2]=\frac{1}{|C^{res}_{N}|}\sum_{c \in C^{res}_{N}} (c\Gamma_1c^{\dagger}) \wedge (c\Gamma_2c^{\dagger}) =\frac{1}{|C^{res}_{N}|}\sum_{c \in C^{res}_{N}} \Bigg[(cc'^{\dagger}(\sigma_z^{f} \wedge \mathbb{I} \wedge \dots \wedge \mathbb{I})c'c^{\dagger}) \wedge (cc'^{\dagger}\gamma c' c^{\dagger})\Bigg] \nonumber
			\end{eqnarray}
			Notice that due to the closure property of the group \(C_N\), \(c.c'^{\dagger} \in C_N\). Also the set of all block-diagonal and anti-block diagonal matrices in the basis \(\mathcal{B^P_A}\) is closed under matrix multiplication. Hence, for any \(c,c' \in C^{res}_{N}\) and \(cc'^{\dagger} \in C_N\) implies \(cc'^{\dagger} \in C^{res}_N\). Let, \(cc'^{\dagger}=\tilde{c}\), then
			\begin{eqnarray}
				\mathcal{T}_{C^{res}_{N}}[\Gamma_1 \wedge \Gamma_2]=\frac{1}{|C^{res}_{N}|}\sum_{\tilde{c} \in C^{res}_{N}} \Bigg[&&(\tilde{c}(\sigma_z^{f} \wedge \mathbb{I} \wedge \dots \wedge \mathbb{I})\tilde{c}^{\dagger}) \wedge (\tilde{c}\gamma \tilde{c}^{\dagger})\Bigg] \nonumber
			\end{eqnarray}
			Notice, for any \(\tilde{c}\) we can always find a \(\tilde{c}'\) such that \[\tilde{c}' (\sigma_z^{f} \wedge \mathbb{I} \wedge \dots \wedge \mathbb{I})\tilde{c}'^{\dagger}=-\tilde{c}(\sigma_z^{f} \wedge \mathbb{I} \wedge \dots \wedge \mathbb{I})\tilde{c}^{\dagger}\] and \(\tilde{c}'\gamma \tilde{c}'^{\dagger}=\tilde{c}\gamma \tilde{c}^{\dagger}\). Hence \(\mathcal{T}_{C^{res}_{N}}[\Gamma_1 \wedge \Gamma_2]=0\). For,  \(\Gamma_1 = \Gamma_2\) 
			\begin{eqnarray}
				\mathcal{T}_{C^{res}_{N}}[\Gamma_1 \wedge \Gamma_1]=&&\frac{1}{|C^{res}_{N}|}\sum_{c \in C^{res}_{N}} c (\sigma_z^{f} \wedge \mathbb{I} \wedge \dots \wedge \mathbb{I})c^{\dagger} \wedge \ c (\sigma_z^{f} \wedge \mathbb{I} \wedge \dots \wedge \mathbb{I})c^{\dagger}
			\end{eqnarray}
			Now for \(c \in C^{res}_{N}\), \(c (\sigma_z^{f} \wedge \mathbb{I} \wedge \dots \wedge \mathbb{I})c^{\dagger}\) corresponds to all the elements in \(\Xi_{BD}^H-\{I, \sigma_z^{f} \wedge \dots \wedge \sigma_z^{f} \}\). Also, irrespective of the element \(c \in C^{res}_{N}\), number of occurrence of each elements of \(\Xi_{BD}^H-\{I, \sigma_z^{f} \wedge \dots \wedge \sigma_z^{f} \}\) in the sum will be same. This can be argued by noticing that all the elements \(c \in C_{N}\) that takes \(c (\sigma_z^{f} \wedge \mathbb{I} \wedge \dots \wedge \mathbb{I})c^{\dagger}\) to the elements in \(\Xi_{BD}^H-\{I, \sigma_z^{f} \wedge \dots \wedge \sigma_z^{f} \}\) are also part of \(c \in C^{res}_{N}\). Hence, if elements of \(C_N\) outputs the same occurrence of elements in \(\Xi_{BD}^H-\{I, \sigma_z^{f} \wedge \dots \wedge \sigma_z^{f} \}\), then so does the elements of \(C^{res}_{N}\). Hence we have, 
			\[ \mathcal{T}_{C^{res}_{N}}[\Gamma_1 \wedge \Gamma_1]=\frac{1}{|\Xi_{BD}^H|-2}\sum_{P \in \Xi_{BD}^H-\{I, \sigma_z^{f} \wedge \dots \wedge \sigma_z^{f} \}}P \wedge P\]
			Similarly, choosing a basis \( \Gamma_1 \wedge \Gamma_2 \in \Xi_{ABD}^H \wedge \Xi_{ABD}^H\) it can be proved that for \( \Gamma_1 \neq \Gamma_2\), we get \(\mathcal{T}_{C^{res}_{N}}[\Gamma_1 \wedge \Gamma_2]=0\) and 
			\[ \mathcal{T}_{C^{res}_{N}}[\Gamma_1 \wedge \Gamma_1]=\frac{1}{|\Xi_{ABD}^H|}\sum_{P \in \Xi_{ABD}^H}P \wedge P\]
			Hence, \(\mathcal{T}_{C^{res}_{N}}[\rho^f]\) can be expressed as the linear combinantion of \(\mathbb{I}_N \wedge \mathbb{I}_N, (\sigma_z^{f} \wedge \dots \wedge \sigma_z^{f} )\wedge (\sigma_z^{f} \wedge \dots \wedge \sigma_z^{f}), H^{|E\rangle \langle E|}_N \) and \(H^{|O\rangle \langle O|}_N\)
			Now notice that \(\mathbb{I}^{|E\rangle \langle E|}+ \mathbb{I}^{|O\rangle \langle O|} = \mathbb{I}\). Using Eq.~(\ref{eq:23}), it can be shown that \[\mathbb{I}^{|E\rangle \langle E|} - \mathbb{I}^{|O\rangle \langle O|} = (\sigma_z^{f} \wedge \dots \wedge \sigma_z^{f} )\wedge (\sigma_z^{f} \wedge \dots \wedge \sigma_z^{f})\]
			Hence, under the twirling \(\mathcal{T}_{C^{res}_{N}}\), any \(N \times N\)-mode fermionic state \(\rho^f\) can be reduced to the linear combination of \(\mathbb{I}^{|E\rangle \langle E|}, \ \mathbb{I}^{|O\rangle \langle O|}, \ H^{|E\rangle \langle E|}_N\) and \(H^{|O\rangle \langle O|}_N\). \(\blacksquare\)
			
			\section{\label{sec:a4}Invariance of Optimal Teleportation Fidelity}

			Notice that, \(F(\mathbb{I} \wedge \mathcal{T}^{\ast}_{C^{res}_{N}}[\mathcal{E}])\) can be written as
			
			\begin{eqnarray}
				=&&\frac{1}{|C^{res}_{N}|}\int_{\psi} d\psi \sum_{c \in C^{res}_{N}} \Bigg( \min_{\{E_m^{AB}\}} \sum_m \sqrt{ \text{tr}((\mathbb{I} \wedge c^{\dagger})[(\mathbb{I} \wedge \mathcal{E} )((\mathbb{I} \wedge c)(|\psi\rangle \langle \psi |)(\mathbb{I} \wedge c^{\dagger}))](\mathbb{I} \wedge c)(E_m^{AB}))\text{tr}(|\psi\rangle \langle \psi |E_m^{AB})} \Bigg)^{2} \nonumber\\
				=&& \frac{1}{|C^{res}_{N}|}\int_{\psi} d\psi \sum_{c \in C^{res}_{N}}  \Bigg(\min_{\{E_m^{AB}\}} \sum_m \sqrt{ \text{tr}((\mathbb{I} \wedge c^{\dagger})[(\mathbb{I} \wedge \mathcal{E})((\mathbb{I} \wedge c)(|\psi\rangle \langle \psi |)(\mathbb{I} \wedge c^{\dagger}))](\mathbb{I} \wedge c)(E_m^{AB}))}\nonumber \\
				&&\sqrt{\text{tr}((\mathbb{I} \wedge c^{\dagger})(\mathbb{I} \wedge c)|\psi\rangle \langle \psi |(\mathbb{I} \wedge c^{\dagger})(\mathbb{I} \wedge c)(E_m^{AB}))}\Bigg)^{2} 
			\end{eqnarray}
			Note that, the measures \(d\psi\) remains invariant under the change of variable \(\psi_{O}'=c^{ABD}\psi_{E}\) and \(\psi_{E}'=c^{ABD}\psi_{O}\) respectively  as the pure state ensemble \(\{\psi_{E},\psi_{O}\}\) as a whole remain invariant. Also, note that the set of POVM elements remains invariant under the operation \(\mathbb{I} \wedge c\). Using the cyclic property of trace, we get the invariance of the teleportation fidelity under restricted clifford twirling operation.
				\section{\label{sec:a5}General Teleportation Channel in FQT}
				\textit{Proof of lemma \ref{lm9}.} 
				Let us first consider the fermionic bi-partite state (global parity preserving) of the system \(\tilde{A}' \tilde{A}\) given by
				
				\begin{eqnarray}
					\rho_{\tilde{A}\tilde{A}' }^f=|\psi\rangle\langle \psi|_{\tilde{A}\tilde{A}' } =&&\sum_{\vec{\lambda}, \vec{\lambda}'}\sum_{\vec{\nu}, \vec{\nu}'}\eta_{\vec{\lambda}, \vec{\nu}}(\eta_{\vec{\lambda}', \vec{\nu}'})^{\ast} f_{(\tilde{A},\nu_1)}^{\dagger}f_{(\tilde{A},\nu_2)}^{\dagger}\dots f_{(\tilde{A},\nu_h)}^{\dagger}f_{(\tilde{A}',\lambda_1)}^{\dagger}f_{(\tilde{A}',\lambda_2)}^{\dagger}\dots f_{(\tilde{A}',\lambda_l)}^{\dagger}|\Omega\rangle\nonumber\\
					&& \langle\Omega|f_{(\tilde{A}',\lambda_{l'}')}\dots f_{(\tilde{A}',\lambda_2')}f_{(\tilde{A}',\lambda_1')}f_{(\tilde{A},\nu_{h'}')}\dots f_{(\tilde{A},\nu_2')}f_{(\tilde{A},\nu_1')}  
					\label{eq:10}
				\end{eqnarray}
                The weyl operator basis in d-dimension  \cite{Bertlmann_2008} is given by \[U_{pq}=\sum_{k=0}^{d-1}e^{\frac{2 \pi i p k}{d}} |k\rangle\langle(k+q) \ \text{mod} \ d| \ \ \ \ \ \ p,q=0,1, \dots, d-1\] 
                For dimension \(d=2^N\), these unitaries can be mapped to \(\{U_{BD}^x, U_{ABD}^x\}\) by corresponding even \(k\) with even basis and odd \(k\) with odd basis.
                
				Without loss of generality, we consider \(U_x\) to be block-diagonal unitary corresponds to outcome \(x\) and defined as
				\begin{eqnarray}
					U_{BD}^x=\sum_{\vec{z}, \vec{y}}R_{\vec{z}, \vec{y}}^{x}f_{z_1}^{\dagger}f_{z_2}^{\dagger}\dots f_{z_s}^{\dagger}|\Omega\rangle \langle \Omega |f_{y_t}\dots f_{y_2}f_{y_1}
				\end{eqnarray}
				with \(P(\vec{z})=P(\vec{y})\). Using this \(U_{BD}^x\) one can express \(M_x\) in fermionic system as 
				\begin{eqnarray}
					M_x=(U_{BD}^x \wedge \mathbb{I})|\psi^{+}\rangle \langle \psi^{+}|_{f}({U_{BD}^x}^{\dagger} \wedge \mathbb{I})
				\end{eqnarray}
				Expressing  \(\omega_{AB}^f \) in terms of fermionic creation and annihilation operator, we get
				\begin{eqnarray}
					\omega_{AB}^f &&=\frac{1}{d}\sum_{\vec{\sigma}, \vec{\sigma}', P(\vec{\sigma}) =P(\vec{\sigma}')} f_{(A,\sigma_1)}^{\dagger}f_{(A,\sigma_2)}^{\dagger}\dots f_{(A,\sigma_j)}^{\dagger}f_{(B,\sigma_1)}^{\dagger}f_{(B,\sigma_2)}^{\dagger}\dots f_{(B,\sigma_j)}^{\dagger}|\Omega\rangle\langle\Omega|f_{(B,\sigma_r')}\dots f_{(B,\sigma_2')}f_{(B,\sigma_1')}f_{(A,\sigma_r')}\dots \nonumber\\
					&&f_{(A,\sigma_2')}f_{(A,\sigma_1')}  +  \frac{\Upsilon}{d}\sum_{\vec{\sigma}, \vec{\sigma}', P(\vec{\sigma}) \neq P(\vec{\sigma}')}f_{(A,\sigma_1)}^{\dagger}f_{(A,\sigma_2)}^{\dagger}\dots f_{(A,\sigma_j)}^{\dagger}f_{(B,\sigma_1)}^{\dagger}f_{(B,\sigma_2)}^{\dagger}\dots f_{(B,\sigma_j)}^{\dagger}|\Omega\rangle\langle\Omega|f_{(B,\sigma_r')}\dots f_{(B,\sigma_2')}f_{(B,\sigma_1')}\nonumber\\
					&&f_{(A,\sigma_r')}\dots f_{(A,\sigma_2')}f_{(A,\sigma_1')}
					\label{eq:9}
				\end{eqnarray}
				Here \(P(.)\) denotes the parity of the vector.

				Now the fermionic singlet state \(|\psi^{+}\rangle \langle \psi^{+}|_{f}\) can be expressed in terms of fermionic creation and annihilation operator using Eq.~(\ref{eq:9}) and setting the value of \(\Upsilon=1\).Then \((M_x)_{\tilde{A}A}\) can be expressed as 
				\begin{eqnarray}
					(M_x)_{\tilde{A}A}=&&\frac{1}{d}\sum_{\vec{z}', \vec{y}'}\sum_{\vec{\pi},\vec{\pi}'}\sum_{\vec{z}, \vec{y}}R_{\vec{z}, \vec{y}}^{x}(R_{\vec{z}', \vec{y}'}^{x})^{\ast}f_{(\tilde{A},z_1)}^{\dagger}f_{(\tilde{A},z_2)}^{\dagger}\dots f_{(\tilde{A},z_s)}^{\dagger}|\Omega\rangle \langle \Omega |f_{(\tilde{A},y_t)}\dots f_{(\tilde{A},y_2)}f_{(\tilde{A},y_1)}f_{(\tilde{A},\pi_1)}^{\dagger}f_{(\tilde{A},\pi_2)}^{\dagger}\dots f_{(\tilde{A},\pi_q)}^{\dagger}\nonumber\\
					&&f_{(A,\pi_1)}^{\dagger}f_{(A,\pi_2)}^{\dagger}\dots f_{(A,\pi_q)}^{\dagger}|\Omega\rangle\langle\Omega|f_{(A,\pi_{q'}')}\dots f_{(A,\pi_2')}f_{(A,\pi_1')}f_{(\tilde{A},\pi_{q'}')}\dots f_{(\tilde{A},\pi_2')}f_{(\tilde{A},\pi_1')}f_{(\tilde{A},y_1')}^{\dagger} f_{(\tilde{A},y_2')}^{\dagger}\dots f_{(\tilde{A},y_{t'}')}^{\dagger}|\Omega\rangle \nonumber\\
					&& \langle \Omega | f_{(\tilde{A},z_{s'}')}\dots f_{(\tilde{A},z_2')}f_{(\tilde{A},z_1')} \nonumber\\
					=&&\sum_{\vec{z}', \vec{\pi}'}\sum_{\vec{z},\vec{\pi}}R_{\vec{z}, \vec{\pi}}^{x}(R_{\vec{z}', \vec{\pi}'}^{x})^{\ast}f_{(\tilde{A},z_1)}^{\dagger}f_{(\tilde{A},z_2)}^{\dagger}\dots f_{(\tilde{A},z_s)}^{\dagger}f_{(A,\pi_1)}^{\dagger}f_{(A,\pi_2)}^{\dagger}\dots f_{(A,\pi_q)}|\Omega\rangle\langle\Omega| f_{(A,\pi_{q'}')}\dots f_{(A,\pi_2')}f_{(A,\pi_1')}f_{(\tilde{A},z_{s'}')} \nonumber \\
					&&\dots f_{(\tilde{A},z_2')}f_{(\tilde{A},z_1')}
					\label{eq:11}
				\end{eqnarray}

				Let \(W^f_{AB}\) be any arbitrary bi-partite \(N \times N\)-mode Fermionic state shared between Alice and Bob. Initially Alice prepare a Fermionic state \(|\psi\rangle \langle \psi |_{\tilde{A}\tilde{A}'}\) locally and wish to teleport part of the state to Bob using the canonical form of the state \(W^f_{AB}\) as resource. In this case, Initial state can be written as, \((|\psi\rangle\langle \psi |_{\tilde{A}\tilde{A}'} \wedge (\omega_{noise}^f)_{AB})\) where \((\omega_{noise}^f)_{AB}=\mathcal{T}_{C^{res}_{N}}[W^f_{AB}]\) with \(\alpha,\ \beta, \ \Upsilon_1\) and \(\omega_{AB}^f\) depends on \(W^f_{AB}\). Upon performing Bell basis measurement by Alice with \(M_x= | \Psi_x\rangle\langle\Psi_x|_{\tilde{A} A}\), the \(x^{th}\) term of the initial state can be written as 
				\begin{eqnarray}
					&&\text{tr}_{\tilde{A} A}\Bigg[(|\psi\rangle\langle \psi |_{\tilde{A}\tilde{A}'} \wedge (\omega_{noise}^f)_{AB})( M_x \wedge \mathbb{I}_{B}\wedge\mathbb{I}_{\tilde{A}'} )\Bigg] \nonumber \\
					&&= \alpha d^2\text{tr}_{\tilde{A} A}\Bigg[\Bigg(|\psi\rangle\langle \psi |_{\tilde{A}\tilde{A}'} \wedge \Bigg(\frac{\mathbb{I}_N \wedge\mathbb{I}_N}{d^2} + \Upsilon_1\frac{\mathbb{S}_z \wedge \mathbb{S}_z}{d^2} \Bigg)\Bigg)\Bigg(M_x \wedge \mathbb{I}_{B}\wedge\mathbb{I}_{\tilde{A}'} \Bigg)\Bigg] 
					+ \beta d \text{tr}_{\tilde{A} A}\Bigg[(|\psi\rangle\langle \psi |_{\tilde{A}\tilde{A}'} \wedge (\omega^f)_{AB})\nonumber \\
					&&(M_x \wedge \mathbb{I}_{B}\wedge\mathbb{I}_{\tilde{A}'} )\Bigg]
				\end{eqnarray}
				Writing each term of the expression using creation and annihilation operators for the noisy part of the resource state, we get 
				\begin{eqnarray}
					&&\text{tr}_{\tilde{A} A}\Bigg[\Bigg(|\psi\rangle\langle \psi |_{\tilde{A}\tilde{A}'} \wedge \Bigg(\frac{\mathbb{I}_N \wedge\mathbb{I}_N}{d^2} + \Upsilon_1\frac{\mathbb{S}_z \wedge \mathbb{S}_z}{d^2}\Bigg) \Bigg)\Bigg(M_x \wedge \mathbb{I}_{B}\wedge\mathbb{I}_{\tilde{A}'} \Bigg)\Bigg] \nonumber \\
					&&=\text{tr}_{\tilde{A} A}\Bigg[\frac{1}{d^3}\Bigg( \sum_{\vec{\sigma}, \vec{d}}\sum_{\vec{\lambda}, \vec{\lambda}'}\sum_{\vec{\nu}, \vec{\nu}'}\eta_{\vec{\lambda}, \vec{\nu}}(\eta_{\vec{\lambda}', \vec{\nu}'})^{\ast} f_{(\tilde{A},\nu_1)}^{\dagger}f_{(\tilde{A},\nu_2)}^{\dagger}\dots f_{(\tilde{A},\nu_h)}^{\dagger}f_{(\tilde{A}',\lambda_1)}^{\dagger}f_{(\tilde{A}',\lambda_2)}^{\dagger}\dots f_{(\tilde{A}',\lambda_l)}^{\dagger}f_{(A,\sigma_1)}^{\dagger}f_{(A,\sigma_2)}^{\dagger}\dots  f_{(A,\sigma_j)}^{\dagger}f_{(B,d_1)}^{\dagger}\nonumber\\
					&&f_{(B,d_2)}^{\dagger}\dots f_{(B,d_g)}^{\dagger}|\Omega\rangle \langle\Omega|f_{(B,d_g)}\dots f_{(B,d_2)}f_{(B,d_1)}f_{(A,\sigma_j)}\dots f_{(A,\sigma_2)}f_{(A,\sigma_1)}f_{(\tilde{A}',\lambda_{l'}')} \dots f_{(\tilde{A}',\lambda_2')}f_{(\tilde{A}',\lambda_1')}f_{(\tilde{A},\nu_{h'}')}\dots f_{(\tilde{A},\nu_2')}\nonumber\\
					&&f_{(\tilde{A},\nu_1')} + \Upsilon_1 \sum_{\vec{\sigma}, \vec{d}, P(\vec{\sigma})=P(\vec{d})}\sum_{\vec{\lambda}, \vec{\lambda}'}\sum_{\vec{\nu}, \vec{\nu}'}\eta_{\vec{\lambda}, \vec{\nu}}(\eta_{\vec{\lambda}', \vec{\nu}'})^{\ast} f_{(\tilde{A},\nu_1)}^{\dagger}f_{(\tilde{A},\nu_2)}^{\dagger}\dots f_{(\tilde{A},\nu_h)}^{\dagger}f_{(\tilde{A}',\lambda_1)}^{\dagger}f_{(\tilde{A}',\lambda_2)}^{\dagger}\dots f_{(\tilde{A}',\lambda_l)}^{\dagger}f_{(A,\sigma_1)}^{\dagger}f_{(A,\sigma_2)}^{\dagger}\dots  f_{(A,\sigma_j)}^{\dagger}\nonumber\\
					&&f_{(B,d_1)}^{\dagger}f_{(B,d_2)}^{\dagger}\dots f_{(B,d_g)}^{\dagger}|\Omega\rangle \langle\Omega|f_{(B,d_g)}\dots f_{(B,d_2)}f_{(B,d_1)}f_{(A,\sigma_j)}\dots f_{(A,\sigma_2)}f_{(A,\sigma_1)}f_{(\tilde{A}',\lambda_{l'}')} \dots f_{(\tilde{A}',\lambda_2')}f_{(\tilde{A}',\lambda_1')}f_{(\tilde{A},\nu_{h'}')}\dots \nonumber\\
					&&f_{(\tilde{A},\nu_2')}f_{(\tilde{A},\nu_1')} - \Upsilon_1 \sum_{\vec{\sigma}, \vec{d}, P(\vec{\sigma}) \neq P(\vec{d})}\sum_{\vec{\lambda}, \vec{\lambda}'}\sum_{\vec{\nu}, \vec{\nu}'}\eta_{\vec{\lambda}, \vec{\nu}}(\eta_{\vec{\lambda}', \vec{\nu}'})^{\ast} f_{(\tilde{A},\nu_1)}^{\dagger}f_{(\tilde{A},\nu_2)}^{\dagger}\dots f_{(\tilde{A},\nu_h)}^{\dagger}f_{(\tilde{A}',\lambda_1)}^{\dagger}f_{(\tilde{A}',\lambda_2)}^{\dagger}\dots f_{(\tilde{A}',\lambda_l)}^{\dagger}f_{(A,\sigma_1)}^{\dagger}f_{(A,\sigma_2)}^{\dagger}\nonumber\\
					&&\dots  f_{(A,\sigma_j)}^{\dagger}f_{(B,d_1)}^{\dagger}f_{(B,d_2)}^{\dagger}\dots f_{(B,d_g)}^{\dagger}|\Omega\rangle \langle\Omega|f_{(B,d_g)}\dots f_{(B,d_2)}f_{(B,d_1)}f_{(A,\sigma_j)}\dots f_{(A,\sigma_2)}f_{(A,\sigma_1)}f_{(\tilde{A}',\lambda_{l'}')} \dots f_{(\tilde{A}',\lambda_2')}f_{(\tilde{A}',\lambda_1')}f_{(\tilde{A},\nu_{h'}')}\nonumber\\
					&&\dots f_{(\tilde{A},\nu_2')}f_{(\tilde{A},\nu_1')}\Bigg)\Bigg(\sum_{\vec{b}, \vec{c}} \sum_{\vec{z}', \vec{\pi}'}\sum_{\vec{z},\vec{\pi}}R_{\vec{z}, \vec{\pi}}^{x}(R_{\vec{z}', \vec{\pi}'}^{x})^{\ast}f_{(\tilde{A},z_1)}^{\dagger}f_{(\tilde{A},z_2)}^{\dagger}\dots f_{(\tilde{A},z_s)}^{\dagger}f_{(A,\pi_1)}^{\dagger}f_{(A,\pi_2)}^{\dagger}\dots  f_{(A,\pi_q)}^{\dagger}f_{(B,c_1)}^{\dagger}f_{(B,c_2)}^{\dagger}\dots f_{(B,c_t)}^{\dagger}\nonumber\\
					&&f_{(\tilde{A}',b_1)}^{\dagger}f_{(\tilde{A}',b_2)}^{\dagger}\dots f_{(\tilde{A}',b_{j'})}^{\dagger} |\Omega\rangle\langle\Omega|f_{(\tilde{A}',b_{j'})}\dots f_{(\tilde{A}',b_2)}f_{(\tilde{A}',b_1)}f_{(B,c_{t})}\dots f_{(B,c_2)}f_{(B,c_1)}f_{(A,\pi_{q'}')}\dots  f_{(A,\pi_2')}f_{(A,\pi_1')}\nonumber\\
					&&f_{(\tilde{A},z_{s'}')}\dots f_{(\tilde{A},z_2')}f_{(\tilde{A},z_1')}\Bigg)\Bigg] 
				\end{eqnarray}
				We follow the partial tracing procedure mentioned in \cite{friis2013} that utilizes the physical consistency condition to eliminate the possibility of any sign ambiguity. Rearranging the modes \(\tilde{A}\tilde{A}'AB \Rightarrow\tilde{A}AB\tilde{A}'\) and using the scaler product definition in wedge-product space we get
				\begin{eqnarray}
					&&\text{tr}_{\tilde{A} A}\Bigg[\Bigg(|\psi\rangle\langle \psi |_{\tilde{A}\tilde{A}'} \wedge \Bigg(\frac{\mathbb{I}_N \wedge\mathbb{I}_N}{d^2} + \Upsilon_1\frac{\mathbb{S}_z \wedge \mathbb{S}_z}{d^2}\Bigg) \Bigg)\Bigg(M_x \wedge \mathbb{I}_{B}\wedge\mathbb{I}_{\tilde{A}'}\Bigg)\Bigg] \nonumber \\
					&&=\frac{1}{d^3}\Bigg[ \sum_{\vec{\sigma}, \vec{d}}\sum_{\vec{\lambda}, \vec{\lambda}'}\sum_{\vec{\nu}, \vec{\nu}'} \eta_{\vec{\lambda}, \vec{\nu}}(\eta_{\vec{\lambda}', \vec{\nu}'})^{\ast} R_{\vec{\nu}', \vec{\sigma}}^{x}(R_{\vec{\nu}, \vec{\sigma}}^{x})^{\ast} f_{(B,d_1)}^{\dagger}f_{(B,d_2)}^{\dagger}\dots f_{(B,d_g)}^{\dagger}f_{(\tilde{A}',\lambda_1)}^{\dagger}f_{(\tilde{A}',\lambda_2)}^{\dagger}\dots f_{(\tilde{A}',\lambda_l)}^{\dagger}|\Omega\rangle \langle\Omega|f_{(\tilde{A}',\lambda_{j'}')} \dots f_{(\tilde{A}',\lambda_2')}\nonumber \\
					&&f_{(\tilde{A}',\lambda_1')}f_{(B,d_g)}\dots f_{(B,d_2)}f_{(B,d_1)} + \Upsilon_1 \sum_{\vec{\sigma}, \vec{d}, P(\vec{\sigma})=P(\vec{d})} \sum_{\vec{\lambda}, \vec{\lambda}'}\sum_{\vec{\nu}, \vec{\nu}'} \eta_{\vec{\lambda}, \vec{\nu}}(\eta_{\vec{\lambda}', \vec{\nu}'})^{\ast} R_{\vec{\nu}', \vec{\sigma}}^{x}(R_{\vec{\nu}, \vec{\sigma}}^{x})^{\ast} f_{(B,d_1)}^{\dagger}f_{(B,d_2)}^{\dagger}\dots f_{(B,d_g)}^{\dagger}f_{(\tilde{A}',\lambda_1)}^{\dagger}\nonumber \\
					&&f_{(\tilde{A}',\lambda_2)}^{\dagger}\dots f_{(\tilde{A}',\lambda_l)}^{\dagger}|\Omega\rangle \langle\Omega|f_{(\tilde{A}',\lambda_{j'}')} \dots f_{(\tilde{A}',\lambda_2')}f_{(\tilde{A}',\lambda_1')}f_{(B,d_g)}\dots f_{(B,d_2)}f_{(B,d_1)} - \Upsilon_1 \sum_{\vec{\sigma}, \vec{d}, P(\vec{\sigma}) \neq P(\vec{d})} \sum_{P(\vec{\lambda})= P(\vec{\lambda}')}\sum_{\vec{\nu}, \vec{\nu}'} \eta_{\vec{\lambda}, \vec{\nu}}(\eta_{\vec{\lambda}', \vec{\nu}'})^{\ast} \nonumber \\
					&& R_{\vec{\nu}', \vec{\sigma}}^{x}(R_{\vec{\nu}, \vec{\sigma}}^{x})^{\ast} f_{(B,d_1)}^{\dagger}f_{(B,d_2)}^{\dagger}\dots f_{(B,d_g)}^{\dagger}f_{(\tilde{A}',\lambda_1)}^{\dagger}f_{(\tilde{A}',\lambda_2)}^{\dagger}\dots f_{(\tilde{A}',\lambda_l)}^{\dagger}|\Omega\rangle \langle\Omega|f_{(\tilde{A}',\lambda_{j'}')} \dots f_{(\tilde{A}',\lambda_2')}f_{(\tilde{A}',\lambda_1')}f_{(B,d_g)}\dots f_{(B,d_2)}f_{(B,d_1)} \nonumber \\
                    &&+\Upsilon_1 \sum_{\vec{\sigma}, \vec{d}, P(\vec{\sigma}) \neq P(\vec{d})} \sum_{P(\vec{\lambda})\neq P(\vec{\lambda}')}\sum_{\vec{\nu}, \vec{\nu}'} \eta_{\vec{\lambda}, \vec{\nu}}(\eta_{\vec{\lambda}', \vec{\nu}'})^{\ast} R_{\vec{\nu}', \vec{\sigma}}^{x}(R_{\vec{\nu}, \vec{\sigma}}^{x})^{\ast} f_{(B,d_1)}^{\dagger}f_{(B,d_2)}^{\dagger}\dots f_{(B,d_g)}^{\dagger}f_{(\tilde{A}',\lambda_1)}^{\dagger}f_{(\tilde{A}',\lambda_2)}^{\dagger}\dots f_{(\tilde{A}',\lambda_l)}^{\dagger}|\Omega\rangle \langle\Omega|\nonumber \\
					&&f_{(\tilde{A}',\lambda_{j'}')} \dots f_{(\tilde{A}',\lambda_2')}f_{(\tilde{A}',\lambda_1')}f_{(B,d_g)}\dots f_{(B,d_2)}f_{(B,d_1)}\Bigg]\label{eq:noisy}
				\end{eqnarray}
				Notice that \(R^{x}\) represents PSSR-respected unitary. Hence, using properties of PSSR-respected unitary matrices, we get
				\begin{eqnarray}
					&&\text{tr}_{\tilde{A} A}\Bigg[\Bigg(|\psi\rangle\langle \psi |_{\tilde{A}\tilde{A}'} \wedge \Bigg(\frac{\mathbb{I}_N \wedge\mathbb{I}_N}{d^2} + \Upsilon_1\frac{\mathbb{S}_z \wedge \mathbb{S}_z}{d^2}\Bigg) \Bigg)\Bigg(M_x \wedge \mathbb{I}_{B}\wedge\mathbb{I}_{\tilde{A}'}\Bigg)\Bigg] \nonumber \\
					&&= \frac{1}{d^3}\Bigg[ \sum_{\vec{d}}\sum_{\vec{\lambda}, \vec{\lambda}'}\sum_{\vec{\nu}} \eta_{\vec{\lambda}, \vec{\nu}}(\eta_{\vec{\lambda}', \vec{\nu}})^{\ast} f_{(B,d_1)}^{\dagger}f_{(B,d_2)}^{\dagger}\dots f_{(B,d_g)}^{\dagger}f_{(\tilde{A}',\lambda_1)}^{\dagger}f_{(\tilde{A}',\lambda_2)}^{\dagger}\dots f_{(\tilde{A}',\lambda_l)}^{\dagger}|\Omega\rangle \langle\Omega|f_{(\tilde{A}',\lambda_{j'}')} \dots f_{(\tilde{A}',\lambda_2')}f_{(\tilde{A}',\lambda_1')}f_{(B,d_g)}\nonumber \\
					&&\dots f_{(B,d_2)}f_{(B,d_1)} + \Upsilon_1 \sum_{\vec{d}} \sum_{\vec{\lambda}, \vec{\lambda}'}\sum_{\vec{\nu}} \eta_{\vec{\lambda}, \vec{\nu}}(\eta_{\vec{\lambda}', \vec{\nu}})^{\ast} f_{(B,d_1)}^{\dagger}f_{(B,d_2)}^{\dagger}\dots f_{(B,d_g)}^{\dagger}f_{(\tilde{A}',\lambda_1)}^{\dagger}f_{(\tilde{A}',\lambda_2)}^{\dagger}\dots f_{(\tilde{A}',\lambda_l)}^{\dagger} |\Omega\rangle \langle\Omega|f_{(\tilde{A}',\lambda_{j'}')} \dots \nonumber \\
					&& f_{(\tilde{A}',\lambda_2')}f_{(\tilde{A}',\lambda_1')}f_{(B,d_g)}\dots f_{(B,d_2)}f_{(B,d_1)} - \Upsilon_1 \sum_{\vec{d}} \sum_{\vec{\lambda}, \vec{\lambda}'}\sum_{\vec{\nu}} \eta_{\vec{\lambda}, \vec{\nu}}(\eta_{\vec{\lambda}', \vec{\nu}})^{\ast}  f_{(B,d_1)}^{\dagger}f_{(B,d_2)}^{\dagger}\dots f_{(B,d_g)}^{\dagger}f_{(\tilde{A}',\lambda_1)}^{\dagger}f_{(\tilde{A}',\lambda_2)}^{\dagger}\dots f_{(\tilde{A}',\lambda_l)}^{\dagger}\nonumber \\
					&&|\Omega\rangle \langle\Omega|f_{(\tilde{A}',\lambda_{j'}')} \dots f_{(\tilde{A}',\lambda_2')}f_{(\tilde{A}',\lambda_1')}f_{(B,d_g)}\dots f_{(B,d_2)}f_{(B,d_1)}\Bigg] 
                    \label{eq:noisy1}
				\end{eqnarray}

                \begin{table}[t]
\caption{Condition on parity in Eq.~(\ref{eq:noisy}) and Eq.~(\ref{eq:noisy1})}
\label{tab:properties}
\begin{ruledtabular}
\begin{tabular}{lcccc}
Unitary(x) &  condition & even parity input state & odd parity input state \\
\hline
Block-diagonal (\(P(\vec{\nu})=P(\vec{\sigma}) \) )             & \(P(\vec{d})=P(\vec{\sigma}) \) & \(P(\vec{\lambda})=P(\vec{\lambda}')=P(\vec{d})\) & \(P(\vec{\lambda}) =P(\vec{\lambda}')\neq P(\vec{d})\) \\
Block-diagonal (\(P(\vec{\nu})=P(\vec{\sigma}) \) )             & \(P(\vec{d})\neq P(\vec{\sigma}) \) & \(P(\vec{\lambda})=P(\vec{\lambda}')\neq P(\vec{d})\) & \(P(\vec{\lambda}) =P(\vec{\lambda}')= P(\vec{d})\) \\
Anti Block-diagonal (\(P(\vec{\nu}) \neq P(\vec{\sigma}) \) )   & \(P(\vec{d})=P(\vec{\sigma}) \) & \(P(\vec{\lambda})=P(\vec{\lambda}')\neq P(\vec{d})\) & \(P(\vec{\lambda}) =P(\vec{\lambda}')= P(\vec{d})\) \\
Anti Block-diagonal (\(P(\vec{\nu}) \neq P(\vec{\sigma}) \) )   & \(P(\vec{d})\neq P(\vec{\sigma}) \) & \(P(\vec{\lambda})=P(\vec{\lambda}')=P(\vec{d})\) & \(P(\vec{\lambda}) =P(\vec{\lambda}')\neq P(\vec{d})\)
\end{tabular}
\end{ruledtabular}
\end{table}
				with \(P(\vec{\lambda})=P(\vec{\lambda}')=P(\vec{d})\) in the second term and \(P(\vec{\lambda})=P(\vec{\lambda}') \neq P(\vec{d})\) in the third term, given the corresponding \(x\) represents block-diagonal unitary and the input state is of even parity. For \(x\) representing anti block-diagonal unitary, \(P(\vec{\lambda})=P(\vec{\lambda}') \neq P(\vec{d})\) satisfies in the second term and \(P(\vec{\lambda})=P(\vec{\lambda}')=P(\vec{d})\) in the third term (see Table~\ref{tab:properties}). 
				
				Based on the outcome \(x\), Bob then applies the corresponding Unitary operation. Hence, the output state due to noisy part of the resource is given by
				\begin{eqnarray}
					(\rho^f_{B\tilde{A}'})_{noisy}&&=\frac{1}{d^3}\Bigg[ \sum_{\vec{p}, \vec{q}}\sum_{\vec{p}', \vec{q}'}\sum_{\vec{d}}\sum_{\vec{\lambda}, \vec{\lambda}'}\sum_{\vec{\nu}} R_{\vec{p}, \vec{q}}^{x}(R_{\vec{p}', \vec{q}'}^{x})^{\ast}\eta_{\vec{\lambda}, \vec{\nu}}(\eta_{\vec{\lambda}', \vec{\nu}})^{\ast}f_{(B,p_1)}^{\dagger}f_{(B,p_2)}^{\dagger}\dots f_{(B,p_m)}^{\dagger}|\Omega\rangle \langle\Omega|f_{(B,q_n)}\dots f_{(B,q_2)}f_{(B,q_1)}\nonumber \\
					&&
					f_{(B,d_1)}^{\dagger}f_{(B,d_2)}^{\dagger}\dots f_{(B,d_g)}^{\dagger}f_{(\tilde{A}',\lambda_1)}^{\dagger}f_{(\tilde{A}',\lambda_2)}^{\dagger}\dots f_{(\tilde{A}',\lambda_l)}^{\dagger}|\Omega\rangle \langle\Omega|f_{(\tilde{A}',\lambda_{j'}')} \dots f_{(\tilde{A}',\lambda_2')}f_{(\tilde{A}',\lambda_1')}f_{(B,d_g)}\dots f_{(B,d_2)}f_{(B,d_1)}\nonumber \\
					&&f_{(B,q_1')}^{\dagger}f_{(B,q_2')}^{\dagger}\dots f_{(B,q_{n'}')}^{\dagger}|\Omega\rangle \langle\Omega| f_{(B,p_{m'}')}\dots f_{(B,p_2')}f_{(B,p_1')} + \Upsilon_1 \sum_{\vec{p}, \vec{q}}\sum_{\vec{p}', \vec{q}'}\sum_{\vec{d}} \sum_{\vec{\lambda}, \vec{\lambda}'}\sum_{\vec{\nu}} R_{\vec{p}, \vec{q}}^{x}(R_{\vec{p}', \vec{q}'}^{x})^{\ast}\eta_{\vec{\lambda}, \vec{\nu}}(\eta_{\vec{\lambda}', \vec{\nu}})^{\ast} \nonumber \\
					&& f_{(B,p_1)}^{\dagger}f_{(B,p_2)}^{\dagger}\dots f_{(B,p_m)}^{\dagger}|\Omega\rangle \langle\Omega|f_{(B,q_n)}\dots f_{(B,q_2)}f_{(B,q_1)}
					f_{(B,d_1)}^{\dagger}f_{(B,d_2)}^{\dagger}\dots f_{(B,d_g)}^{\dagger}f_{(\tilde{A}',\lambda_1)}^{\dagger}f_{(\tilde{A}',\lambda_2)}^{\dagger}\dots f_{(\tilde{A}',\lambda_l)}^{\dagger}\nonumber \\
					&&|\Omega\rangle \langle\Omega|f_{(\tilde{A}',\lambda_{j'}')} \dots f_{(\tilde{A}',\lambda_2')}f_{(\tilde{A}',\lambda_1')}f_{(B,d_g)}\dots f_{(B,d_2)}f_{(B,d_1)}f_{(B,q_1')}^{\dagger}f_{(B,q_2')}^{\dagger}\dots f_{(B,q_{n'}')}^{\dagger}|\Omega\rangle \langle\Omega|f_{(B,p_{m'}')}\dots f_{(B,p_2')}\nonumber \\
					&&f_{(B,p_1')} - \Upsilon_1 \sum_{\vec{p}, \vec{q}}\sum_{\vec{p}', \vec{q}'}\sum_{\vec{d}} \sum_{\vec{\lambda}, \vec{\lambda}'}\sum_{\vec{\nu}} R_{\vec{p}, \vec{q}}^{x}(R_{\vec{p}', \vec{q}'}^{x})^{\ast}\eta_{\vec{\lambda}, \vec{\nu}}(\eta_{\vec{\lambda}', \vec{\nu}})^{\ast}  f_{(B,p_1)}^{\dagger}f_{(B,p_2)}^{\dagger}\dots f_{(B,p_m)}^{\dagger}|\Omega\rangle \langle\Omega|f_{(B,q_n)}\dots f_{(B,q_2)}\nonumber \\
					&&f_{(B,q_1)}f_{(B,d_1)}^{\dagger}f_{(B,d_2)}^{\dagger}\dots f_{(B,d_g)}^{\dagger}f_{(\tilde{A}',\lambda_1)}^{\dagger}f_{(\tilde{A}',\lambda_2)}^{\dagger}\dots f_{(\tilde{A}',\lambda_l)}^{\dagger}|\Omega\rangle \langle\Omega|f_{(\tilde{A}',\lambda_{j'}')} \dots f_{(\tilde{A}',\lambda_2')}f_{(\tilde{A}',\lambda_1')}f_{(B,d_g)}\dots f_{(B,d_2)}\nonumber \\
					&&f_{(B,d_1)}f_{(B,q_1')}^{\dagger}f_{(B,q_2')}^{\dagger}\dots f_{(B,q_{n'}')}^{\dagger}|\Omega\rangle \langle\Omega|f_{(B,p_{m'}')}\dots f_{(B,p_2')}f_{(B,p_1')}\Bigg] \nonumber \\
					&&=\frac{1}{d^3}\Bigg[ \sum_{\vec{p}}\sum_{\vec{\lambda}, \vec{\lambda}'}\sum_{\vec{\nu}} \eta_{\vec{\lambda}, \vec{\nu}}(\eta_{\vec{\lambda}', \vec{\nu}})^{\ast} f_{(B,p_1)}^{\dagger}f_{(B,p_2)}^{\dagger}\dots f_{(B,p_m)}^{\dagger}f_{(\tilde{A}',\lambda_1)}^{\dagger}f_{(\tilde{A}',\lambda_2)}^{\dagger}\dots f_{(\tilde{A}',\lambda_l)}^{\dagger}|\Omega\rangle \langle\Omega|f_{(\tilde{A}',\lambda_{j'}')} \dots f_{(\tilde{A}',\lambda_2')}\nonumber \\
					&&f_{(\tilde{A}',\lambda_1')}f_{(B,p_m)}\dots f_{(B,p_2)}f_{(B,p_1)} + \Upsilon_1 \sum_{\vec{d}} \sum_{\vec{\lambda}, \vec{\lambda}'}\sum_{\vec{\nu}} \eta_{\vec{\lambda}, \vec{\nu}}(\eta_{\vec{\lambda}', \vec{\nu}})^{\ast}  f_{(B,p_1)}^{\dagger}f_{(B,p_2)}^{\dagger}\dots f_{(B,p_m)}^{\dagger}f_{(\tilde{A}',\lambda_1)}^{\dagger}f_{(\tilde{A}',\lambda_2)}^{\dagger}\dots\nonumber \\
					&& f_{(\tilde{A}',\lambda_l)}^{\dagger}|\Omega\rangle \langle\Omega|f_{(\tilde{A}',\lambda_{j'}')} \dots f_{(\tilde{A}',\lambda_2')}f_{(\tilde{A}',\lambda_1')}f_{(B,p_m)}\dots f_{(B,p_2)}f_{(B,p_1)} - \Upsilon_1 \sum_{\vec{d}} \sum_{\vec{\lambda}, \vec{\lambda}'}\sum_{\vec{\nu}} \eta_{\vec{\lambda}, \vec{\nu}}(\eta_{\vec{\lambda}', \vec{\nu}})^{\ast} f_{(B,p_1)}^{\dagger}f_{(B,p_2)}^{\dagger}\nonumber \\
					&&\dots f_{(B,p_m)}^{\dagger}f_{(\tilde{A}',\lambda_1)}^{\dagger}f_{(\tilde{A}',\lambda_2)}^{\dagger}\dots f_{(\tilde{A}',\lambda_l)}^{\dagger} |\Omega\rangle \langle\Omega|f_{(\tilde{A}',\lambda_{j'}')} \dots f_{(\tilde{A}',\lambda_2')}f_{(\tilde{A}',\lambda_1')}f_{(B,p_m)}\dots f_{(B,p_2)}f_{(B,p_1)}\Bigg]\label{eq:noisy2}
				\end{eqnarray}
				
				with \(P(\vec{\lambda})=P(\vec{\lambda}')=P(\vec{p})\) in the second term and \(P(\vec{\lambda})=P(\vec{\lambda}') \neq P(\vec{p})\) in the third term of Eq.~(\ref{eq:noisy2}) irrespective of \(x\). Hence,
				\begin{eqnarray}
					(\rho^f_{B\tilde{A}'})_{noisy}= \frac{\mathbb{I}_N\wedge\rho^f_{\tilde{A}'} }{d^3} + \Upsilon_1\frac{(\mathbb{I}_N \wedge\rho^f_{\tilde{A}'})(\mathbb{S}_z \wedge \mathbb{S}_z)}{d^3} \nonumber 
				\end{eqnarray}
				Summing over all \(x\) we get the noisy part of Eq.~(\ref{eq:25}) for even input states. For odd input state, notice that, \(P(\vec{\lambda})=P(\vec{\lambda}') \neq P(\vec{p})\) in the second term and \(P(\vec{\lambda})=P(\vec{\lambda}') = P(\vec{p})\) in the third term irrespective of \(x\). Hence, 
				\begin{eqnarray}
					(\rho^f_{B\tilde{A}'})_{noisy}= \frac{\mathbb{I}_N \wedge\rho^f_{\tilde{A}'}}{d^3} - \Upsilon_1\frac{(\mathbb{I}_N \wedge\rho^f_{\tilde{A}'})(\mathbb{S}_z \wedge \mathbb{S}_z)}{d^3} \nonumber 
				\end{eqnarray}
				In this case also, summing over all \(x\) we get the noisy part of Eq.~(\ref{eq:25}) for odd input states. Proceeding similarly for the exact part, we get
				\begin{eqnarray}
					&&\text{tr}_{\tilde{A} A}\Bigg[(|\psi\rangle\langle \psi |_{\tilde{A}\tilde{A}'} \wedge (\omega^f)_{AB})
					(M_x \wedge \mathbb{I}_{B}\wedge\mathbb{I}_{\tilde{A}'})\Bigg] \nonumber\\
					&&=\text{tr}_{\tilde{A} A}\Bigg[\frac{1}{d^2}\Bigg( \sum_{\vec{\sigma}, \vec{\sigma}', P(\vec{\sigma}) =P(\vec{\sigma}')}\sum_{\vec{\lambda}, \vec{\lambda}'}\sum_{\vec{\nu}, \vec{\nu}'}\eta_{\vec{\lambda}, \vec{\nu}}(\eta_{\vec{\lambda}', \vec{\nu}'})^{\ast} f_{(\tilde{A},\nu_1)}^{\dagger}f_{(\tilde{A},\nu_2)}^{\dagger}\dots f_{(\tilde{A},\nu_h)}^{\dagger}f_{(\tilde{A}',\lambda_1)}^{\dagger}f_{(\tilde{A}',\lambda_2)}^{\dagger}\dots f_{(\tilde{A}',\lambda_l)}^{\dagger}f_{(A,\sigma_1)}^{\dagger}f_{(A,\sigma_2)}^{\dagger}\dots  \nonumber\\
					&&f_{(A,\sigma_j)}^{\dagger}f_{(B,\sigma_1)}^{\dagger}f_{(B,\sigma_2)}^{\dagger}\dots f_{(B,\sigma_j)}^{\dagger}|\Omega\rangle \langle\Omega|f_{(B,\sigma_r')}\dots f_{(B,\sigma_2')}f_{(B,\sigma_1')}f_{(A,\sigma_r')}\dots f_{(A,\sigma_2')}f_{(A,\sigma_1')}f_{(\tilde{A}',\lambda_{l'}')} \dots f_{(\tilde{A}',\lambda_2')}f_{(\tilde{A}',\lambda_1')}\nonumber\\
					&&f_{(\tilde{A},\nu_{h'}')}\dots f_{(\tilde{A},\nu_2')}f_{(\tilde{A},\nu_1')} + \Upsilon \sum_{\vec{\sigma}, \vec{\sigma}', P(\vec{\sigma}) \neq P(\vec{\sigma}')}\sum_{\vec{\lambda}, \vec{\lambda}'}\sum_{\vec{\nu}, \vec{\nu}'}\eta_{\vec{\lambda}, \vec{\nu}}(\eta_{\vec{\lambda}', \vec{\nu}'})^{\ast} f_{(\tilde{A},\nu_1)}^{\dagger}f_{(\tilde{A},\nu_2)}^{\dagger}\dots f_{(\tilde{A},\nu_h)}^{\dagger}f_{(\tilde{A}',\lambda_1)}^{\dagger}f_{(\tilde{A}',\lambda_2)}^{\dagger}\dots\nonumber\\
					&& f_{(\tilde{A}',\lambda_l)}^{\dagger}f_{(A,\sigma_1)}^{\dagger}f_{(A,\sigma_2)}^{\dagger}\dots f_{(A,\sigma_j)}^{\dagger} f_{(B,\sigma_1)}^{\dagger}f_{(B,\sigma_2)}^{\dagger}\dots f_{(B,\sigma_j)}^{\dagger}|\Omega\rangle \langle\Omega|f_{(B,\sigma_r')}\dots f_{(B,\sigma_2')}f_{(B,\sigma_1')}f_{(A,\sigma_r')}\dots f_{(A,\sigma_2')}f_{(A,\sigma_1')}\nonumber\\
					&&f_{(\tilde{A}',\lambda_{l'}')} \dots f_{(\tilde{A}',\lambda_2')}f_{(\tilde{A}',\lambda_1')}f_{(\tilde{A},\nu_{h'}')}\dots f_{(\tilde{A},\nu_2')}f_{(\tilde{A},\nu_1')}\Bigg)\Bigg(\sum_{\vec{b}, \vec{c}} \sum_{\vec{z}', \vec{\pi}'}\sum_{\vec{z},\vec{\pi}}R_{\vec{z}, \vec{\pi}}^{x}(R_{\vec{z}', \vec{\pi}'}^{x})^{\ast}f_{(\tilde{A},z_1)}^{\dagger}f_{(\tilde{A},z_2)}^{\dagger}\dots f_{(\tilde{A},z_s)}^{\dagger}\nonumber\\
					&&f_{(A,\pi_1)}^{\dagger}f_{(A,\pi_2)}^{\dagger}\dots  f_{(A,\pi_q)}^{\dagger}f_{(B,c_1)}^{\dagger}f_{(B,c_2)}^{\dagger}\dots f_{(B,c_t)}^{\dagger}f_{(\tilde{A}',b_1)}^{\dagger}f_{(\tilde{A}',b_2)}^{\dagger}\dots f_{(\tilde{A}',b_{j'})}^{\dagger} |\Omega\rangle\langle\Omega|f_{(\tilde{A}',b_{j'})}\dots f_{(\tilde{A}',b_2)}f_{(\tilde{A}',b_1)}f_{(B,c_{t})}\dots \nonumber\\  &&f_{(B,c_2)}f_{(B,c_1)}f_{(A,\pi_{q'}')}\dots f_{(A,\pi_2')}f_{(A,\pi_1')}f_{(\tilde{A},z_{s'}')}\dots f_{(\tilde{A},z_2')}f_{(\tilde{A},z_1')}\Bigg)\Bigg] \nonumber \\
					&&= \frac{1}{d^2}\Bigg( \sum_{\vec{\sigma}, \vec{\sigma}', P(\vec{\sigma}) =P(\vec{\sigma}')}\sum_{\vec{\lambda}, \vec{\lambda}'}\sum_{\vec{\nu}, \vec{\nu}'} R_{\vec{\nu'}, \vec{\sigma'}}^{x}(R_{\vec{\nu}, \vec{\sigma}}^{x})^{\ast}\eta_{\vec{\lambda}, \vec{\nu}}(\eta_{\vec{\lambda}', \vec{\nu}'})^{\ast} f_{(B,\sigma_1)}^{\dagger}f_{(B,\sigma_2)}^{\dagger}\dots f_{(B,\sigma_j)}^{\dagger}f_{(\tilde{A}',\lambda_1)}^{\dagger}f_{(\tilde{A}',\lambda_2)}^{\dagger}\dots f_{(\tilde{A}',\lambda_l)}^{\dagger}\nonumber\\
					&&|\Omega\rangle \langle\Omega|f_{(\tilde{A}',\lambda_{l'}')} \dots f_{(\tilde{A}',\lambda_2')}f_{(\tilde{A}',\lambda_1')}f_{(B,\sigma_r')}\dots f_{(B,\sigma_2')}f_{(B,\sigma_1')} + \Upsilon \sum_{\vec{\sigma}, \vec{\sigma}', P(\vec{\sigma}) \neq P(\vec{\sigma}')}\sum_{\vec{\lambda}, \vec{\lambda}'}\sum_{\vec{\nu}, \vec{\nu}'}R_{\vec{\nu'}, \vec{\sigma'}}^{x}(R_{\vec{\nu}, \vec{\sigma}}^{x})^{\ast}\eta_{\vec{\lambda}, \vec{\nu}}(\eta_{\vec{\lambda}', \vec{\nu}'})^{\ast}\nonumber\\
					&&  f_{(B,\sigma_1)}^{\dagger}f_{(B,\sigma_2)}^{\dagger}\dots f_{(B,\sigma_j)}^{\dagger}f_{(\tilde{A}',\lambda_1)}^{\dagger}f_{(\tilde{A}',\lambda_2)}^{\dagger}\dots f_{(\tilde{A}',\lambda_l)}^{\dagger}|\Omega\rangle \langle\Omega|f_{(\tilde{A}',\lambda_{l'}')} \dots f_{(\tilde{A}',\lambda_2')}f_{(\tilde{A}',\lambda_1')}f_{(B,\sigma_r')}\dots f_{(B,\sigma_2')}f_{(B,\sigma_1')}\Bigg)
				\end{eqnarray}

				Again, based on the outcome \(x\), Bob then applies the corresponding Unitary operation. Summing over all possible outcome gives us the the output state
				
				\begin{eqnarray}
					(\rho^f_{B\tilde{A}'})_{exact}=&&\sum_{\vec{\lambda},\vec{\lambda}',\vec{\nu},\vec{\nu}', P(\vec{\nu})= P(\vec{\nu}')}f_{(B,\nu_1)}^{\dagger}f_{(B,\nu_2)}^{\dagger}\dots f_{(B,\nu_h)}^{\dagger}f_{(\tilde{A}',\lambda_1)}^{\dagger}f_{(\tilde{A}',\lambda_2)}^{\dagger}\dots f_{(\tilde{A}',\lambda_l)}^{\dagger}|\Omega\rangle \langle\Omega| f_{(\tilde{A}',\lambda_{j'}')} \dots f_{(\tilde{A}',\lambda_2')}f_{(\tilde{A}',\lambda_1')}f_{(B,\nu_m')}\nonumber \\
					&& \dots f_{(B,\nu_2')} f_{(B,\nu_1')} + \Upsilon \sum_{\vec{\lambda},\vec{\lambda}',\vec{\nu},\vec{\nu}', P(\vec{\nu}) \neq P(\vec{\nu}')}f_{(B,\nu_1)}^{\dagger}f_{(B,\nu_2)}^{\dagger}\dots f_{(B,\nu_h)}^{\dagger}f_{(\tilde{A}',\lambda_1)}^{\dagger}f_{(\tilde{A}',\lambda_2)}^{\dagger}\dots f_{(\tilde{A}',\lambda_l)}^{\dagger}|\Omega\rangle \langle\Omega| f_{(\tilde{A}',\lambda_{j'}')} \dots\nonumber\\
					&& f_{(\tilde{A}',\lambda_2')}f_{(\tilde{A}',\lambda_1')}f_{(B,\nu_m')}\dots f_{(B,\nu_2')} f_{(B,\nu_1')}\nonumber\\
					=&&\text{loc-p}(|\psi\rangle \langle \psi |_{B\tilde{A}'}) + \Upsilon \ \overline{\text{loc-p}}(|\psi\rangle \langle \psi |_{B\tilde{A}'})
				\end{eqnarray}
				
				Hence, the output state of the teleportation channel can be expressed as in Eq.~(\ref{eq:25}). \(\blacksquare\)
				
				To find the teleportation fidelity of the give teleportation channel for the exact case i.e. to prove \(F(\mathbb{I} \wedge \mathcal{E})_{exact}=1\), notice,
				\begin{eqnarray}
					F(\mathbb{I} \wedge \mathcal{E}) =&&\int d\psi  \Bigg( \min_{\{E_m^{B\tilde{A}'}\}} \sum_m \sqrt{\text{tr}(|\psi\rangle \langle \psi |E_m^{B\tilde{A}'}) \text{tr}((\mathbb{I} \wedge \mathcal{E}_{tel}(|\psi\rangle \langle \psi |))E_m^{B\tilde{A}'})}\Bigg)^{2} \nonumber\\
					=&&\int d\psi  \Bigg( \min_{\{E_m^{B\tilde{A}'}\}} \sum_m \sqrt{\text{tr}(\text{loc-p}(|\psi\rangle \langle \psi |)E_m^{B\tilde{A}'}) \text{tr}(\text{loc-p}(|\psi\rangle \langle \psi |)E_m^{B\tilde{A}'})}\Bigg)^{2} \nonumber \\
					=&&\int d\psi \Bigg(  \min_{\{E_m^{B\tilde{A}'}\}} \sum_m \text{tr}(\text{loc-p}(|\psi\rangle \langle \psi |)E_m^{B\tilde{A}'}) \Bigg)^{2} =\int d\psi=1 
				\end{eqnarray}  
				where we have used the relation \(\sum_m E_m^{B\tilde{A}'}=\mathbb{I}\) and \(\text{tr}(\text{loc-p}(|\psi\rangle \langle \psi |))=1\). 
				
				For the noisy part of the channel, the entanglement fidelity can be derived as follows. First note that, for pure fermionic state with entanglement, the random ensemble of pure states become random ensemble of local-parity preserving states given by \(\{\psi_{EE},\psi_{EO},\psi_{OE},\psi_{OO}\}\). With a probablity of $0.25$, one can choose any one of the four possible local-parity preserving state space with a normalized Haar measure within that subspace. Then for noisy part of the resource state one has the fidelity expression given by,
				\begin{eqnarray}
					F(\mathbb{I} \wedge \mathcal{E}) =\frac{1}{4}\Bigg[&&\int d\psi_{EE}  \Bigg( \min_{\{E_m^{B\tilde{A}'}\}} \sum_m \sqrt{\text{tr}\Bigg(\Bigg(\frac{\mathbb{I}_N \wedge\rho^f_{\tilde{A}'}}{d}+\Upsilon_1\frac{(\mathbb{S}_z\wedge\rho^f_{\tilde{A}'} )}{d} \Bigg)E_m^{B\tilde{A}'}\Bigg) \text{tr}(|\psi_{EE}\rangle \langle \psi_{EE} |E_m^{B\tilde{A}'})} \Bigg)^{2}\nonumber \\
					&&+  \int d\psi_{OO}\Bigg(\min_{\{E_m^{B\tilde{A}'}\}} \sum_m \sqrt{\text{tr}\Bigg(\Bigg(\frac{\mathbb{I}_N \wedge\rho^f_{\tilde{A}'}}{d}-\Upsilon_1\frac{(\mathbb{S}_z\wedge\rho^f_{\tilde{A}'} )}{d} \Bigg)E_m^{B\tilde{A}'}\Bigg)\text{tr}(|\psi_{OO}\rangle \langle \psi_{OO} |E_m^{B\tilde{A}'})}\Bigg)^{2}\nonumber \\
					&&+\int d\psi_{EO}  \Bigg( \min_{\{E_m^{B\tilde{A}'}\}} \sum_m \sqrt{\text{tr}\Bigg(\Bigg(\frac{\mathbb{I}_N\wedge\rho^f_{\tilde{A}'} }{d}+\Upsilon_1\frac{(\mathbb{S}_z\wedge\rho^f_{\tilde{A}'} )}{d} \Bigg)E_m^{B\tilde{A}'}\Bigg) \text{tr}(|\psi_{EO}\rangle \langle \psi_{EO} |E_m^{B\tilde{A}'})} \Bigg)^{2}\nonumber \\
					&&+  \int d\psi_{OE}\Bigg(\min_{\{E_m^{B\tilde{A}'}\}} \sum_m \sqrt{\text{tr}\Bigg(\Bigg(\frac{\mathbb{I}_N\wedge\rho^f_{\tilde{A}'} }{d}-\Upsilon_1\frac{(\mathbb{S}_z\wedge\rho^f_{\tilde{A}'} )}{d} \Bigg)E_m^{B\tilde{A}'}\Bigg)\text{tr}(|\psi_{OE}\rangle \langle \psi_{OE} |E_m^{B\tilde{A}'})}\Bigg)^{2}\Bigg]\nonumber
				\end{eqnarray}
				
				Minimum value of the integrands \cite{nielsen_chuang_2010} is achieved if \(\{E_m^{B\tilde{A}'}\}\) is chosen to be equal to the input. Hence, we get \begin{eqnarray}
					F(\mathbb{I} \wedge \mathcal{E})_{noisy}= &&\frac{1}{4}\Bigg[\int d\psi_{EE}\Bigg(\sqrt{\left\langle \psi_{EE}\left|\frac{\mathbb{I}_N\wedge\rho^f_{\tilde{A}'} }{d}+\Upsilon_1\frac{(\mathbb{S}_z\wedge\rho^f_{\tilde{A}'} )}{d}\right|\psi_{EE}\right\rangle}\Bigg)^{2}\nonumber\\
					&&+ \int d\psi_{OO}\Bigg(\sqrt{\left\langle \psi_{OO}\left|\frac{\mathbb{I}_N\wedge\rho^f_{\tilde{A}'} }{d}-\Upsilon_1\frac{(\mathbb{S}_z\wedge\rho^f_{\tilde{A}'} )}{d}\right|\psi_{OO}\right\rangle} \Bigg)^{2}\nonumber \\
					&&+\int d\psi_{EO}\Bigg(\sqrt{\left\langle \psi_{EO}\left|\frac{\mathbb{I}_N\wedge\rho^f_{\tilde{A}'} }{d}+\Upsilon_1\frac{(\mathbb{S}_z\wedge\rho^f_{\tilde{A}'} )}{d}\right|\psi_{EO}\right\rangle}\Bigg)^{2}\nonumber\\
					&&+ \int d\psi_{OE}\Bigg(\sqrt{\left\langle \psi_{OE}\left|\frac{\mathbb{I}_N\wedge\rho^f_{\tilde{A}'} }{d}-\Upsilon_1\frac{(\mathbb{S}_z\wedge\rho^f_{\tilde{A}'} )}{d}\right|\psi_{OE}\right\rangle} \Bigg)^{2}\Bigg]\nonumber \\
					=&&\frac{1}{4}\Bigg[\int d\psi_{EE}\frac{\text{tr}({\rho^f_{\tilde{A}'}}^2)}{d}(1+\Upsilon_1)+\int d\psi_{OO}\frac{\text{tr}({\rho^f_{\tilde{A}'}}^2)}{d}(1+\Upsilon_1)+\int d\psi_{EO}\frac{\text{tr}({\rho^f_{\tilde{A}'}}^2)}{d}(1+\Upsilon_1)\nonumber \\
					&&+\int d\psi_{OE}\frac{\text{tr}({\rho^f_{\tilde{A}'}}^2)}{d}(1+\Upsilon_1)\Bigg] 
					\label{eq:41}
				\end{eqnarray}
				
				To solve the Haar integral over the random ensemble of pure states with local-parity, we use the following encoding strategy:
				
				\textit{Example:-}Let us consider a state \[|\Psi_{EE}\rangle=\alpha|E_1,E_1\rangle+\beta|E_2,E_2\rangle\] Such state can always be mapped to a two-qubit state of the form \(\alpha|00\rangle+\beta|11\rangle)\). Similar mapping can always be done from \(d \times d\) dimensional fermionic state to \(\frac{d}{2} \times \frac{d}{2}\) dimensional state of SQT. 
				
				Extending this strategy for all the integral in Eq.(~\ref{eq:41}) and averaging via Hilbert-schmidt measure \cite{Karol2001} we get 
				\begin{eqnarray}
					F(\mathbb{I} \wedge \mathcal{E})_{noisy}=\frac{(1+\Upsilon_1)}{d}\times\frac{4d}{(d^2+4)}=\frac{4(1+\Upsilon_1)}{d^2+4}
					\label{eq:42}
				\end{eqnarray}

			\end{widetext}


			\nocite{*}
			
			\bibliography{apssamp}

@PREAMBLE{
	"\providecommand{\noopsort}[1]{}" 
	# "\providecommand{\singleletter}[1]{#1}%" 
}

@ARTICLE{PhysRevA.104.032411,
	author       = {Vidal, Nicetu Tibau and Bera, Mohit Lal and Riera, Arnau and Lewenstein, Maciej and Bera, Manabendra Nath},
	year         = "2021",
	journal      = "Phys.\ Rev.\ A",
	volume       = "104",
	pages        = "032411",
    url = {https://link.aps.org/doi/10.1103/PhysRevA.104.032411}
}

@article{PhysRev.88.101,
	author = {Wick, G. C. and Wightman, A. S. and Wigner, E. P.},
	journal = {Phys. Rev.},
	volume = {88},
	pages = {101--105},
	year = {1952},
    url = {https://link.aps.org/doi/10.1103/PhysRev.88.101}
}

@article{SSRfriis,
	author = {Nicolai Friis },
	journal = {New J. Phys.},
	volume = {18},
	pages = {033014},
	year = {2016},
    url={https://dx.doi.org/10.1088/1367-2630/18/3/033014}
}

@article{wolf1,
	author = {Mari-Carmen Bañuls and  J Ignacio Cirac and  Michael M Wolf},
	journal = {J. Phys.: Conf. Ser.},
	volume = {171},
	pages = {012032},
	year = {2009},
    url = {https://dx.doi.org/10.1088/1742-6596/171/1/012032},
}

@article{BENATTI20201,
	author = {F. Benatti and R. Floreanini and F. Franchini and U. Marzolino},
	journal = {Phys. Rep.},
	volume = {878},
	pages = {1},
	year = {2020},
    url={https://doi.org/10.1016/j.physrep.2020.07.003}
}

@article{D'Ariano_2014_2,
	author = {G. M. D'Ariano and F. Manessi and P. Perinotti and A. Tosini},
	journal = {Europhys. Lett.},
	volume = {107},
	pages = {20009},
	year = {2014},
    url = {https://dx.doi.org/10.1209/0295-5075/107/20009},
}

@ARTICLE{Dariano2014,
	author       =  {G.M. D’Ariano and F. Manessi and P. Perinotti and A. Tosini},
	
	year         = {2014},
	journal      = {Int. J. Mod. Phys. A},
	volume       = {29},
	pages        = {1430025},
    url = {https://doi.org/10.1142/S0217751X14300257}
}

@article{SSR2,
	author       = " M. Johansson", 
	title        = "Comment on 'Reasonable fermionic quantum information theories require relativity'", 
	journal        = "	arXiv:1610.00539 ", 
	
	month        = "", 
	year         = "",
    url={https://doi.org/10.48550/arXiv.1610.00539}
}

@article{f_teleport,
	author = {Debarba, Tiago and Iemini, Fernando and Giedke, Geza and Friis, Nicolai},
	journal = "Phys. \ Rev. \ A",
	volume = "101",
	pages = "052326",
	year = "2020",
    url = {https://link.aps.org/doi/10.1103/PhysRevA.101.052326}
}

@article{werner_teleport,
	author = {R F Werner},
	journal = {J. Phys. A: Math. Gen.},
	volume = {34},
	pages = {7081},
	year = {2001},
    url = {https://dx.doi.org/10.1088/0305-4470/34/35/332}
}

@article{dariano2010,
	author = {Chiribella, Giulio and D'Ariano, Giacomo Mauro and Perinotti, Paolo},
	journal = {Phys. Rev. A},
	volume = {84},
	pages = {012311},
	year = {2011},
    url = {https://link.aps.org/doi/10.1103/PhysRevA.84.012311}
}

@article{werner89,
	author = {Werner, Reinhard F.},
	journal = {Phys. Rev. A},
	volume = {40},
	pages = {4277},
	year = {1989},
    url = {https://link.aps.org/doi/10.1103/PhysRevA.40.4277}
}

@article{horodecki99,
	author = {Horodecki, Micha\l{} and Horodecki, Pawe\l{}},
	journal = {Phys. Rev. A},
	volume = {59},
	pages = {4206},
	year = {1999},
    url = {https://link.aps.org/doi/10.1103/PhysRevA.59.4206}
}

@article{horodecki99_2,
	author = {Horodecki, Micha\l{} and Horodecki, Pawe\l{} and Horodecki, Ryszard},
	journal = {Phys. Rev. A},
	volume = {60},
	pages = {1888},
	year = {1999},
    url = {https://link.aps.org/doi/10.1103/PhysRevA.60.1888}
}

@ARTICLE{epr,
	author       = "A. Einstein and {\relax Yu} Podolsky and N. Rosen", 
	collaboration = "EPR",
	year         = "1935", 
	journal      = "Phys.\ Rev.", 
	volume       = "47", 
	pages        = "777",
    url = {https://link.aps.org/doi/10.1103/PhysRev.47.777}
}

@ARTICLE{DiVincenzo_2002,
	author={DiVincenzo, D P. and Leung, D W. and Terhal, B M.},
	journal={IEEE Trans. Inf. Theory}, 
	title={Quantum data hiding}, 
	year={2002},
	volume={48},
	pages={580},
url={https://doi.org/10.1109/18.985948}
}

@book{nielsen_chuang_2010, place={Cambridge}, title={Quantum Computation and Quantum Information: 10th Anniversary Edition}, publisher={Cambridge University Press}, author={Nielsen, Michael A. and Chuang, Isaac L.}, year={2010}}

@book{Halmos1976,
	title     = "Measure Theory",
	author    = "Halmos, Paul R",
	publisher = "Springer",
	year      =  1976,
	
}

@article{friis2013,
	author = {Friis, Nicolai and Lee, Antony R. and Bruschi, David Edward},
	journal = {Phys. Rev. A},
	volume = {87},
	pages = {022338},
	year = {2013},
    url = {https://link.aps.org/doi/10.1103/PhysRevA.87.022338}
}

@article{bennett1993,
	author = {Bennett, Charles H. and Brassard, Gilles and Cr\'epeau, Claude and Jozsa, Richard and Peres, Asher and Wootters, William K.},
	journal = {Phys. Rev. Lett.},
	volume = {70},
	pages = {1895},
	year = {1993},
    url = {https://link.aps.org/doi/10.1103/PhysRevLett.70.1895}
}

@article{Boumeester1997,
	author = {Boumeester et al., D.},
	journal = {Nature (London)},
	volume = {390},
	pages = {575},
	year = {1997},
    url={https://doi.org/10.1038/37539}
}

@article{knill2001,
	author = {Knill et al., E.},
	journal = {Nature (London)},
	volume = {409},
	pages = {46},
	year = {2001},
    url={https://doi.org/10.1038/35051009}
}

@article{Gottesman1999,
	author = {Gottesman, Daniel and Chuang, Isaac L.},
	journal = {Nature (London)},
	volume = {402},
	pages = {390},
	year = {1999},
    url={https://doi.org/10.1038/46503}
}

@article{Langenfeld2021,
	author = {Langenfeld, Stefan and Welte, Stephan and Hartung, Lukas and Daiss, Severin and Thomas, Philip and Morin, Olivier and Distante, Emanuele and Rempe, Gerhard},
	journal = {Phys. Rev. Lett.},
	volume = {126},
	pages = {130502},
	year = {2021},
    url = {https://link.aps.org/doi/10.1103/PhysRevLett.126.130502}
}

@article{Avis2023,
	author = {Avis, Guus and Rozp\ifmmode \mbox{\k{e}}\else \k{e}\fi{}dek, Filip and Wehner, Stephanie},
	journal = {Phys. Rev. A},
	volume = {107},
	pages = {012609},
	year = {2023},
	month = {Jan},
    url = {https://link.aps.org/doi/10.1103/PhysRevA.107.012609}
}

@article{NISQ2022,
	author = {Bharti, Kishor and Cervera-Lierta, Alba and Kyaw, Thi Ha and Haug, Tobias and Alperin-Lea, Sumner and Anand, Abhinav and Degroote, Matthias and Heimonen, Hermanni and Kottmann, Jakob S. and Menke, Tim and Mok, Wai-Keong and Sim, Sukin and Kwek, Leong-Chuan and Aspuru-Guzik, Al\'an},
	journal = {Rev. Mod. Phys.},
	volume = {94},
	pages = {015004},
	year = {2022},
    url = {https://link.aps.org/doi/10.1103/RevModPhys.94.015004}
}

@article{Lohmann,
	author = {Lohmann, Bernd and Blum Karl},
	journal = {New J. Phys.},
	volume = {21},
	pages = {033025},
	year = {2019},
    url = {https://dx.doi.org/10.1088/1367-2630/ab045b},
}

@article{Ghirardi2004,
	author = {Ghirardi, GianCarlo and Marinatto, Luca},
	journal = {Phys. Rev. A},
	volume = {70},
	pages = {012109},
	year = {2004},
    url = {https://link.aps.org/doi/10.1103/PhysRevA.70.012109}
}

@article{Balachandran,
	author = {Balachandran, A. P. and Govindarajan, T. R. and de Queiroz, Amilcar R. and Reyes-Lega, A. F.},
	journal = {Phys. Rev. Lett.},
	volume = {110},
	pages = {080503},
	year = {2013},
	url = {https://link.aps.org/doi/10.1103/PhysRevLett.110.080503}
}

@article{zanardi,
	author = {Zanardi, Paolo},
	journal = {Phys. Rev. A},
	volume = {65},
	pages = {042101},
	year = {2002},
    url = {https://link.aps.org/doi/10.1103/PhysRevA.65.042101}
}

@article{Schliemann,
	author = {Schliemann, John and Cirac, J. Ignacio and Ku\ifmmode \acute{s}\else \'{s}\fi{}, Marek and Lewenstein, Maciej and Loss, Daniel},
	journal = {Phys. Rev. A},
	volume = {64},
	
	pages = {022303},
	numpages = {9},
	year = {2001},
    url = {https://link.aps.org/doi/10.1103/PhysRevA.64.022303}
}

@article{Iemini2013,
	author = {Iemini, Fernando and Vianna, Reinaldo O.},
	journal = {Phys. Rev. A},
	volume = {87},
	issue = {2},
	pages = {022327},
	numpages = {8},
	year = {2013},
	month = {Feb},
	publisher = {American Physical Society},
	doi = {10.1103/PhysRevA.87.022327},
	url = {https://link.aps.org/doi/10.1103/PhysRevA.87.022327}
}

@article{Li2001,
	author = {Li, Y. S. and Zeng, B. and Liu, X. S. and Long, G. L.},
	journal = {Phys. Rev. A},
	volume = {64},
	issue = {5},
	pages = {054302},
	numpages = {4},
	year = {2001},
	month = {Oct},
	publisher = {American Physical Society},
	doi = {10.1103/PhysRevA.64.054302},
	url = {https://link.aps.org/doi/10.1103/PhysRevA.64.054302}
}

@article{moriya2002,
	author = {Moriya, Hajime},
	journal = {Lett. Math. Phys.},
	volume = {60},
	pages = {109},
	year = {2002},
    url={https://doi.org/10.1023/A:1016158125660}
}

@article{Benatti14,
	author = {Benatti, F. and Floreanini, R. and Marzolino, U.},
	journal = {Phys. Rev. A},
	volume = {89},
	pages = {032326},
	year = {2014},
    url = {https://link.aps.org/doi/10.1103/PhysRevA.89.032326}
}

@article{Gigena17,
	author = {Gigena, N. and Rossignoli, R.},
	journal = {Phys. Rev. A},
	volume = {95},
	pages = {062320},
	year = {2017},
    url = {https://link.aps.org/doi/10.1103/PhysRevA.95.062320}
}

@article{Esposito23,
	author = {Ptaszy\ifmmode \acute{n}\else \'{n}\fi{}ski, Krzysztof and Esposito, Massimiliano},
	journal = {Phys. Rev. Lett.},
	volume = {130},
	pages = {150201},
	numpages = {6},
	year = {2023},
    url = {https://link.aps.org/doi/10.1103/PhysRevLett.130.150201}
}

@Article{Qiao2020,
	author={Qiao, Haifeng
	and Kandel, Yadav P.
	and Manikandan, Sreenath K.
	and Jordan, Andrew N.
	and Fallahi, Saeed
	and Gardner, Geoffrey C.
	and Manfra, Michael J.
	and Nichol, John M.},
	journal={Nat. Commun.},
	year={2020},
	volume={11},
	pages={3022},
    url={https://doi.org/10.1038/s41467-020-16745-0}
}

@Article{Steffen2013,
	author={Steffen, L.
	and Salathe, Y.
	and Oppliger, M.
	and Kurpiers, P.
	and Baur, M.
	and Lang, C.
	and Eichler, C.
	and Puebla-Hellmann, G.
	and Fedorov, A.
	and Wallraff, A.},
	journal={Nature (London)},
	year={2013},
	volume={500},
	number={7462},
	pages={319},
    url={https://doi.org/10.1038/nature12422}
}

@article{TAN2023115565,
	journal = {Physica E},
	volume = {147},
	pages = {115565},
	year = {2023},
	author = {Xiao-Dong Tan and Ya-Feng Song and Li-Jun Li and Le Zhang},
    url= {https://doi.org/10.1016/j.physe.2022.115565}
}

@article{Oszmaniec2022,
	author = {Oszmaniec, Micha\l{} and Dangniam, Ninnat and Morales, Mauro E.S. and Zimbor\'as, Zolt\'an},
	journal = {PRX Quantum},
	volume = {3},
	issue = {2},
	pages = {020328},
	numpages = {54},
	year = {2022},
	month = {May},
	url = {https://link.aps.org/doi/10.1103/PRXQuantum.3.020328}
}

@article{vidal2002,
	author = {Vidal, G. and Werner, R. F.},
	journal = {Phys. Rev. A},
	volume = {65},
	issue = {3},
	pages = {032314},
	numpages = {11},
	year = {2002},
    url = {https://link.aps.org/doi/10.1103/PhysRevA.65.032314}
	
}

@article{Alfsen1963,
	author = {Alfsen, E.M.},
	journal = {Math. Scand.},
	pages = {106},
	url = {http://eudml.org/doc/165837},
	volume = {12},
	year = {1963},
}

@UNPUBLISHED{Gleason,
	author       = " Jonathan Gleason", 
	title        = "Existence and uniqueness of Haar measure", 
	note         = "	preprint ", 
	
	year         = "2010",
}

@Article{Ding2021,
	author={Ding, Lexin
	and Mardazad, Sam
	and Das, Sreetama
	and Szalay, Szil{\'a}rd
	and Schollw{\"o}ck, Ulrich
	and Zimbor{\'a}s, Zolt{\'a}n
	and Schilling, Christian},
	journal={J. Chem. Theory Comput.},
	year={2021},
	volume={17},
	pages={79},
    url={https://doi.org/10.1021/acs.jctc.0c00559}
	
}

@Article{moore1991,
	author={Moore, Gregory and Read, Nicholas },
	journal={Nucl. Phys. B 360, 362 (1991).},
	year={1991},
	volume={360},
	pages={362},
    url={https://doi.org/10.1016/0550-3213(91)90407-O}
}

@article{Nayak2013,
	author = {Willett, R. L. and Nayak, C. and Shtengel, K. and Pfeiffer, L. N. and West, K. W.},
	journal = {Phys. Rev. Lett.},
	volume = {111},
	issue = {18},
	pages = {186401},
	numpages = {5},
	year = {2013},
    url = {https://link.aps.org/doi/10.1103/PhysRevLett.111.186401}
}

@article{Nayak2005,
	author = {Das Sarma, Sankar and Freedman, Michael and Nayak, Chetan},
	journal = {Phys. Rev. Lett.},
	volume = {94},
	issue = {16},
	pages = {166802},
	numpages = {4},
	year = {2005},
    url = {https://link.aps.org/doi/10.1103/PhysRevLett.94.166802}
}

@article{Kitaev_2001,
	year = {2001},
	volume = {44},
	number = {10S},
	pages = {131},
	author = {A Yu Kitaev},
	journal = {Phys. Usp.},
    url = {https://dx.doi.org/10.1070/1063-7869/44/10S/S29},
}

@article{Mourik2012,
	author = {V. Mourik  and K. Zuo  and S. M. Frolov  and S. R. Plissard  and E. P. A. M. Bakkers  and L. P. Kouwenhoven },
	journal = {Science},
	volume = {336},
	number = {6084},
	pages = {1003},
	year = {2012},
    url= {https://www.science.org/doi/abs/10.1126/science.1222360}
}

@article{sarma2010,
	author = {Lutchyn, Roman M. and Sau, Jay D. and Das Sarma, S.},
	journal = {Phys. Rev. Lett.},
	volume = {105},
	issue = {7},
	pages = {077001},
	numpages = {4},
	year = {2010},
    url = {https://link.aps.org/doi/10.1103/PhysRevLett.105.077001}
}

@article{tanaka2024,
	author = {Tanaka, Yukio and Tamura, Shun and Cayao, Jorge},
	journal = {Progress of Theoretical and Experimental Physics},
	volume = {2024},
	number = {8},
	pages = {08C105},
	year = {2024},
    url={https://doi.org/10.1093/ptep/ptae065}
}

@article{pan2021,
	author = {Huang, He-Liang and Naro\ifmmode \dot{z}\else \.{z}\fi{}niak, Marek and Liang, Futian and Zhao, Youwei and Castellano, Anthony D. and Gong, Ming and Wu, Yulin and Wang, Shiyu and Lin, Jin and Xu, Yu and Deng, Hui and Rong, Hao and Dowling, Jonathan P. and Peng, Cheng-Zhi and Byrnes, Tim and Zhu, Xiaobo and Pan, Jian-Wei},
	journal = {Phys. Rev. Lett.},
	volume = {126},
	pages = {090502},
	numpages = {7},
	year = {2021},
    url = {https://link.aps.org/doi/10.1103/PhysRevLett.126.090502}
}

@article{zhou2023,
	author = {Xu, Cheng-Qian and Zhou, D. L.},
	journal = {Phys. Rev. A},
	volume = {108},
	issue = {5},
	pages = {052221},
	numpages = {7},
	year = {2023},
    url = {https://link.aps.org/doi/10.1103/PhysRevA.108.052221}
}

@article{BONDERSON2017,
	journal = {Ann. Phys. (NY)},
	volume = {385},
	pages = {399},
	year = {2017},
	author = {Parsa Bonderson and Christina Knapp and Kaushal Patel},
    url={https://doi.org/10.1016/j.aop.2017.07.018}
	
}

@article{Dankert2009,
	author = {Dankert, Christoph and Cleve, Richard and Emerson, Joseph and Livine, Etera},
	journal = {Phys. Rev. A},
	volume = {80},
	issue = {1},
	pages = {012304},
	numpages = {6},
	year = {2009},
	month = {Jul},
	publisher = {American Physical Society},
	doi = {10.1103/PhysRevA.80.012304},
	url = {https://link.aps.org/doi/10.1103/PhysRevA.80.012304}
}

@article{Karol2001,
	doi = {10.1088/0305-4470/34/35/335},
	url = {https://dx.doi.org/10.1088/0305-4470/34/35/335},
	year = {2001},
	month = {aug},
	publisher = {},
	volume = {34},
	number = {35},
	pages = {7111},
	author = {Karol Zyczkowski and Hans-Jürgen Sommers},
	journal = {J. Phys. A: Math. Gen.},		
}

@article{Verstraete2001,
	author = {Verstraete, Frank and Dehaene, Jeroen and DeMoor, Bart},
	journal = {Phys. Rev. A},
	volume = {64},
	issue = {1},
	pages = {010101},
	numpages = {4},
	year = {2001},
	month = {Jun},
	publisher = {American Physical Society},
	doi = {10.1103/PhysRevA.64.010101},
	url = {https://link.aps.org/doi/10.1103/PhysRevA.64.010101}
}

@article{horodecki2000,
	author = {Badzia\ifmmode \mbox{\c{}}\else \c{}\fi{}g, Piotr and Horodecki, Micha\l{} and Horodecki, Pawe\l{} and Horodecki, Ryszard},
	journal = {Phys. Rev. A},
	volume = {62},
	issue = {1},
	pages = {012311},
	numpages = {7},
	year = {2000},
	month = {Jun},
	publisher = {American Physical Society},
	doi = {10.1103/PhysRevA.62.012311},
	url = {https://link.aps.org/doi/10.1103/PhysRevA.62.012311}
}

@article{ghosal2025,
  author = {Ghosal, Arkaprabha and Ghai, Jatin and Saha, Tanmay and Ghosh, Sibasish and Alimuddin, Mir},
  journal = {Phys. Rev. Lett.},
  volume = {134},
  issue = {16},
  pages = {160803},
  numpages = {7},
  year = {2025},
  month = {Apr},
  publisher = {American Physical Society},
  doi = {10.1103/PhysRevLett.134.160803},
  url = {https://link.aps.org/doi/10.1103/PhysRevLett.134.160803}
}

@article{pssr1968,
	author = {Hegerfeldt, G. C. and Kraus, K. and Wigner, E. P.},
	journal = {J. Math. Phys.},
	volume = {9},
	number = {12},
	pages = {2029},
	year = {1968},
	url = {https://doi.org/10.1063/1.1664539},
}

@article{mzm2015,
	author = {Sarma, Sankar Das and Freedman, Michael and Nayak, Chetan},
	journal = {npj Quantum Inf},
	number = {1},
	pages = {15001},
	url = {https://doi.org/10.1038/npjqi.2015.1},
	volume = {1},
	year = {2015},
}

@article{Jaynes1,
	author = {Jaynes, E. T.},
	journal = {Phys. Rev.},
	volume = {106},
	issue = {4},
	pages = {620--630},
	numpages = {0},
	year = {1957},
	month = {May},
	url = {https://link.aps.org/doi/10.1103/PhysRev.106.620}
}

@article{Jaynes2,
	author = {Jaynes, E. T.},
	journal = {Phys. Rev.},
	volume = {108},
	issue = {2},
	pages = {171--190},
	numpages = {0},
	year = {1957},
	month = {Oct},
	url = {https://link.aps.org/doi/10.1103/PhysRev.108.171}
}

@article{zhou2022,
	author = {Xu, Cheng-Qian and Zhou, D. L.},
	journal = {Phys. Rev. A},
	volume = {106},
	issue = {1},
	pages = {012413},
	numpages = {10},
	year = {2022},
	month = {Jul},
	publisher = {American Physical Society},
	doi = {10.1103/PhysRevA.106.012413},
	url = {https://link.aps.org/doi/10.1103/PhysRevA.106.012413}
}

@article{HORODECKI199621,
	journal = {Phys. Lett. A},
	volume = {222},
	number = {1},
	pages = {21},
	year = {1996},
	doi = {https://doi.org/10.1016/0375-9601(96)00639-1},
	url = {https://www.sciencedirect.com/science/article/pii/0375960196006391},
	author = {Ryszard Horodecki and Michał Horodecki and Paweł Horodecki},
}

@article{gong2024,
	author = {Gong, Neng-Fei and Cai, Dun-Bo and Huang, Zhi-Guo and Qian, Ling and Zhang, Run-Qing and Hu, Xiao-Min and Liu, Bi-Heng and Wang, Tie-Jun},
	journal = {Phys. Rev. Appl.},
	volume = {22},
	issue = {5},
	pages = {054045},
	numpages = {26},
	year = {2024},
	doi = {10.1103/PhysRevApplied.22.054045},
	url = {https://link.aps.org/doi/10.1103/PhysRevApplied.22.054045}
}

@article{galler2021,
  author = {Galler, Anna and Thunstr\"om, Patrik},
  journal = {Phys. Rev. Res.},
  volume = {3},
  issue = {3},
  pages = {033120},
  numpages = {17},
  year = {2021},
  month = {Aug},
  publisher = {American Physical Society},
  doi = {10.1103/PhysRevResearch.3.033120},
  url = {https://link.aps.org/doi/10.1103/PhysRevResearch.3.033120}
}

@article{small1999,
  title = {Quantum Information Processing Using Quantum Dot Spins and Cavity QED},
  author = {Imamog\ifmmode\bar\else\textasciimacron\fi{}lu, A. and Awschalom, D. D. and Burkard, G. and DiVincenzo, D. P. and Loss, D. and Sherwin, M. and Small, A.},
  journal = {Phys. Rev. Lett.},
  volume = {83},
  issue = {20},
  pages = {4204--4207},
  numpages = {0},
  year = {1999},
  month = {Nov},
  publisher = {American Physical Society},
  doi = {10.1103/PhysRevLett.83.4204},
  url = {https://link.aps.org/doi/10.1103/PhysRevLett.83.4204}
}

@article{wolf2007,
  author = {Ba\~nuls, Mari-Carmen and Cirac, J. Ignacio and Wolf, Michael M.},
  journal = {Phys. Rev. A},
  volume = {76},
  issue = {2},
  pages = {022311},
  numpages = {13},
  year = {2007},
  month = {Aug},
  publisher = {American Physical Society},
  doi = {10.1103/PhysRevA.76.022311},
  url = {https://link.aps.org/doi/10.1103/PhysRevA.76.022311}
}

@article{Bertlmann_2008,
url = {https://dx.doi.org/10.1088/1751-8113/41/23/235303},
year = {2008},
month = {may},
publisher = {},
volume = {41},
number = {23},
pages = {235303},
author = {Bertlmann, Reinhold A and Krammer, Philipp},
journal = {J. Phys. A: Math. Theor.},
}

@misc{inprep,
  note = {manuscript in preparation}
}

@article{
pnas2304294120,
author = {D. González-Cuadra  and D. Bluvstein  and M. Kalinowski  and R. Kaubruegger  and N. Maskara  and P. Naldesi  and T. V. Zache  and A. M. Kaufman  and M. D. Lukin  and H. Pichler  and B. Vermersch  and Jun Ye  and P. Zoller },
journal = {Proceedings of the National Academy of Sciences},
volume = {120},
number = {35},
pages = {e2304294120},
year = {2023},
doi = {10.1073/pnas.2304294120},
URL = {https://www.pnas.org/doi/abs/10.1073/pnas.2304294120},
eprint = {https://www.pnas.org/doi/pdf/10.1073/pnas.2304294120}
}

@article{zkpl-hh28,
 author = {Ott, Robert and Gonz\'alez-Cuadra, Daniel and Zache, Torsten V. and Zoller, Peter and Kaufman, Adam M. and Pichler, Hannes},
  journal = {Phys. Rev. Lett.},
  volume = {135},
  issue = {9},
  pages = {090601},
  numpages = {7},
  year = {2025},
  month = {Aug},
  publisher = {American Physical Society},
  doi = {10.1103/zkpl-hh28},
  url = {https://link.aps.org/doi/10.1103/zkpl-hh28}
}

@misc{schuckert2025,
      author={Alexander Schuckert and Eleanor Crane and Alexey V. Gorshkov and Mohammad Hafezi and Michael J. Gullans},
      year={2025},
      eprint={2411.08955},
      archivePrefix={arXiv},
      primaryClass={quant-ph},
      url={https://arxiv.org/abs/2411.08955}, 
}
			
		\end{document}